\documentclass[a4paper, 11pt]{article}
\usepackage{graphicx}
\usepackage{natbib}
\usepackage{amsmath, amsfonts, amsthm, bm, mathrsfs}
\usepackage{multicol}
\usepackage{color}

\usepackage{algorithm}
\usepackage{algorithmic}
\usepackage{natbib}
\usepackage{booktabs} % 表の罫線の制御用

\usepackage[margin=15mm]{geometry} % 余白少なめの原稿に設定

\usepackage{authblk} % \affilの使用のため

\newtheorem{Theo}{Theorem}
\newtheorem{Lemm}{Lemma}
\newtheorem{prop}{Proposition}

\newtheorem{Cond}{Condition}
\newtheorem{Assumption}{Assumption}

\theoremstyle{definition}
\newtheorem{example}{Example}

\theoremstyle{remark}
\newtheorem{Rem}{Remark}

\newcommand{\bs}{\boldsymbol}
\newcommand{\mb}{\mathbf}

\newcommand{\mac}{\mathcal}
\newcommand{\mbb}{\mathbb}
\newcommand{\mcr}{\mathscr}

\newcommand{\ra}{\rightarrow}

\newcommand{\lan}{\langle}
\newcommand{\ran}{\rangle}
\newcommand{\fr}{\frac{1}{2}}

\DeclareMathOperator{\tr}{tr}

\DeclareMathOperator{\Var}{Var}
\DeclareMathOperator{\Cov}{Cov}
\DeclareMathOperator{\Corr}{Corr}
\DeclareMathOperator{\dom}{dom}

\DeclareMathOperator{\Span}{span}

\DeclareMathOperator{\im}{Im}
\DeclareMathOperator{\myspan}{span}

\allowdisplaybreaks

\def\ba#1\ea{\begin{align*}#1\end{align*}} %\ba = \begin{algin*}, \ea = \end{align*}
	\def\banum#1\eanum{\begin{align}#1\end{align}} %\banum = \begin{algin}, \eanum

\newif\ifdraft
\draftfalse % 出版モード

\begin{document}

\title{Functional Multiple-Set Canonical Correlation Analysis Revisited: From Finite-Dimensional Samples to Infinite-Dimensional Populations}
\author[1, 3, 4]{Michio Yamamoto} % 滋賀大のも含めたもの
% \author[1, 3]{Michio Yamamoto}
\author[2, 3]{Yoshikazu Terada}
\affil[1]{\small Graduate School of Human Sciences, The University of Osaka}
\affil[2]{\small Graduate School of Engineering Science, The University of Osaka}
\affil[3]{\small RIKEN AIP}
\affil[4]{\small Data Science and AI Innovation Research Promotion Center, Shiga University}

\maketitle

\begin{abstract}
  We develop a population-level formulation of functional multiple-set canonical correlation analysis (P-FMCCA) 
  for multivariate functional data in an infinite-dimensional Hilbert space.
  Since covariance operators for functional data are typically compact, 
  the inverse covariance operators that appear in the formal extension of multiple-set canonical correlation analysis (MCCA) 
  are generally unbounded and are not defined on the whole Hilbert space.
  We therefore provide sufficient conditions under which the proposed population formulation is well-defined 
  and show that the resulting constrained maximization problem is characterized 
  by an eigenvalue problem for a Hilbert--Schmidt extension of the relevant correlation operator.
  We further establish a canonical decomposition induced by the P-FMCCA components 
  and introduce the associated truncated canonical representation.
  In addition, we formulate functional homogeneity analysis at the population level 
  and show that the finite-dimensional equivalence between homogeneity analysis and MCCA 
  does not generally carry over to the infinite-dimensional setting.
  Finally, we prove that, for the finite-rank truncated canonical representation, 
  functional homogeneity analysis admits an explicit characterization in terms of the P-FMCCA components, 
  thereby providing a population-level counterpart of the classical finite-dimensional correspondence.
\end{abstract}

\vspace{25pt}

\noindent Keywords: Functional data analysis; Functional canonical correlation analysis; Multiple-set canonical correlation analysis; Canonical decomposition; Homogeneity analysis

\section{Introduction}

Longitudinal, spatial, spectral, and other densely observed data are increasingly collected in the form of curves or functions over continua such as time, space, or wavelength. 
Functional data analysis provides a natural framework for such data by regarding discretely observed measurements as realizations of underlying smooth stochastic processes \citep{RamsaySilverman2005,HorvathKokoszka2012}.
This framework has been widely used to extend statistical methodology from finite-dimensional vectors to functional observations. 
While many functional data methods are designed for situations in which each observational unit is represented by a single function, modern applications often involve several related functions observed for the same subject or experimental unit. 
Examples include multiple longitudinal biomarkers measured jointly in clinical studies, multimodal neuroimaging signals, and several functional measurements obtained under different experimental conditions \citep{HwangEtAl2012,ChiouMuller2014,HappGreven2018,SortEtAl2024}.
Such observations are often referred to as multivariate functional data. 
For these data, it is not sufficient to describe the variation of each functional variable separately; one also needs to characterize the association structure among the functions.

Canonical correlation analysis (CCA) is a classical method for studying associations between two sets of variables \citep{Hotelling1936}. 
Its extensions to more than two sets, such as multiple-set canonical correlation analysis (MCCA) and generalized canonical correlation analysis (GCCA), provide a natural framework for integrating information from several blocks of variables \citep{Horst1961,Carroll1968,Kettenring1971}. 
These methods construct low-dimensional components that summarize the association structure among multiple variable sets. 
Therefore, extending CCA and MCCA-type methods to multivariate functional data is an important step toward understanding dependence among multiple functional variables.

Functional CCA for pairs of curves has been studied from several complementary perspectives. 
\cite{leurgans1993canonical} showed that the naive unregularized sample formulation breaks down in infinite-dimensional settings and proposed a roughness-penalized approach. 
At the population level, \cite{he2003functional} investigated the existence of functional canonical correlations and weight functions for pairs of square-integrable stochastic processes and established a canonical decomposition. 
\cite{cupidon2008some} studied canonical correlations and variates in infinite-dimensional Hilbert spaces through regularization, whereas \cite{eubank2008canonical} developed a formulation based on reproducing kernel Hilbert spaces. 
A related covariance-based approach is functional singular component analysis, which uses the singular system of the cross-covariance operator to characterize dependence between paired functional variables \citep{yang2011functional}. 

For more than two functional variables, \citet{HwangEtAl2012} introduced functional multiple-set canonical correlation analysis (FMCCA), a functional analogue of MCCA. 
Their method seeks weight functions for several sets of curves so that the resulting canonical variates are highly associated across the sets. 
Computationally, the method is implemented by representing both data and weight functions through finite-dimensional basis expansions. 
\citet{GoreckiEtAl2020} proposed GCCA for multivariate functional data, also reducing the problem  to an ordinary finite-dimensional GCCA problem through basis representation.
More recently, \citet{SortEtAl2024} proposed functional generalized canonical correlation analysis for multiple longitudinal variables, using regularization to accommodate sparse and irregular longitudinal observations. 

Despite this substantial literature on bivariate functional CCA, a population theory remains incomplete for the multiple-set criterion underlying FMCCA. 
In particular, the bivariate results above do not directly provide conditions for the well-definedness of the single block operator on the product Hilbert space that arises under the global block-diagonal normalization used by FMCCA.
% Despite these important developments, a basic population-level question remains insufficiently understood. 
Functional data are usually assumed to reside in infinite-dimensional Hilbert spaces, and covariance operators on such spaces are typically compact. 
Hence, their inverses are generally unbounded and cannot be applied to arbitrary elements of the underlying Hilbert space. 
As a consequence, the usual formal analogy with finite-dimensional CCA or MCCA does not by itself guarantee that the population canonical variates are well-defined.
\citet{HwangEtAl2012} and \citet{GoreckiEtAl2020} avoid this difficulty by working with finite-dimensional basis representations.
The approach of \citet{SortEtAl2024}, on the other hand, uses a regularized formulation to make the relevant constraint operators invertible. 
However, without such finite-dimensional reduction or explicit regularization, it remains unclear under what conditions a population version of CCA for multiple functional variables is well-defined in the original infinite-dimensional space.

The purpose of this paper is to provide a population-level formulation of functional multiple-set canonical correlation analysis 
for multivariate functional data and to investigate its theoretical structure. 
We call the proposed framework population functional multiple-set canonical correlation analysis (P-FMCCA), 
emphasizing that it provides a population-level and infinite-dimensional counterpart of the sample-based FMCCA introduced by \citet{HwangEtAl2012}. 
The proposed framework formulates canonical variates directly in the population, rather than as quantities defined only from a finite sample. 
The resulting optimization problem is closely related to an operator eigenvalue problem, but its validity depends on whether the relevant inverse covariance operators are well-defined. 
Thus, a central task of this paper is to identify conditions under which the population formulation is mathematically meaningful in an infinite-dimensional space.

The contributions of this paper are fourfold.
First, we define population functional multiple-set canonical correlation analysis on an infinite-dimensional Hilbert space.
We provide sufficient conditions under which the relevant operator formulation is well-defined, 
and the resulting constrained maximization problem is characterized by an eigenvalue problem.
Second, we establish a canonical decomposition for multivariate functional data 
and introduce the corresponding truncated canonical representation.
This representation clarifies how successive canonical components capture cross-functional dependence and 
provides an interpretable low-dimensional representation of the data.
Third, we formulate a population version of homogeneity analysis for multivariate functional data (FHOMALS) 
and examine its relationship with P-FMCCA.
We show that the classical finite-dimensional equivalence between homogeneity analysis and MCCA does not automatically carry over to the relationship between FHOMALS and P-FMCCA in the infinite-dimensional setting.
Fourth, we characterize the FHOMALS solution for the finite-rank truncated canonical representation.
We show that, for this truncated process, FHOMALS is equivalent to P-FMCCA defined from the same covariance structure, thereby establishing a population-level counterpart of the finite-dimensional equivalence between homogeneity analysis and MCCA.

The remainder of this paper is organized as follows. 
Section~\ref{sec:P-FMCCA} introduces the notation, data structure, and formulation of P-FMCCA. 
Section~\ref{sec:existence} gives sufficient conditions for the well-definedness of P-FMCCA and shows that the population problem reduces to an eigenvalue problem under these conditions.
Section~\ref{sec:canonical_decomposition} develops the canonical decomposition and discusses its properties. 
Section~\ref{sec:homogeneity_analysis} formulates FHOMALS and clarifies its relationship with P-FMCCA, including the explicit FHOMALS solution for the truncated canonical representation.
Section~\ref{sec:concluding-remarks} concludes with a discussion of theoretical implications and possible extensions.

\section{Population formulation of functional multiple-set canonical correlation analysis}
\label{sec:P-FMCCA}

% Then, we formulate CCA for multivariate functional data in the population.
% We also discuss how the proposed population formulation relates to the sample-based FMCCA of \citet{HwangEtAl2012}.

We first introduce the notation and data structure used in the proposed framework.
% Then, we formulate P-FMCCA at the population level and relate it to the sample-based formulation of \citet{HwangEtAl2012}.
Then, we review the sample-based formulation of \citet{HwangEtAl2012}, which serves as the starting point for our population-level formulation of P-FMCCA.
Proofs of all propositions are provided in the Supplementary Material.

\subsection{Notation and data structure}
\label{subsec:notation-data_structure}

We first introduce the notation and data structure used in the proposed framework.
For any positive integer $M$, let $[M]$ denote the set $\{1,\dots,M\}$.
Let $P\in\mbb{N}$ denote the number of functional variables.
For $p\in[P]$, let $\mac{T}_p$ be a compact interval in $\mbb{R}$ and write $\mac{T}^P:=\prod_{p=1}^P\mac{T}_p$.
Let $L_2(\mac{T}_p)$ denote the Hilbert space of square-integrable functions on $\mac{T}_p$ with respect to the Lebesgue measure $\mu_p$.
With respect to $\mu_p$, the inner product on $L_2(\mac{T}_p)$ is defined by $\lan x,y\ran_{(p)}=\int_{\mac{T}_p} x(t)y(t)d\mu_p(t)$, which induces the norm $\|x\|_{(p)}=\lan x,x\ran_{(p)}^{\frac{1}{2}}$.
We define the product Hilbert space $L_2^P:=\prod_{p=1}^P L_2(\mac{T}_p)$, equipped with the inner product $\lan \bs{x},\bs{y}\ran_P
=\sum_{p=1}^P\lan x_p,y_p\ran_{(p)}$ for $\bs{x}=(x_1,\dots,x_P)^\top$ and $\bs{y}=(y_1,\dots,y_P)^\top$.
The associated norm is given by $\|\bs{x}\|_P=\lan \bs{x},\bs{x}\ran_P^{\frac{1}{2}}$.
The product measure of $\mu_1,\dots,\mu_P$ is denoted by $\mu^P$.
% Denote by $L_2(\mac{T}_p)$ the space of equivalence classes of square integrable functions on $\mac{T}_p$ with respect to the Lebesgue measure $\mu_p$, and set $L_2^P:=L_2^P(\mac{T}^P):=\prod_{p=1}^P L_2(\mac{T}_p)$.
% The product measure of $\mu_1,\dots,\mu_P$ is denoted by $\mu^P$.
% With respect to $\mu_p$, the inner product on $L_2(\mac{T}_p)$ is defined by
% $
%   \lan x,y\ran_{p}
%   =\int_{\mac{T}_p} x(t)y(t)d\mu_p(t),\ x,y\in L_2(\mac{T}_p),
% $
% % \ba
% %   \lan x,y\ran_{p}
% %   =\int_{\mac{T}_p} x(t)y(t)d\mu_p(t),\ x,y\in L_2(\mac{T}_p),
% % \ea
% which induces the norm $\|x\|_{p}=\lan x,x\ran_{p}^{\frac{1}{2}}$.
For notational simplicity, throughout this paper, we assume that the domains $\mac{T}_p$ and measures $\mu_p$ do not depend on $p$, and we write them as  $\mac{T}$ and $\mu$, respectively.
Also, every integral is taken over $\mac {T}$, and we suppress the domain $\mac{T}$ in the integral symbol.
Accordingly, we omit the subscript and write the inner product and the norm simply as $\lan x,y\ran$ and $\|x\|$, respectively.
% Define an inner product on $L_2^P$ by $\lan \bs{x},\bs{y}\ran_P
%   =\sum_{p=1}^P\lan x_p,y_p\ran$
% % \ba
% %   \lan \bs{x},\bs{y}\ran_P
% %   =\sum_{p=1}^P\lan x_p,y_p\ran,
% % \ea
% where $\bs{x}=(x_1,\dots,x_P)^\top$ and $\bs{y}=(y_1,\dots,y_P)^\top$.
Similarly, we denote $L_2(\mac{T})$ as $L_2$.

We consider multivariate functional data, that is, each observation is a vector of functions, $\bs{X}(\bs{t})=(X_1(t_1),\dots,X_P(t_P))^\top\in L_2^P$ with $\bs{t}=(t_1,\dots,t_P)^\top\in\mac{T}^P$.
Here, for $p\in[P]$, $X_p$ is a second-order stochastic process defined on $\mac{T}$ that is jointly measurable.
Without loss of generality, we assume that the mean function is zero: $\mbb{E}[\bs{X}(\bs{t})]
  =(\mbb{E}[X_1(t_1)],\dots,\mbb{E}[X_P(t_P)])^\top
  =0$
% \ba
%   \mbb{E}[\bs{X}(\bs{t})]
%   =(\mbb{E}[X_1(t_1)],\dots,\mbb{E}[X_P(t_P)])^\top
%   =0
% \ea
for all $\bs{t}\in\mac{T}^P$.
We introduce a tensor product notation: for any $\theta,\phi\in L_2$, an operator $\theta\otimes\phi:L_2\ra L_2$ is given by $(\theta\otimes\phi)(h)
  =\lan h,\theta\ran\phi$, for $h\in L_2.$
Also, for any $\bs{\theta},\bs{\phi}\in L_2^P$, an operator $\bs{\theta}\otimes_P\bs{\phi}:L_2^P\ra L_2^P$ is defined as $(\bs{\theta}\otimes_P\bs{\phi})(\bs{h})=\lan\bs{h},\bs{\theta}\ran_P\bs{\phi}$ for $\bs{h}=(h_1,\dots,h_P)^\top\in L_2^P$.

For $\bs{s},\bs{t}\in\mac{T}^P$, define the matrix of covariance functions $C(\bs{s},\bs{t}):=\mbb{E}[\bs{X}(\bs{s})\bs{X}(\bs{t})^\top]\in\mbb{R}^{P\times P}$ with elements $C_{ij}(s_i,t_j)
  =\Cov(X_i(s_i),X_j(t_j))
  =\mbb{E}[X_i(s_i)X_j(t_j)]$.
% \ba
%   C_{ij}(s_i,t_j)
%   =\Cov(X_i(s_i),X_j(t_j))
%   =\mbb{E}[X_i(s_i)X_j(t_j)].
% \ea
Then, the cross-covariance operator $\mac{V}_{pq}$ for variables $p$ and $q$ is defined as $\mac{V}_{pq}h(s)
 =\mbb{E}[\left<h,X_{q}\right>X_{p}]
 =\int C_{pq}(s,t)h(t)d\mu(t)$
% \begin{equation*}
%  \mac{V}_{pq}h(t)
%  =\mbb{E}[\left<h,X_{q}\right>X_{p}]
%  =\int C_{pq}(t,s)h(s)d\mu(s)
% %  =\mbb{E}[X_{q}\otimes X_{p}](h)
% \end{equation*}
for $h\in L_2$.

Define the covariance operator $\mac{V}:L_2^P\ra L_2^P$ componentwise by
% $
%   (\mac{V}\bs{h})(\bs{s})
%   =\int C(\bs{s},\bs{t})\bs{h}(\bs{t})d\mu^P(\bs{t})
% $
% for $\bs{h}\in L_2^P$.
% follows:
% \ba
%   (\mac{V}\bs{h})(\bs{t})
%   &=\int C(\bs{s},\bs{t})\bs{h}(\bs{s})d\mu(\bs{s}),\ 
%   \text{for}\ \bs{h}\in L_2^P.
%   % &=\int_{\mac{T}}
%   % \begin{bmatrix}
%   %   \sum_{q=1}^{P}\mbb{E}[X_{1}(t_1)X_{q}(s_q)]h_{q}(s_q)\\
%   %   \vdots\\
%   %   \sum_{q=1}^{P}\mbb{E}[X_{P}(t_P)X_{q}(s_q)]h_{q}(s_q)
%   % \end{bmatrix}
%   % d\mu(\bs{s})
% \ea
% Note that the $p$th component of $\mac{V}\bs{h}$ is given by
\ba
  (\mac{V}\bs{h}(s_p))_{p}
  =\sum_{q=1}^P\mac{V}_{pq}h_q(s_p)
  =\sum_{q=1}^P\int C_{pq}(s_p,t_q)h_{q}(t_q)d\mu(t_q),
  \qquad p\in[P].
\ea
Here, for a vector $\bs{v}$, $(\bs{v})_p$ denotes its $p$th component.
We assume that $\mac{V}$ is not the zero operator; otherwise, $\bs{X}$ is degenerate at zero in $L_2^P$, and the correlation-based analysis described below would be meaningless.

\begin{Assumption}
 \label{assumption:square_integrable} 
 Stochastic processes $\{X_p\}_{p=1,\dots,P}$ are mean-square continuous.
\end{Assumption}
The following proposition shows that this assumption ensures that the covariance operator of $\bs{X}$ admits an eigenvalue decomposition.
The corresponding eigenvectors, together with a complete orthonormal system (CONS) of the null space of $\mac{V}$, form a CONS of $L_2^P$.
This assumption also ensures that each $X_p(t)$ has a Karhunen-Lo\`eve (KL) expansion, which is used in deriving the theoretical results below.
Note also that this assumption implies that $X_p$ has the finite second moment, i.e., $\mbb{E}[\|X_p\|^2]<\infty$, which is equivalent to $\mbb{E}[\|\bs{X}\|_P^2]<\infty$.
We assume that Assumption \ref{assumption:square_integrable} holds throughout this paper.

\begin{prop}
  \label{prop:V-HS}  
  % Under Assumption \ref{assumption:square_integrable}, t
  The covariance operators $\{\mac{V}_{pq}\}_{1\le p,q\le P}$ and $\mac{V}$ are trace-class. 
\end{prop}
% Proofs for all propositions are given in the Supplementary Material.

Under Assumption \ref{assumption:square_integrable}, $X_{p}$, the $p$th element of $\bs{X}\in L_{2}^{P}$, has the following univariate KL expansion:
$
  X_{p}
  =\sum_{i=1}^{\infty}\xi_{pi}\phi_{pi},
$
% \begin{equation*}
%   X_{p}
%   =\sum_{i=1}^{\infty}\xi_{pi}\phi_{pi}
% \end{equation*}
where $\{\xi_{pi}\}_{i\ge 1}$ is a sequence of real zero-mean random variables such that $\mbb{E}[\xi_{pi}\xi_{pj}]=\zeta_{pi}\delta_{ij}$, 
and $\{(\zeta_{pi},\phi_{pi})\}_{i\ge1}$ denotes the positive eigenvalue-eigenfunction pairs of $\mac{V}_{pp}$.
%% そもそも、KL展開は2乗期待値の収束で定義されるので、その意味でも、\xi_{pi}=0 a.s. となる係数\xi_{pi}は、KL展開の定義上、0 a.s. となるので、KL展開の和に含める必要はない。

For a densely defined linear operator $\mac{B}$ between Hilbert spaces, 
$\mac{B}^*$ denotes its Hilbert-space adjoint whenever it is well-defined.
For a set $S\subset L_2^P$, we denote by $\overline{S}$ the closure of $S$ in $L_2^P$.

\subsection {Sample formulation of FMCCA}
\label{subsec:sample-formulation-FMCCA}

In this section, we briefly review the sample-based formulation of FMCCA proposed by \citet{HwangEtAl2012}. 
Following their formulation, we treat the data as fixed observed functions in this section, rather than as random elements.

Suppose that there are $P$ sets of functions, each consisting of $I$ sample curves. 
Let $z_{ip}(t)$ denote the $i$th function in the $p$th set, where $t \in \mathcal{T}$ for a compact interval $\mathcal{T}$. 
In practice, the functional observation $z_{ip}$ is typically obtained by smoothing discretely observed measurements $y_{ijp}$, $j = 1, \dots, J_{ip}$.
For each $p = 1, \ldots, P$, let $\alpha_p(t)$ be a weight function, and define
$
  \eta_{ip} 
  =\lan z_{ip},\alpha_p\ran
  =\int_{\mathcal{T}} \alpha_p(t) z_{ip}(t)\, dt,
$
% \ba
% \eta_{ip} 
% =\int_{\mathcal{T}} \alpha_p(t) z_{ip}(t)\, dt,
% \ea
which is the canonical score of the $i$th sample curve in the $p$th set.

\citet{HwangEtAl2012} defined FMCCA through the criterion
\ba
  \phi(\bs{\alpha})
  =\sum_{i=1}^I
  \left(
    \sum_{p=1}^P \eta_{ip}
  \right)^2,
\ea
to be maximized with respect to $\bs{\alpha}=(\alpha_1,\ldots,\alpha_P)^\top$, subject to the normalization constraints
$
  \sum_{i=1}^I \eta_{ip}^2 = 1
$
for $p = 1, \ldots, P$.
% \ba
%   \sum_{i=1}^I \eta_{ip}^2 = 1,
%   \qquad p = 1, \ldots, P.
% \ea
That is, the criterion first sums the canonical scores over the $P$ sets for each subject $i$, then squares the resulting sum, and finally aggregates these squared sums over all subjects.
This is equivalent to maximizing the variance of the sum of canonical variates over the $P$ sets, subject to the normalization constraints on the canonical variates for each set.

The above optimization gives the first canonical component.
Subsequent canonical components are obtained sequentially by imposing additional orthogonality constraints with respect to the previously obtained canonical scores.
More precisely, for the $k$th component, the score $\eta_{ip}^{(k)}$ is required to satisfy$\sum_{i=1}^I \eta_{ip}^{(k)}\eta_{ip}^{(\ell)}=0$, where $p=1,\ldots,P$ and $\ell=1,\ldots,k-1$.
Equivalently, this condition can be written as an empirical covariance-orthogonality constraint on the corresponding weight functions.

\citet{HwangEtAl2012} additionally imposed roughness penalties on the weight functions to control their smoothness. 
In the present paper, however, our interest lies in the theoretical structure of the population version, and therefore we restrict attention to the unpenalized formulation.

To make computation feasible, \citet{HwangEtAl2012} use finite-dimensional basis expansions for the observed functions $z_{ip}$ and the corresponding weight functions $\alpha_{kp}$.
Under this representation, the functional optimization problem above is reduced to a finite-dimensional matrix optimization problem. 

Although the normalization constraints are written above in a setwise scalar form, the computational formulation of \citet{HwangEtAl2012} is expressed as a single generalized eigenvalue problem after the basis expansion.
In that formulation, the normalization is imposed through one block-diagonal quadratic form for the stacked coefficient vector, rather than through separate eigenvalue problems for the individual sets.
This is the same normalization structure as the matrix formulation of regularized GCANO in \citet{TakaneEtAl2008}.
Accordingly, the population constraint introduced in the next subsection can be viewed as the infinite-dimensional counterpart of the global normalization underlying the matrix-based FMCCA formulation.

\subsection{Population formulation of FMCCA}
\label{subsec:population-formulation-FMCCA}

In this section, we formulate a population version of functional multiple-set canonical correlation analysis (P-FMCCA) for multivariate functional data.
% Let $\eta_{kp}=\left<\alpha_{kp},X_{p}\right>$ be the $k$th canonical variate for the $p$th variable, where $\alpha_{kp}\in L_2$ denotes the $k$th weight function for the $p$th variable.
% We write $\bs{\alpha}_{k}=(\alpha_{k1},\dots,\alpha_{kP})^{\top}$ and $\eta_{k}=\left<\bs{\alpha}_{k},\bs{X}\right>_P=\sum_{p=1}^P\lan\alpha_{kp},X_p\ran=\sum_{p=1}^P\eta_{kp}$.
For a generic weight function $\bs{\alpha}=(\alpha_1,\dots,\alpha_P)^\top\in L_2^P$, define $\eta(\bs{\alpha})=\lan\bs{\alpha},\bs{X}\ran_P$ and $\eta_p(\alpha_p)=\lan\alpha_p,X_p\ran$.
Then, the first P-FMCCA component is obtained by maximizing the variance of $\eta(\bs{\alpha})$ with respect to $\bs{\alpha}$:
\banum
 \ell(\bs{\alpha})
 &=\mbb{E}\left[\eta(\bs{\alpha})^2\right]
 =\mbb{E}\left[\left(\sum_{p=1}^{P}\eta_p(\alpha_p)\right)^{2}\right]
 \label{eq:criterion_population}
\eanum
subject to the normalization constraint
\banum
  \mbb{E}\left[\sum_{p=1}^{P}\eta_p(\alpha_p)^{2}\right]
 =1.
  \label{eq:constraint_population}
\eanum
Since P-FMCCA is intended to characterize cross-functional association among the functional variables, we exclude the trivial case in which $\mac{V}_{pq}=0$ for all $(p,q)\in[P]^2$ with $p\neq q$.
In this case, all cross-covariances between linear scores from distinct functional variables vanish, and under the normalization constraint \eqref{eq:constraint_population}, the criterion \eqref{eq:criterion_population} is identically equal to 1.
Hence, throughout the remainder of the paper, we assume that $\mac{V}_{pq}\neq 0$ for at least one pair $(p,q)$ with $p\neq q$.
We denote a maximizer by $\bs{\alpha}_1$, whenever such a maximizer exists, and define $\eta_1=\lan\bs{\alpha}_1,\bs{X}\ran_P$ and $\rho_1=\ell(\bs{\alpha}_1)$.

Consistent with the global block-diagonal normalization discussed in the preceding subsection, we use the normalization constraint~\eqref{eq:constraint_population} as the population counterpart of the matrix-based FMCCA formulation in \citet{HwangEtAl2012}.
The same global convention is used for subsequent components: the orthogonality constraints are imposed on the aggregated canonical variates $\eta(\bs{\alpha})=\lan\bs{\alpha},\bs{X}\ran_P$, rather than separately on the setwise scores $\eta_p(\alpha_p)$.
Thus, for $k\ge 2$, the $k$th weight function $\bs{\alpha}_k$ is obtained by maximizing $\ell(\bs{\alpha})$ subject to the normalization constraint~\eqref{eq:constraint_population} and the additional orthogonality constraints
\ba
 \mbb{E}[\eta(\bs{\alpha})\eta_{l}]=0\ \
	\text{for $l=1,\dots,k-1$}.
	% \label{eq:constraint_population2}
\ea
We then define $\eta_k=\lan\bs{\alpha}_k,\bs{X}\ran_P$ and $\rho_k=\ell(\bs{\alpha}_k)$, which are called the $k$th canonical variate and the $k$th canonical criterion value, respectively.
The triplet $(\rho_k,\eta_k,\bs{\alpha}_k)$ is called the $k$th canonical component.

Note that the criterion (\ref{eq:criterion_population}) can be expanded as follows:
\ba
  \ell(\bs{\alpha})
  &=\mbb{E}\left[\left(\sum_{p=1}^{P}\left<X_{p},\alpha_{p}\right>\right)^{2}\right]
  =\sum_{p=1}^{P}\sum_{q=1}^{P}\left<\alpha_{p}, \mac{V}_{pq}\alpha_{q}\right>
  =\left<\bs{\alpha},\mac{V}\bs{\alpha}\right>_{P}.
\ea
% \begin{align*}
%   \ell(\bs{\alpha}_{k})
%   &=\mbb{E}\left[\left(\sum_{p=1}^{P}\left<X_{p},\alpha_{kp}\right>\right)^{2}\right]
%   % &=\mbb{E}\left[\left(\sum_{p=1}^{P}\left<X_{p},\alpha_{kp}\right>\right)
%   % \left(\sum_{q=1}^{P}\left<X_{q},\alpha_{kq}\right>\right)\right]\\
%   % &=\sum_{p=1}^{P}\sum_{q=1}^{P}
%   % \mbb{E}[\left<X_{p},\alpha_{kp}\right>\left<X_{q},\alpha_{kq}\right>]\\
%   % &=\sum_{p=1}^{P}\sum_{q=1}^{P}\mbb{E}[\left<\alpha_{kp}, \lan X_q,\alpha_{kq}\ran X_p\right>]\\
%   =\sum_{p=1}^{P}\sum_{q=1}^{P}\left<\alpha_{kp}, \mac{V}_{pq}\alpha_{kq}\right>
%   % &=\sum_{p=1}^{P}\left<\alpha_{kp},\sum_{q=1}^{P}\mac{V}_{pq}\alpha_{kq}\right>\\
%   =\left<\bs{\alpha}_{k},\mac{V}\bs{\alpha}_{k}\right>_{P}.
% \end{align*}
Here, we define an operator $\mac{W}$ as $\mac{W}\bs{y}=(\mac{V}_{11}y_{1},\dots,\mac{V}_{PP}y_{P})^{\top}$ for $\bs{y}\in L_{2}^{P}$.
\begin{prop}
  \label{prop:W-HS}
  The operator $\mac{W}$ is trace-class.
\end{prop}
The operator $\mac{W}$ is self-adjoint and nonnegative.
Using the operator $\mac{W}$, the constraint (\ref{eq:constraint_population}) can be written as $\lan\bs{\alpha},\mac{W}\bs{\alpha}\ran_{P}=1$. 
Here, we define $\tilde{\bs{\alpha}}=\mac{W}^{\frac{1}{2}}\bs{\alpha}$, i.e., $\bs{\alpha}=\mac{W}^{-\frac{1}{2}}\tilde{\bs{\alpha}}$ if we assume the well-definedness of $\mac{W}^{-\frac{1}{2}}\tilde{\bs{\alpha}}$.
With the change of variables $\tilde{\bs{\alpha}}=\mac{W}^{\frac{1}{2}}\bs{\alpha}$, and whenever $\mac{W}^{-\frac{1}{2}}\tilde{\bs{\alpha}}$ is well-defined, the P-FMCCA optimization problem (\ref{eq:criterion_population}) under the constraint (\ref{eq:constraint_population}) can be rewritten as the following constrained maximization problem over $\tilde{\bs{\alpha}}$:
\begin{equation}
  \ell(\tilde{\bs{\alpha}})
  =\left<\tilde{\bs{\alpha}},\mac{W}^{-\frac{1}{2}}\mac{V}\mac{W}^{-\frac{1}{2}}\tilde{\bs{\alpha}}\right>_{P}
  \label{eq:criterion_l1}
\end{equation}
subject to $\|\tilde{\bs{\alpha}}\|_P=1$.
% \begin{align}
%   \left<\tilde{\bs{\alpha}},\tilde{\bs{\alpha}}\right>_{P}
%   =1.
%   \label{eq:constraint_l1}
% \end{align}
% Additional canonical variates and a weight function, i.e., $\eta_l$ and $\bs{\alpha}_{l}$ for $l\ge2$, can be obtained by solving the maximization problem in (\ref{eq:criterion_population}) with the constraint (\ref{eq:constraint_population}) and $\left<\bs{\alpha}_{k},\mac{W}\bs{\alpha}_{l}\right>_{P}=\delta_{kl}$.
% We write the value of $\ell(\tilde{\bs{\alpha}}_{k})$ corresponding to the $k$th weight function as $\rho_{k}$, which is called the $k$th canonical criterion value.
% The triplet $(\rho_{k},\eta_{k},\bs{\alpha}_{k})$ is called the $k$th canonical component.
Note, however, that $\mac{W}^{-\frac{1}{2}}\tilde{\bs{\alpha}}$ is not necessarily well-defined in general.
The domain issues associated with $\mac{W}^{-\frac{1}{2}}$ and the additional conditions ensuring well-defined P-FMCCA weight functions are discussed in the next section.
% The conditions ensuring its well-definedness will be given in the next section (Condition \ref{cond:cond2}).

Under the normalization constraint $\|\tilde{\bs{\alpha}}\|_P=1$,
% constraint \eqref{eq:constraint_l1}, 
we have $\ell(\tilde{\bs{\alpha}})=1+\lan\tilde{\bs{\alpha}},\mac{W}^{-\frac{1}{2}}\mac{V}_{\star}\mac{W}^{-\frac{1}{2}}\tilde{\bs{\alpha}}\ran$,
% \ba
%   % \lan\tilde{\bs{\alpha}}_k,\mac{W}^{-\frac{1}{2}}\mac{V}\mac{W}^{-\frac{1}{2}}\tilde{\bs{\alpha}}_k\ran
%   % &=\lan\tilde{\bs{\alpha}}_k,\tilde{\bs{\alpha}}_k+\mac{W}^{-\frac{1}{2}}\mac{V}_{\star}\mac{W}^{-\frac{1}{2}}\tilde{\bs{\alpha}}_k\ran\\
%   % &=1+\lan\tilde{\bs{\alpha}}_k,\mac{W}^{-\frac{1}{2}}\mac{V}_{\star}\mac{W}^{-\frac{1}{2}}\tilde{\bs{\alpha}}_k\ran.
% \ea
where $\mac{V}_\star\bs{z}=(\sum_{q\neq1}\mac{V}_{1q}z_{q},\dots,\sum_{q\neq{P}}\mac{V}_{Pq}z_{q})^{\top}$ for $\bs{z}\in L_{2}^{P}$.
Therefore, the P-FMCCA optimization problem formally reduces to the eigenvalue problem for the operator $\mac{R}:=\mac{W}^{-1/2}\mac{V}_{\star}\mac{W}^{-1/2}$.
Because additional conditions are required to guarantee the well-definedness of this operator and of the corresponding weight functions, one might ask whether this formal reduction can be justified.
As shown in Theorem~\ref{theo:optimization-P-FMCCA} in the next section, the answer is affirmative after replacing $\mathcal{R}$ by its Hilbert--Schmidt extension $\tilde{\mathcal{R}}$ under suitable conditions.

Later, Theorem~\ref{theo:optimization-P-FMCCA} shows that the canonical weight functions obtained from the eigenvalue problem for $\tilde{\mathcal{R}}$ satisfy a generalized eigenvalue relation of the form $\mac{V}\bs{\alpha}_l=\rho_l\mac{W}\bs{\alpha}_l$.
For such components, the orthogonality condition $\mbb{E}[\eta(\bs{\alpha})\eta_l]=0$ is therefore equivalent to $\lan\bs{\alpha},\mac{W}\bs{\alpha}_l\ran_P=0$.
This equivalence is used in the proof of Theorem~\ref{theo:optimization-P-FMCCA}.

\section{Well-definedness and solutions of the P-FMCCA problem}
\label{sec:existence}

% In this section, we provide the main theoretical results, which ensure the existence and well-definedness of the canonical variates.
% In the following, we proceed based on the discussion in \citet{he2003functional}, in which the existence of the canonical variates is examined for the case of two functional variables.

% First, we provide the condition for the existence of $\mac{W}^{-\frac{1}{2}}$.
% We will consider the well-definedness of P-FMCCA componentwise, through the univariate Karhunen-Lo\`eve (KL) expansion of each process $X_p$.
% Our approach is to consider a subset of $L_2^P$, on which the inverse of a compact operator can be defined.
% Although this approach is essentially the same as that adopted by \citet{he2003functional}, in our case, we need to consider the case of multiple variables, i.e., $P>2$.

In this section, we establish conditions under which the P-FMCCA formulation introduced above is well-defined and leads to a well-posed eigenvalue problem.
Our analysis is motivated by the work of \citet{he2003functional}, who studied the well-definedness of FCCA for a pair of square-integrable stochastic processes.
Related bivariate population formulations include the regularized operator approach of \cite{cupidon2008some} and the RKHS construction of \cite{eubank2008canonical}.
The present setting, however, is not obtained by a direct pairwise application of these bivariate formulations.
Because P-FMCCA involves more than two functional variables, the relevant object is a single operator on the product Hilbert space $L_2^P$, constructed from the block-diagonal covariance operator $\mac{W}$ and the off-diagonal covariance operator $\mathcal{V}_\star$.
Thus, the main difficulty is to identify conditions under which $\mac{W}^{-\frac{1}{2}}$ can be used on an appropriate domain, the formal operator
$
\mathcal{R}
=W^{-\frac{1}{2}}\mathcal{V}_\star W^{-\frac{1}{2}}
$
admits a Hilbert--Schmidt extension to $L_2^P$, and the resulting eigenfunctions yield well-defined canonical weight functions.

To this end, we use the univariate KL expansions of the component processes $X_p$.
This componentwise representation allows us to express the required conditions in terms of the cross-covariances of the KL scores across all pairs of variables.
We first characterize the domain on which $W^{-\frac{1}{2}}$ is defined, then give a sufficient condition for the formal operator $\mathcal{R}$ to be extended to $L_2^P$, and finally impose a stronger condition ensuring that the eigenfunctions of this extension lead to well-defined P-FMCCA weight functions.

Following \citet[Theorem IV.8.2]{gohberg2012basic}, the domain of $\mac{W}^{-\frac{1}{2}}$ can be characterized by
\banum
 F_{W}
 =\left\{\bs{z}\in L_{2}^{P}\,\bigg|\,
 \forall p\in[P];
  \sum_{l=1}^{\infty}\zeta_{pl}^{-1}|\left<z_{p},\phi_{pl}\right>|^{2}<\infty\wedge
  z_p\perp\ker(\mac{V}_{pp})
  \right\}
  \label{eq:F_W_univariate}
\eanum
where $\{\zeta_{pl},\phi_{pl}\}$ are the nonzero eigenvalues and eigenvectors of $\mac{V}_{pp}$.

Defining
\banum
 F_{W}^{-1}=\left\{\bs{h}\in L_{2}^{P}\,\bigg|\,
 \exists\bs{z}\in F_{W}; \forall p\in[P];
						 h_{p}=\sum_{l=1}^{\infty}\zeta_{pl}^{-\frac{1}{2}}\left<z_{p},\phi_{pl}\right>\phi_{pl}
						\right\},
 \label{eq:F_W_inverse_univariate}
\eanum
we find that $\mac{W}^{\frac{1}{2}}$ is a one-to-one mapping from the vector space $F_{W}^{-1}\subset L_{2}^{P}$ onto the vector space $F_{W}$.
Here, restricting the domain of the operator $\mac{W}^{\frac{1}{2}}$ to the subset $F_{W}^{-1}$, we can define the inverse for $\bs{z}\in F_{W}$ as
\ba
  \mac{W}^{-\frac{1}{2}}\bs{z}
  =\left(\sum_{l=1}^{\infty}\zeta_{1l}^{-\frac{1}{2}}\left<z_{1},\phi_{1l}\right>\phi_{1l},
  \dots,\sum_{l=1}^{\infty}\zeta_{Pl}^{-\frac{1}{2}}\left<z_{P},\phi_{Pl}\right>\phi_{Pl}\right)^{\top}.
\ea
Let $\text{dom}(\mathcal{U})$ refer to the domain on which the operator $\mathcal{U}$ is defined.
Then, based on the sets $F_{W}$ and $F_{W}^{-1}$ defined above, the following condition will ensure the well-definedness of $\mac{R}$, that is, $\text{dom}(\mac{R})=F_{W}$:
\banum
  \mac{V}_{\star}\mac{W}^{-\frac{1}{2}}(F_{W})\subset F_{W}
  \label{eq:cond1}
\eanum
for $F_W$ defined as above.

%%%%%%%%%%%%%%%%%%%%%%%%%%%%%%%%%%%%%%
%%         Condition 1
%%%%%%%%%%%%%%%%%%%%%%%%%%%%%%%%%%%%%%
% \begin{Cond}
%  \label{cond:cond1} For $F_{W}$ defined as above,
%  $\mac{V}_{\star}\mac{W}^{-\frac{1}{2}}(F_{W})\subset F_{W}$.
% \end{Cond}
To ensure that the inclusion (\ref{eq:cond1}) holds, we need to impose a condition on the cross-covariances of the KL scores $\xi_{pi}$ and $\xi_{qj}$ for $p\neq q$.
\begin{Cond}
 \label{cond:cond1} 
 For $(p,q)\in\{1,\dots,P\}^{2}$ with $p\neq q$,
\banum
  \sum_{i=1}^{\infty}\sum_{j=1}^{\infty}\frac{(\mbb{E}[\xi_{pi}\xi_{qj}])^{2}}{\zeta_{pi}\zeta_{qj}}<\infty.
  \label{eq:sufficient_condition1}
\eanum
\end{Cond}
Note that the condition (\ref{eq:sufficient_condition1}) is equivalent to 
\ba
	\sum_{i=1}^{\infty}\sum_{j=1}^{\infty}\left(\Corr(\xi_{pi},\xi_{qj})\right)^{2}<\infty.
\ea

% The following proposition provides a sufficient condition  under which Condition \ref{cond:cond1} holds for the data.

% We can see that Condition \ref{cond:cond1} implies the well-definedness of the operator $\mac{R}$.

The following proposition shows that Condition~\ref{cond:cond1} is sufficient for the inclusion~\eqref{eq:cond1} to hold, and therefore for the well-definedness of $\mac{R}$.

%%%%%%%%%%%%%%%%%%%%%%%%%%%%%%%%%%%%%%
%%            Proposition 1
%%%%%%%%%%%%%%%%%%%%%%%%%%%%%%%%%%%%%%
\begin{prop}
  \label{prop:existence_condition1} 
  The inclusion~\eqref{eq:cond1} holds if Condition \ref{cond:cond1} is satisfied.
\end{prop}

\vspace{10pt}

Under Condition \ref{cond:cond1}, $\mac{R}$ extends to a Hilbert--Schmidt operator on $L_2^P$.
\begin{prop}
  \label{prop:extend_R}
  If Condition \ref{cond:cond1} holds, there exists a self-adjoint Hilbert--Schmidt operator $\tilde{\mac{R}}:L_2^P\to L_2^P$ such that $\tilde{\mac{R}}|_{F_{W}}=\mac{R}$.
\end{prop}

In what follows, $\tilde{\mac{R}}$ denotes the Hilbert--Schmidt extension constructed in the proof of Proposition~\ref{prop:extend_R}; here $\bs{e}_{pj}\in L_2^P$ denotes the element whose $p$th component is $\phi_{pj}$ and whose other components are zero. 
The extension is defined on $\overline{\myspan\{\bs{e}_{pj}:p\in[P],j\ge1\}}$ and set to be zero on its orthogonal complement.
% In what follows, $\tilde{\mathcal R}$ denotes the Hilbert--Schmidt extension constructed in the proof of Proposition~\ref{prop:extend_R}; that is, $\tilde{\mac{R}}$ is first defined on $\overline{\myspan\{\bs{e}_{pj}:p\in[P],j\ge1\}}$ where $\{\bs{e}_{pj}\}$ is an orthonormal system consisting of $\{\phi_{jk}\}$and then extended by zero on its orthogonal complement.

By the nontriviality assumption introduced in Section \ref{subsec:population-formulation-FMCCA}, the cross-covariance operators are not all zero.
As a consequence of the construction of $\tilde{\mac{R}}$, this implies $\tilde{\mac{R}}\neq0$.
Moreover, since $\tilde{\mac{R}}$ is self-adjoint with zero diagonal blocks, $\tilde{\mac{R}}\neq0$ implies that it has at least one positive eigenvalue.

%%%%%%%%%%%%%%%%%%%%%%%%%%%%%%%%%%%%%%%%%%%
%%               Condition 2
%%%%%%%%%%%%%%%%%%%%%%%%%%%%%%%%%%%%%%%%%%%
As with the operator $\mac{R}$, the weight function $\bs{\alpha}_{k}=\mac{W}^{-\frac{1}{2}}\tilde{\bs{\alpha}}_{k}$ is not well-defined for the functions in the whole space $L_{2}^{P}$. 
The following condition ensures the well-definedness of $\bs{\alpha}_{k}$, which will be shown in Theorem \ref{theo:optimization-P-FMCCA}.

% \begin{Cond}
%  \label{cond:cond2}

%  For $F_{W}$ defined as above, $\tilde{\bs{\alpha}}_{k}\in F_{W}$.
% \end{Cond}

\begin{Cond}
 \label{cond:cond2}
  For $(p,q)\in[P]^2$ with $p\neq q$,
  \banum
    \sum_{i=1}^{\infty}\sum_{j=1}^{\infty}\frac{(\mbb{E}[\xi_{pi}\xi_{qj}])^{2}}{\zeta_{pi}^{2}\zeta_{qj}}<\infty.
    \label{eq:sufficient_condition2}
  \eanum
\end{Cond}
Because Condition~\ref{cond:cond2} is imposed on all ordered pairs $(p,q)$ with $p\ne q$, the corresponding condition with the roles of $p$ and $q$ interchanged is also included.
Condition \ref{cond:cond2} implies Condition \ref{cond:cond1}, and hence the Hilbert-Schmidt extension $\tilde{\mac{R}}$ in Proposition \ref{prop:extend_R} is well-defined under Condition \ref{cond:cond2}.

% \mymemo{
%   【NOTE】$p=q$の場合は，$\mbb{E}[\xi_{pi}\xi_{pj}]=\zeta_{pi}\delta_{ij}$を用いて，
%   \ba
%     \sum_{i=1}^{\infty}\sum_{j=1}^{\infty}\frac{(\mbb{E}[\xi_{pi}\xi_{pj}])^{2}}{\zeta_{pi}^{2}\zeta_{pj}}
%     =\sum_{i=1}^\infty\frac{(\mbb{E}[\xi_{pi}^2])^2}{\zeta_{pi}^3}
%     =\sum_{i=1}^\infty\frac{1}{\zeta_{pi}}
%   \ea
%   を得る．
%   一般に，$\zeta_{pi}\ra0$であるので，この級数は発散してしまうため，$p=q$の場合は除外しておく必要がある．
% }

\begin{example}
  \label{example:existence_condition1}

  We provide an example showing that Condition~\ref{cond:cond1} does not imply Condition~\ref{cond:cond2}.
  This example is motivated by Example 4.4 of \citet{he2003functional}, adapted to the present multiple-set setting.
  
  For each $p\in[P]$, let $\{\phi_{pi}\}_{i\ge1}$ be a CONS of $L_2$.
  We consider centered stochastic processes of the form $X_p=\sum_{i=1}^{\infty}\xi_{pi}\phi_{pi}$, $p\in[P]$.
  Set $\zeta_i=\mbb{E}[\xi_{pi}^2]=1/i^2$ for $i\in\mbb{N}$.
  % Thus, $\sum_{i=1}^{\infty}\zeta_i<\infty$.
  We specify the covariance structure of the KL scores by
  \ba
    \mbb{E}[\xi_{pi}\xi_{qj}]
    =\begin{cases}
      \zeta_i\delta_{ij}, & p=q,\\[4pt]
      \frac{1}{(i+1)^2}\frac{1}{(j+1)^2}, & p\ne q.
    \end{cases}
  \ea
  Note that we assume that the marginal eigenvalues are the same across $p\in[P]$, i.e., $\zeta_{pi}=\zeta_i$ for all $p\in[P]$ and $i\ge1$.

  We first verify Condition~\ref{cond:cond1}.
  Let
  \ba
    C^2
    :=
    \sum_{i=1}^{\infty}\frac{1}{\zeta_i}\frac{1}{(i+1)^4}
    =
    \sum_{i=1}^{\infty}\frac{i^2}{(i+1)^4}
    <\infty.
  \ea

  For $p,q\in[P]$ with $p\ne q$, we have
  % \ba
  %   \mbb{E}[\xi_{pi}\xi_{qj}]
  %   =\lambda b_i b_j.
  % \ea
  % Therefore,
  \ba
    \sum_{i=1}^{\infty}\sum_{j=1}^{\infty}
    \frac{(\mbb{E}[\xi_{pi}\xi_{qj}])^2}{\zeta_i\zeta_j}
    % &=\sum_{i=1}^{\infty}\sum_{j=1}^{\infty} \frac{b_i^2b_j^2}{\zeta_i\zeta_j}\\
    =
    \left(
      \sum_{i=1}^{\infty} \frac{1}{\zeta_i}\frac{1}{(i+1)^4}
    \right)
    \left(
      \sum_{j=1}^{\infty} \frac{1}{\zeta_j}\frac{1}{(j+1)^4}
    \right)
    =C^4
    <\infty.
  \ea
  Thus, Condition~\ref{cond:cond1} holds.

  On the other hand, Condition~\ref{cond:cond2} does not hold.
  Indeed, 
  \ba
    \sum_{i=1}^{\infty}\sum_{j=1}^{\infty}
    \frac{(\mbb{E}[\xi_{pi}\xi_{qj}])^2}{\zeta_i^2\zeta_j}
    % &=\sum_{i=1}^{\infty}\sum_{j=1}^{\infty}\frac{b_i^2b_j^2}{\zeta_i^2\zeta_j}\\
    =
    \left(
      \sum_{i=1}^{\infty}\frac{1}{\zeta_i^2}\frac{1}{(i+1)^4}
    \right)
    \left(
      \sum_{j=1}^{\infty}\frac{1}{\zeta_j}\frac{1}{(j+1)^4}
    \right).
  \ea
  Since
  \ba
    \sum_{i=1}^{\infty}\frac{1}{\zeta_i^2}\frac{1}{(i+1)^4}
    =\sum_{i=1}^{\infty}\frac{i^4}{(i+1)^4}
    =\infty,
  \ea
  whereas
  \ba
  0<\sum_{j=1}^{\infty}
    \frac{1}{\zeta_j}\frac{1}{(j+1)^4}
    =C^2
    <\infty,
  \ea
  the product diverges.
  Therefore, Condition~\ref{cond:cond2} fails.

  This example shows that Condition~\ref{cond:cond1} does not imply Condition~\ref{cond:cond2}.
  Thus, although Condition~\ref{cond:cond1} is sufficient to ensure the well-definedness of the operator $\mac{R}$ on $F_W$, it is not strong enough to guarantee the additional domain condition required for the P-FMCCA weight function.
\end{example}

\begin{example}
  \label{example:existence_condition2}
  We present an example that satisfies Condition~\ref{cond:cond2}, motivated by Example 4.7 of \citet{he2003functional} and adapted to the present multiple-set setting.
  As in Example \ref{example:existence_condition1}, for each $p\in[P]$, consider the KL expansion $X_p=\sum_{i=1}^{\infty}\xi_{pi}\phi_{pi}$.
  Let $\zeta_{pi}=\mbb{E}[\xi_{pi}^2]=1/i^2$ for $p\in[P]$, and assume that for $p,q\in[P]$ with $p\neq q$, 
  \ba
    \mbb{E}[\xi_{pi}\xi_{qj}]
    =\frac{1}{(i+1)^3(j+1)^3}
  \ea
  for $i,j\ge1$.
  Then, we have
  \ba
    \sum_{i,j=1}^\infty\frac{(\mbb{E}[\xi_{pi}\xi_{qj}])^{2}}{\zeta_{pi}^{2}\zeta_{qj}}
    &=\sum_{i,j=1}^\infty\frac{i^4j^2}{(i+1)^6(j+1)^6}
    % <\frac{(i+1)^4(j+1)^2}{(i+1)^6(j+1)^6}
    <\sum_{i,j=1}^\infty\frac{1}{(i+1)^2(j+1)^4}\\
    &=\left(\sum_{i=1}^{\infty}\frac{1}{(i+1)^2}\right)\left(\sum_{j=1}^{\infty}\frac{1}{(j+1)^4}\right)
    <\left(\sum_{i=1}^{\infty}\frac{1}{i^2}\right)^2
    <\infty.
  \ea
  % Therefore, we have
  % \ba
  %   \sum_{i,j=1}^\infty\frac{(\mbb{E}[\xi_{pi}\xi_{qj}])^{2}}{\zeta_{pi}^{2}\zeta_{qj}}
  %   &=\left(\sum_{i=1}^{\infty}\frac{1}{(i+1)^2}\right)\left(\sum_{j=1}^{\infty}\frac{1}{(j+1)^4}\right)
  %   % <\left(\sum_{i=1}^{\infty}\frac{1}{i^2}\right)\left(\sum_{j=1}^{\infty}\frac{1}{j^4}\right)\\
  %   % &<\left(\sum_{i=1}^{\infty}\frac{1}{i^2}\right)\left(\frac{1}{j^2}\right)
  %   <\left(\sum_{i=1}^{\infty}\frac{1}{i^2}\right)^2
  %   % =\left(\frac{\pi^2}{6}-1\right)^2
  %   <\infty.
  % \ea
  Thus, the condition (\ref{eq:sufficient_condition2}) holds.
\end{example}

The next theorem establishes two consequences of Condition~\ref{cond:cond2}.
First, it shows that eigenfunctions of $\tilde{\mac{R}}$ lead to well-defined P-FMCCA weight functions.
Second, it shows that, under the same condition, the population optimization problem is characterized by the eigenvalue problem for $\tilde{\mac{R}}$.

% Note that if $\tilde{R}$ has no positive eigenvalues, then there is no P-FMCCA component whose canonical criterion value exceeds the baseline value of $1$, as discussed in Remark~\ref{rem:baseline_value}.
% Hence, in what follows, we restrict attention to the nontrivial case in which $\tilde{R}$ has at least one positive eigenvalue.

\begin{Theo}
  \label{theo:optimization-P-FMCCA}
  Assume that the vector-valued stochastic process $\bs{X}$ satisfies Condition~\ref{cond:cond2}.
  % Let $\{\tilde{\rho}_{k},\tilde{\bs{\alpha}}_{k}\}_{k\in\mbb{N}}$ be the non-zero eigenvalue-eigenfunction pairs of the operator $\tilde{\mac{R}}$, arranged in decreasing order of eigenvalues.
  Let $r\in\mbb{N}\cup\{\infty\}$ denote the number of positive eigenvalues of $\tilde{\mac{R}}$, counted with multiplicity.
  For any fixed finite integer $K$ satisfying $1\le K\le r$, let $\tilde{\rho}_1\ge\tilde{\rho}_2\ge\dots\ge\tilde{\rho}_K>0$ be the first $K$ positive eigenvalues of $\tilde{\mac{R}}$, and let $\tilde{\bs{\alpha}}_1,\dots,\tilde{\bs{\alpha}}_K$ be the corresponding eigenfunctions.
  Then, the following assertions hold:
  \begin{itemize}
    \item[(a)] The eigenfunctions belong to the domain $F_W$ of $\mac{W}^{-\frac{1}{2}}$: $\tilde{\bs{\alpha}}_k\in F_W$ for all $k\in[K]$.
    \item[(b)] The canonical criterion values satisfy $\Var(\eta_k)=\rho_k=1+\tilde{\rho}_k$, and corresponding canonical weight functions can be chosen as $\bs{\alpha}_k=\mac{W}^{-\frac{1}{2}}\tilde{\bs{\alpha}}_k$ for all $k\in[K]$.
    \item[(c)] The resulting canonical variates are mutually uncorrelated: $\Corr(\eta_k,\eta_l)=\delta_{kl}$ for $k,l\in[K]$.
  \end{itemize}
\end{Theo}

Theorem~\ref{theo:optimization-P-FMCCA} can be viewed as a multiple-set analogue of the eigenvalue characterization and well-definedness result for FCCA of pairs of square-integrable stochastic processes studied by \citet{he2003functional}.
It justifies the formal eigenvalue reduction of P-FMCCA by showing that the population problem is characterized by the Hilbert--Schmidt extension $\tilde{\mac{R}}$.
In finite-dimensional implementations, this corresponds to computing the eigenvalues and eigenfunctions of the finite-dimensional analogue of $\mac{R}$.

\begin{Rem}
  The equality $\bs{\alpha}_k=\mac{W}^{-\frac{1}{2}}\tilde{\bs{\alpha}}_k$ in Theorem~\ref{theo:optimization-P-FMCCA}(b) specifies the representative of the weight function lying in $(\ker(\mac{W}))^\perp=F_W^{-1}$.
  If $h\in\ker(\mac{W})$, then $\langle h,X\rangle_P=0$ almost surely.
  Hence, adding such an $h$ to a weight function does not change the corresponding canonical variate.
  Throughout this paper, we use the representative  in $(\ker(\mac{W}))^\perp$.
\end{Rem}

\begin{Rem}
  \label{rem:baseline_value}
  The restriction to positive eigenvalues in Theorem~\ref{theo:optimization-P-FMCCA} is natural from the viewpoint of the P-FMCCA criterion.
  Under the normalization constraint, the transformed objective can be written as $\lan \tilde{\bs{\alpha}},(I+\tilde{\mac{R}})\tilde{\bs{\alpha}}\ran_P
  =1+\lan \tilde{\bs{\alpha}},\tilde{\mac{R}}\tilde{\bs{\alpha}}\ran_P$.
  % \ba
  %   \lan \tilde{\bs{\alpha}},(I+\tilde{\mac{R}})\tilde{\bs{\alpha}}\ran_P
  %   =1 + \lan \tilde{\bs{\alpha}},\tilde{\mac{R}}\tilde{\bs{\alpha}}\ran_P .
  % \ea
  Hence, if $\tilde{\bs{\alpha}}$ is an eigenfunction of $\tilde{\mac{R}}$ with eigenvalue $\tilde{\rho}$, the associated criterion value is $1+\tilde{\rho}$.
  The baseline value $1$ is the value obtained from the marginal normalization alone; it does not include any positive aggregate cross-covariance contribution among the different sets.
  Therefore, eigenfunctions with $\tilde{\rho}>0$ are precisely the directions that increase the P-FMCCA criterion above this baseline and yield a positive aggregate cross-covariance contribution among the functional variables.
  By contrast, eigenfunctions with $\tilde{\rho}=0$ do not increase the criterion, whereas eigenfunctions with $\tilde{\rho}<0$ decrease it and represent directions with negative aggregate cross-covariance contribution.
  For this reason, the P-FMCCA components are defined using the positive eigenvalues of $\tilde{\mac{R}}$.
  In practical implementations, this corresponds to retaining only components whose canonical criterion values exceed the baseline value $1$.
\end{Rem}

\begin{example}[Example \ref{example:existence_condition1} continued]
\label{example:existence-continued}
  We continue Example \ref{example:existence_condition1} to show that Condition~\ref{cond:cond1} alone is not sufficient to ensure the eigenfunctions of $\tilde{\mac{R}}$ belong to $F_W$.
  
  We now derive the explicit form of the operator $\tilde{\mac{R}}$ associated with the covariance structure in Example \ref{example:existence_condition1}.
  For $p\in[P]$, define
  % \ba
  %   a_i
  %   :=\frac{b_i}{\sqrt{\zeta_i}}
  %   =\frac{i}{(i+1)^2},
  %   \qquad g_p
  %   :=\sum_{i=1}^{\infty}a_i\phi_{pi}.
  % \ea
  \ba
    g_p
    :=\sum_{i=1}^{\infty}\frac{i}{(i+1)^2}\phi_{pi}.
  \ea
  Then
  \ba
    \|g_p\|^2
    =\sum_{i=1}^{\infty}\left\{\frac{i}{(i+1)^2}\right\}^2
    =C^2.
  \ea
  For $p,q\in[P]$ with $p\ne q$, the cross-covariance operator $\mac{V}_{pq}$ is given, for $h\in L_2$, by
  \ba
    \mac{V}_{pq}h
    =\sum_{i=1}^{\infty}\sum_{j=1}^{\infty}\mbb{E}[\xi_{pi}\xi_{qj}]\lan h,\phi_{qj}\ran\phi_{pi}
    =\sum_{i=1}^{\infty}\sum_{j=1}^{\infty}\frac{1}{(i+1)^2}\frac{1}{(j+1)^2}
    \lan h,\phi_{qj}\ran\phi_{pi}.
  \ea
  For the diagonal block, we have, for $h\in L_2$,
  \ba
    \mac{V}_{pp}h
    =\sum_{i=1}^{\infty}\zeta_i\lan h,\phi_{pi}\ran\phi_{pi}.
  \ea
  Moreover, whenever the following series is well defined in $L_2$,
  \ba
    \mac{V}_{pp}^{-\frac12}h
    =\sum_{i=1}^{\infty}\zeta_i^{-\frac12}\lan h,\phi_{pi}\ran\phi_{pi}.
  \ea
  For $\bs{z}\in F_W$, direct calculation gives, for $p\in[P]$,
  % Let $\bs{z}=(z_1,\ldots,z_P)^\top\in F_W$.
  % Then $\mac{W}^{-\frac{1}{2}}\bs{z}$ is well-defined, and using the off-diagonal operator $\mac{V}_{\star}$, we have
  % \ba
  %   (\mac{V}_{\star}\mac{W}^{-\frac12}\bs{z})_p
  %   =\sum_{q\ne p}\mac{V}_{pq}\mac{V}_{qq}^{-\frac12}z_q.
  % \ea
  % For each $q\ne p$,
  % \ba
  %   \mac{V}_{pq}\mac{V}_{qq}^{-\frac12}z_q
  %   % &=\sum_{i=1}^{\infty}\sum_{j=1}^{\infty}b_i b_j
  %   % \left\lan
  %   %   \mac{V}_{qq}^{-\frac12}z_q,\phi_{qj}
  %   % \right\ran
  %   % \phi_{pi}\\
  %   % &=\sum_{i=1}^{\infty}\sum_{j=1}^{\infty}b_i b_j\zeta_j^{-\frac12}\lan z_q,\phi_{qj}\ran\phi_{pi}\\
  %   =
  %   \left(
  %     \sum_{j=1}^{\infty}\zeta_j^{-\frac12}\frac{1}{(j+1)^2}\lan z_q,\phi_{qj}\ran
  %   \right)
  %   \left(
  %     \sum_{i=1}^{\infty}\frac{1}{(i+1)^2}\phi_{pi}
  %   \right).
  % \ea
  % Applying $\mac{V}_{pp}^{-\frac12}$ to the $p$th component gives
  \ba
    (\mac{R}\bs{z})_p
    =(\mac{W}^{-\frac12}\mac{V}_{\star}\mac{W}^{-\frac12}\bs{z})_p
    % &=\sum_{q\ne p}
    % \left(
    %   \sum_{j=1}^{\infty}\zeta_j^{-\frac12}\frac{1}{(j+1)^2}\lan z_q,\phi_{qj}\ran
    % \right)
    % \left(
    %   \sum_{i=1}^{\infty}\zeta_i^{-\frac12}\frac{1}{(i+1)^2}\phi_{pi}
    % \right)
    % &=\sum_{q\ne p}\lan z_q,g_q\ran g_p\\
    =g_p\sum_{q\ne p}\lan z_q,g_q\ran.
  \ea
  % Thus, on $F_W$, the operator $\mac{R}=\mac{W}^{-\frac{1}{2}}\mac{V}_{\star}\mac{W}^{-\frac{1}{2}}$ satisfies $(\mac{R}\bs{z})_p
  %   = g_p\sum_{q\ne p}\lan z_q,g_q\ran$ with $p\in[P]$.
  % % \ba
  %   (\mac{R}\bs{z})_p
  %   = g_p\sum_{q\ne p}\lan z_q,g_q\ran,
  %   \qquad p=1,\ldots,P.
  % \ea
  Now define an operator $\mac{S}:L_2^P\to L_2^P$ by $(\mac{S}\bs{z})_p
    = g_p\sum_{q\ne p}\lan z_q,g_q\ran$ with $p\in[P]$.
  % \ba
  %   (\mac{S}\bs{z})_p
  %   = g_p\sum_{q\ne p}\lan z_q,g_q\ran,
  %   \qquad p=1,\ldots,P.
  % \ea
  This operator is bounded and has finite rank, because the range of $\mac{S}$ is spanned by the finite set $\{(g_1,0,\ldots,0)^\top,\ldots,(0,\ldots,0,g_P)^\top\}$.
  Hence, $\mac{S}$ is a Hilbert--Schmidt operator.
  The preceding calculation shows that $\mac{S}$ coincides with $\mac{R}$ on $F_W$.
  Moreover, $\mac{S}$ coincides with the Hilbert--Schmidt extension $\tilde{\mac{R}}$ constructed in Proposition~\ref{prop:extend_R}.
  Indeed, for each $q\in[P]$ and $j\ge1$,
  \ba
    \mac{S}\bs{e}_{qj}
    =\sum_{p\neq q}\sum_{i=1}^{\infty}\frac{1}{(i+1)^2}\frac{1}{(j+1)^2}\bs{e}_{pi}
    =\tilde{\mac{R}}\bs{e}_{qj},
  \ea
  and both operators vanish on the orthogonal complement of $\overline{\myspan\{\bs{e}_{qj}:q\in[P],\,j\ge1\}}$.
  Hence, in this example, the extension $\tilde{\mac{R}}$ is explicitly given by $\mac{S}$.
  Consequently, for every $\bs{z}\in L_2^P$, $(\tilde{\mac{R}}\bs{z})_p
    = g_p\sum_{q\ne p}\lan z_q,g_q\ran$ for $p\in[P]$.
  % \ba
  %   (\tilde{\mac{R}}\bs{z})_p
  %   = g_p\sum_{q\ne p}\lan z_q,g_q\ran,
  %   \qquad p=1,\ldots,P.
  % \ea
  
  Now define $\tilde{\bs{\alpha}}=(\tilde{\alpha}_1,\ldots,\tilde{\alpha}_P)^\top$ by $\tilde{\alpha}_p
    =\frac{1}{\sqrt{P}\,C}g_p$ for $p\in[P]$.
  % \ba
  %   \tilde{\alpha}_p
  %   =\frac{1}{\sqrt{P}\,c}g_p,
  %   \qquad p=1,\ldots,P.
  % \ea
  Then
  \ba
    \|\tilde{\bs{\alpha}}\|_P^2
    =\sum_{p=1}^P\|\tilde{\alpha}_p\|^2
    =\sum_{p=1}^P\frac{1}{PC^2}\|g_p\|^2
    =1.
  \ea
  Also we have
  % Furthermore,
  % \ba
  %   \lan\tilde{\alpha}_q,g_q\ran
  %   =\frac{1}{\sqrt{P}\,C}\|g_q\|^2
  %   =\frac{C}{\sqrt P}.
  % \ea
  % Thus,
  \ba
    (\tilde{\mac{R}}\tilde{\bs{\alpha}})_p
    =g_p\sum_{q\ne p}\lan \tilde{\alpha}_q,g_q\ran
    % &=\lambda g_p\sum_{q\ne p}\frac{c}{\sqrt P}\\
    =(P-1)\frac{C}{\sqrt P}g_p
    =(P-1)C^2\tilde{\alpha}_p.
  \ea
  Therefore, $\tilde{\mac{R}}\tilde{\bs{\alpha}}
    =(P-1)C^2\tilde{\bs{\alpha}}$.
  % \ba
  %   \tilde{\mac{R}}\tilde{\bs{\alpha}}
  %   =\lambda(P-1)C^2\tilde{\bs{\alpha}}.
  % \ea
  Hence, $\tilde{\bs{\alpha}}$ is an eigenfunction of $\tilde{\mac{R}}$ corresponding to the positive eigenvalue $(P-1)C^2>0$.
  % \ba
  %   \lambda(P-1)C^2>0.
  % \ea

  Finally, we show that this eigenfunction does not belong to $F_W$.
  For each $p\in[P]$, we have
  % \ba
  %   \lan\tilde{\alpha}_p,\phi_{pi}\ran
  %   =\frac{1}{\sqrt{P}\,C}a_i
  %   =\frac{1}{\sqrt{P}\,C}\frac{i}{(i+1)^2}.
  % \ea
  % Thus,
  \ba
    \sum_{i=1}^{\infty}\zeta_i^{-1}|\lan\tilde{\alpha}_p,\phi_{pi}\ran|^2
    &=\frac{1}{PC^2}\sum_{i=1}^{\infty}\zeta_i^{-1}\left\{\frac{i}{(i+1)^2}\right\}^2
    % &=\frac{1}{Pc^2}\sum_{i=1}^{\infty}\frac{b_i^2}{\zeta_i^2}\\
    =\frac{1}{PC^2}\sum_{i=1}^{\infty}\frac{i^4}{(i+1)^4}
    =\infty.
  \ea
  Therefore, $\tilde{\bs{\alpha}}\notin F_W$.
  % $\tilde{\alpha}_p\notin F_W$ for every $p\in[P]$, and hence
  % \ba
  %   \tilde{\bs{\alpha}}\notin F_W.
  % \ea
  
  Taken together, Example~\ref{example:existence_condition1} and the preceding discussion show that Condition~\ref{cond:cond1} is sufficient for the existence of a Hilbert--Schmidt extension $\tilde{\mac{R}}$, but it does not guarantee that the eigenfunctions of $\tilde{\mac{R}}$ belong to $F_W$.
  In particular, an eigenfunction of $\tilde{\mac{R}}$ corresponding to a positive eigenvalue may fail to yield a well-defined P-FMCCA weight function through $\mac{W}^{-\frac12}\tilde{\bs{\alpha}}$ unless the stronger Condition~\ref{cond:cond2} is imposed.
\end{example}

\section{Canonical decomposition for multivariate processes}
\label{sec:canonical_decomposition}

This section develops a canonical decomposition associated with P-FMCCA.
The aim is to separate the part of the multivariate functional process explained by the first $K$ P-FMCCA components from the remaining component.
The construction is based on the eigenfunctions of the Hilbert--Schmidt extension $\tilde{\mac{R}}$ introduced in the previous section.
It also provides the key tool for relating P-FMCCA to the functional homogeneity formulation discussed in Section~\ref{sec:homogeneity_analysis}.

% This section introduces a decomposition of multivariate functional data into canonical correlation components. 
% For the univariate case ($P = 1$), the method reduces to ordinary functional principal component analysis; in the bivariate case ($P=2$) it coincides with the canonical decomposition for pairs of random processes described by \citet{he2003functional}.

% Since $\tilde{\mac{R}}$ is a compact, self-adjoint operator on $L_2^P$.
% Hence, it admits an eigenvalue decomposition \citep[Theorem~4.2.4]{hsing2015theoretical}; this decomposition yields the solutions to the P-FMCCA optimization problem, namely the maximizers of the criterion \eqref{eq:criterion_l1} under the normalization constraint.

%% $\mac{W}^{-\frac{1}{2}}$はHilbert--Schmidtではないので，以下のような拡張は考えられないことに注意！ (2025/06/24)
% Similarly, we extend the domain of $\mac{W}^{-\frac{1}{2}}$ to $L_2^P$.
% \begin{prop}
%   \label{prop:extend_W}
%   If $\bs{X}$ satisfies (\ref{eq:sufficient_condition_univariate}), then there exists a Hilbert--Schmidt operator $\mac{R}':L_2^P\to L_2^P$ such that $\mac{R}'|_{F_{W}}=\mac{W}^{-\frac{1}{2}}$.
% \end{prop}
% Henceforth, we understand $\mac{W}^{-\frac{1}{2}}$ in the $\mac{R}'$ sense.

We now develop a canonical decomposition that is analogous to the decomposition for bivariate functional canonical correlation analysis in Theorem~5.1 of \citet{he2003functional}, but adapted to the global-normalization structure of P-FMCCA. 
First, consider the ordinary finite-dimensional multivariate setting rather than the infinite-dimensional functional data; that is, for a positive integer $D>1$, let $\bs{X}\in\mbb{R}^D$.
For $\bs{x},\bs{y}\in\mbb{R}^D$, the inner product on $\mbb{R}^D$ is expressed by $\lan\bs{x},\bs{y}\ran_E$.
In this case, $\mac{V}$ is the variance-covariance matrix of $\bs{X}$, and $\mac{W}$ is a diagonal matrix with $\Var(X_p)$ as the $p$th diagonal element.
Also, $\bar{\mac{R}}=\mac{W}^{-\frac{1}{2}}\mac{V}\mac{W}^{-\frac{1}{2}}$ represents the correlation matrix of $\bs{X}$, and $\tilde{\bs{\alpha}}_k$ represents the eigenvector corresponding to the $k$th eigenvalue of the correlation matrix.
Here, the weight vector $\bs{\alpha}_k$ is calculated as $\bs{\alpha}_k=\mac{W}^{-\frac{1}{2}}\tilde{\bs{\alpha}}_k$.
Let $\mac{P}$ be the orthogonal projection matrix onto the linear span of $\{\tilde{\bs{\alpha}}_k, k\in[D]\}$.
Since $\bar{\mac{R}}$ is precisely the variance-covariance matrix of the standardized data $\mac{W}^{-\frac{1}{2}}\bs{X}$, we have $\mac{W}^{-\frac{1}{2}}\bs{X}\in\myspan\{\tilde{\bs{\alpha}}_k, k\in[D]\}$.
Then, we can express the data $\bs{X}$ using the canonical components as follows:
\ba
  \bs{X}
  % &=\mac{W}^{\frac{1}{2}}\mac{W}^{-\frac{1}{2}}\bs{X}
  =\mac{W}^{\frac{1}{2}}\mac{P}\mac{W}^{-\frac{1}{2}}\bs{X}
  =\mac{W}^{\frac{1}{2}}\sum_{k=1}^{D}\lan\mac{W}^{-\frac{1}{2}}\bs{X},\tilde{\bs{\alpha}}_{k}\ran_E\tilde{\bs{\alpha}}_{k}
  % =\mac{W}^{\frac{1}{2}}\sum_{k=1}^{D}\lan\bs{X},\mac{W}^{-\frac{1}{2}}\tilde{\bs{\alpha}}_{k}\ran_E\tilde{\bs{\alpha}}_{k}\\
  % &=\mac{W}^{\frac{1}{2}}\sum_{k=1}^{D}\lan\bs{X},\bs{\alpha}_{k}\ran_E\tilde{\bs{\alpha}}_{k}
  % =\mac{W}^{\frac{1}{2}}\sum_{k=1}^{D}\eta_k\tilde{\bs{\alpha}}_{k}
  % =\mac{W}^{\frac{1}{2}}\sum_{k=1}^{D}\eta_k\mac{W}^{\frac{1}{2}}\bs{\alpha}_{k}
  =\sum_{k=1}^{D}\eta_k\mac{W}\bs{\alpha}_{k}.
\ea

Next, let us extend this to the case of infinite-dimensional multivariate functional data.
For the infinite-dimensional case, for fixed $K\in\mbb{N}$, we write
\banum
  \bs{X}_{(K)}
  :=\sum_{k=1}^K\eta_k\mac{W}\bs{\alpha}_k.
  \label{eq:canonical_decomposition}
\eanum
We refer to $\bs{X}_{(K)}$ as the truncated canonical representation of $\bs{X}$, which corresponds to the finite-rank component constructed from the first $K$ canonical components.
Note that, unlike the finite-dimensional case, $\bs{X}_{(K)}$ does not necessarily converge to $\bs{X}$ even when $K\to\infty$, as is shown in Theorem \ref{theo:canonical_decomposition}.

\begin{Rem}[Projection interpretation of $X_{(K)}$]
  Let $\mac{P}_K$ denote the orthogonal projection onto 
  $\operatorname{span}\{\tilde{\bs{\alpha}}_1,\ldots,\tilde{\bs{\alpha}}_K\}$.
  % that is the linear space spanned by $\{\tilde{\bs{\alpha}}_1,\ldots,\tilde{\bs{\alpha}}_K\}$.
  In finite dimensions, under the positive definiteness of $\mac{V}_{pp}$ for $p\in[P]$, the truncated representation can be written as
  $\mac{W}^{\frac{1}{2}}\mac{P}_K \mac{W}^{-\frac{1}{2}}\bs{X}$.
  In the present infinite-dimensional setting, however, this expression must be interpreted with care because $\mac{W}^{-\frac{1}{2}}$ is not defined on the whole space even under the positive definiteness of $\mac{V}_{pp}$ for $p\in[P]$.
  Nevertheless, define the finite-rank operator
  $
  \tilde{\mac{P}}_K
  =\sum_{k=1}^K \bs{\alpha}_k\otimes_P \tilde{\bs{\alpha}}_k .
  $
  Then, $\tilde{\mac{P}}_K$ is Hilbert--Schmidt on $L_2^P$.
  Moreover, for any $\bs{z}\in F_W$,
  \ba
    \tilde{\mac{P}}_K \bs{z}
    =\sum_{k=1}^K \langle \bs{z},\bs{\alpha}_k\rangle_P\tilde{\bs{\alpha}}_k
    =\sum_{k=1}^K 
      \langle \mac{W}^{-\frac{1}{2}}\bs{z} ,\tilde{\bs{\alpha}}_k\rangle_P
      \tilde{\bs{\alpha}}_k
    =\mac{P}_K\mac{W}^{-\frac{1}{2}}\bs{z} .
  \ea
  Thus, $\tilde{\mac{P}}_K$ is a finite-rank extension of $\mac{P}_K\mac{W}^{-\frac{1}{2}}$ from $F_W$ to $L_2^P$.
  Furthermore,
  \[
  \mac{W}^{\frac{1}{2}}\tilde{\mac{P}}_K\bs{X}
  =
  \sum_{k=1}^K
  \langle \bs{X},\bs{\alpha}_k\rangle_P \mac{W}^{\frac{1}{2}}\tilde{\bs{\alpha}}_k
  =
  \sum_{k=1}^K \eta_k \mac{W}\bs{\alpha}_k
  =
  \bs{X}_{(K)}.
  \]
  Hence, $\bs{X}_{(K)}$ can be viewed as the reconstruction obtained by projecting the marginally normalized representation of $\bs{X}$ onto the first $K$ canonical directions and mapping the result back to the original scale.
\end{Rem}
\vspace{10pt}

%% 元のProposition 5 は、この論文中に必要な性質でもないので、PropositionではなくRemarkとして紹介し、Supplementary Materialでの証明も省略することとした。（2026/05/28）
%% 上記の式展開等の細かな部分は supp_detailed.tex を参照のこと。
% Let $\mac{P}_K$ be the projection operator onto $\myspan\{\tilde{\bs{\alpha}}_{k},k\in[K]\}$ that is the linear space spanned by $\{\tilde{\bs{\alpha}}_{1},\dots,\tilde{\bs{\alpha}}_{K}\}$.
% The truncated canonical representation $\bs{X}_{(K)}$ can be rewritten using the projection operator $\mac{P}_K$.
% \begin{prop}
%   \label{prop:canonical_decomposition}
%   If $\bs{X}$ satisfies Condition \ref{cond:cond2}, then there exists a Hilbert--Schmidt operator $\tilde{\mac{P}}_K:L_2^P\to L_2^P$ such that $\tilde{\mac{P}}_K|_{F_W}=\mac{P}_K\mac{W}^{-\frac{1}{2}}$, implying that $\bs{X}_{(K)}=\mac{W}^{\frac{1}{2}}\tilde{\mac{P}}_K\bs{X}$.
% \end{prop}

Also, we define the residual component as $\bs{X}_{(K)}^{\perp}
  :=\bs{X}-\bs{X}_{(K)}$.
Then, the following theorem provides the properties of the canonical decomposition.

%%%%%%%%%%%%%%%%%%%%%%%%%%%%%%%%%%%%%%%%%%%
%%            Theorem 2
%%%%%%%%%%%%%%%%%%%%%%%%%%%%%%%%%%%%%%%%%%%
\begin{Theo}
 \label{theo:canonical_decomposition} 
 Assume that the vector-valued stochastic process $\bs{X}$ satisfies Condition~\ref{cond:cond2}.
 Let $r\in\mbb{N}\cup\{\infty\}$ be the number of positive eigenvalues of $\tilde{\mac{R}}$, counted with multiplicity.
 If $r<\infty$, let $\tilde{\rho}_1\ge\tilde{\rho}_2\ge\cdots\ge\tilde{\rho}_r>0$ be the positive eigenvalues of $\tilde{\mac{R}}$.
 If $r=\infty$, let $\tilde{\rho}_1\ge\tilde{\rho}_2\ge\cdots$ be the positive eigenvalues of $\tilde{\mac{R}}$, with $\tilde{\rho}_k>0$ for all $k\in\mbb{N}$.
 Let $\{(\rho_k,\eta_k,\bs{\alpha}_k)\}$ be the corresponding canonical components of $\bs{X}$ where $k\in[r]$ if $r<\infty$ and $k\in\mbb{N}$ if $r=\infty$.
 % k\in[r] としても良いかもしれないが、この記号は自然数Mに対して[M]=\{1,\dots,M\}と定義していたので、r=\inftyのときは厳密には使えないので、あえて 1\le k\le rのような表記としている。
 % with $\bs{\alpha}_k\in L_2^P$ for all $k$
 Set $\tilde{\bs{\alpha}}_k=\mac{W}^{\frac{1}{2}}\bs{\alpha}_k$, for each corresponding index $k$.
 For any fixed finite integer $K$ with $1\le K\le r$, define the truncated canonical representation $\bs{X}_{(K)}$ and the residual component $\bs{X}_{(K)}^{\perp}$ as above.
 Then, the following assertions hold.
 \begin{enumerate}
    \item[(a)] $\bs{X}_{(K)}$ and $\bs{X}$ share the same first $K$ canonical components $(\rho_{k},\eta_{k},\bs{\alpha}_{k})$, $k\in[K]$.
    \item[(b)] If $K<r$, then the canonical components of $\bs{X}_{(K)}^\perp$ are given by the subsequent canonical components of $\bs{X}$.
    That is, if $r<\infty$, then for $l\in[r-K]$, the $l$th canonical component of $\bs{X}_{(K)}^\perp$ is the same as the $(K+l)$th canonical component of $\bs{X}$.
    If $r=\infty$, this statement holds for every $l\in\mbb{N}$.
    % then, for each $l$ such that $l\in[r-K]$, the $l$th canonical component of $\bs{X}_{(K)}^{\perp}$ is the same as the $(K+l)$th canonical component of $\bs{X}$, $l\ge1$.
    % When $r=\infty$, this statement holds for every $l\in\mbb{N}$.
    \item[(c)] $\bs{X}_{(K)}$ and $\bs{X}_{(K)}^{\perp}$ are uncorrelated, that is, the cross-covariance operator $\mbb{E}[\bs{X}_{(K)}\otimes_P\bs{X}_{(K)}^{\perp}]$ is zero.
    % $\Corr(X_{(K)p},X_{(K)q}^{\perp})=0$ for all $p,q$.
    \item[(d)] If $r=\infty$, then, as $K\ra\infty$, we have
    \ba
      \bs{X}_{(K)}\ra \bs{X}_{(\infty)}
      :=\sum_{k=1}^{\infty}\eta_{k}\mac{W}\bs{\alpha}_{k}
    \ea
    where the convergence is in the mean squared norm $\mbb{E}[\|\cdot\|_{P}^{2}]$.
    % Let $\mac{P}_{\infty}$ be the projection operator onto $\overline{\myspan\{\tilde{\bs{\alpha}}_{k},k\ge1\}}$, and then, there exists a finite linear operator $\tilde{\mac{P}}_{\infty}:L_2^P\to L_2^P$ such that $\tilde{\mac{P}}_{\infty}|_{F_W}=\mac{P}_{\infty}\mac{W}^{-\frac{1}{2}}$, implying that $\bs{X}_{(\infty)}=\mac{W}^{\frac{1}{2}}\tilde{\mac{P}}_{\infty}\bs{X}$.
    Furthermore, $\bs{X}_{(\infty)}$ and $\bs{X}$ share the same canonical components, and $\bs{X}_{(\infty)}$ and $\bs{X}_{(\infty)}^\perp:=\bs{X}-\bs{X}_{(\infty)}$ are uncorrelated, that is, $\mbb{E}[\bs{X}_{(\infty)}\otimes_P\bs{X}_{(\infty)}^\perp]=0$.
    \item[(e)] If $\bs{X}\in F_W$ a.s. and $\mac{W}^{-\frac{1}{2}}\bs{X}\in\overline{\Span\{\tilde{\bs{\alpha}}_k, 1\le k\le r\}}$ a.s., where the index set is interpreted as $\{1,\dots,r\}$ if $r<\infty$ and as $\mbb{N}$ if $r=\infty$, then the canonical representation based on canonical components corresponding to positive eigenvalues reconstructs $\bs{X}$.
    Specifically, if $r<\infty$, then $\bs{X}=\bs{X}_{(r)}$ a.s., whereas if $r=\infty$, then $\bs{X}=\bs{X}_{(\infty)}$ a.s.
  \end{enumerate}
  If some canonical criterion values have multiplicity greater than one, the statements are understood up to the choice of an orthonormal basis within the corresponding eigenspaces.
\end{Theo}

\begin{Rem}
  In Theorem~\ref{theo:canonical_decomposition}, the statement that two processes share the same canonical components is understood with respect to the original $\mac{W}$-normalization.
  That is, when the canonical components of $\bs{X}_{(K)}$, $\bs{X}^\perp_{(K)}$, or $\bs{X}_{(\infty)}$ are considered, the objective function is computed from the process under consideration, whereas the normalization and orthogonality constraints are imposed through the block-diagonal covariance operator $\mac{W}$ of the original process $\bs{X}$.
  Thus, the theorem does not claim that these processes have the same canonical components when their own block-diagonal covariance operators are used for normalization.
\end{Rem}

\begin{Rem}[Relation to the bivariate canonical decomposition]
  Theorem~\ref{theo:canonical_decomposition} is motivated by the canonical decomposition for pairs of square-integrable stochastic processes studied by \citet{he2003functional}.
  The analogy lies in the decomposition into a finite-rank component generated by the first canonical directions and a residual component carrying the subsequent canonical components, together with the uncorrelatedness between these two components.
  However, the present decomposition is not identical to the bivariate FCCA decomposition, even when $P=2$.
  In the bivariate FCCA formulation, the two canonical variables are normalized separately and the $k$th component is represented by a pair of canonical variables.
  In contrast, P-FMCCA imposes a single global normalization through the block-diagonal covariance operator $\mac{W}$, and the $k$th canonical variate is the aggregated score $\eta_k=\lan \bs{\alpha}_k,\bs{X}\ran_P$.
  Thus, for $P=2$, Theorem~\ref{theo:canonical_decomposition} should be understood as a global-normalization counterpart of the decomposition of \citet{he2003functional}, rather than as its literal special case.
  This viewpoint is useful because it clarifies how the operator-theoretic construction underlying the bivariate decomposition can be generalized to the multiple-set setting: by stacking the $P$ functional variables in the product Hilbert space, the marginal covariance structure is represented by the block-diagonal operator $\mac{W}$, while the cross-covariance structure is represented by the off-diagonal operator $\mac{V}_\star$.
  The resulting canonical decomposition is then obtained from the spectral analysis of the normalized operator $\mac{R}$.
\end{Rem}

\begin{Rem}[Relation to the sample estimator of \citet{HwangEtAl2012}]
  If the observed sample curves are exactly representable in a finite-dimensional basis space and $K$ is chosen so that $\bs{X}_{(K)}=\bs{X}$, then the present functional formulation reduces to the ordinary finite-dimensional setting. 
  Accordingly, under these conditions, the estimator given by $\{(\rho_k,\eta_k, \bs{\alpha}_k)\}_{k=1}^K$ coincides with the unpenalized finite-dimensional FMCCA estimator under the same global block-diagonal normalization.
\end{Rem}

\section{Functional homogeneity analysis and its relationship to P-FMCCA}
\label{sec:homogeneity_analysis}

Homogeneity analysis, also known as optimal scaling \citep{Gifi1990,de2009gifi}, is a multivariate method that seeks a low-dimensional Euclidean representation of data by optimally quantifying variables so that objects with similar profiles are located close to each other.
In the ordinary multivariate setting, it has been shown that homogeneity analysis for multiple sets of variables is equivalent to multiple-set canonical correlation analysis (MCCA; \citealp{van1988homogeneity,TakaneEtAl2008,HwangEtAl2012}).
This equivalence provides a finite-dimensional benchmark for extending the concept of homogeneity analysis to functional data and for examining its relationship with population functional multiple-set canonical correlation analysis (P-FMCCA).

\subsection{Homogeneity analysis for multivariate data}
\label{subsec:homogeneity_analysis_multivariate}

In this subsection, we briefly review homogeneity analysis for ordinary multivariate data, following \citet{de2009gifi}, and refer to it as HOMALS in what follows.
This review provides the finite-dimensional benchmark for the functional extension developed in the next subsection.

For simplicity, we restrict attention to the case in which all variables are quantitative and have already been represented by numerical data matrices.
The more general HOMALS framework also covers categorical variables through optimal quantification, but this aspect is not needed for the population-level functional formulation considered here.
For each $p\in[P]$, let $\mb{X}_p$ be the $I \times J_p$ data matrix for the $p$th set of variables, where $I$ is the number of objects and $J_p$ is the number of variables in the $p$th set. 
We assume that each $\mb{X}_p$ is column-wise standardized. 
Let $J=\sum_{p=1}^P J_p$, and define the concatenated data matrix $\mb{X}=[\mb{X}_1,\ldots,\mb{X}_P]$, which is an $I \times J$ matrix.

Let $\mb{F}$ be an $I \times K$ matrix of object scores where $K$ is the number of dimensions, and let $\mb{B}_p$ be a $J_p \times K$ matrix of quantifications (or weights) for the $p$th dataset. 
Write $\mb{B} = [\mb{B}_1^\top,\ldots,\mb{B}_P^\top]^\top$ so that $\mb{B}$ is partitioned conformably with $\mb{X}$. 
Then, the homogeneity criterion is defined by
\banum
  \psi(\mb{F},\mb{B}) 
  = \sum_{p=1}^P \|\mb{F} - \mb{X}_p \mb{B}_p\|_F^2
  \label{eq:homogeneity_criterion}
\eanum
where $\|\cdot\|_F$ denotes the Frobenius norm.
The HOMALS optimization problem is to minimize $\psi(\mb{F},\mb{B})$ over $\mb{F}$ and $\mb{B}$ subject to the normalization constraint $\mb{F}^\top \mb{F} = \mb{I}_K$.

Let $\mb{D}_p = \mb{X}_p^\top \mb{X}_p$, and define the block diagonal matrix $\mb{D} = \mathrm{diag}(\mb{D}_1,\ldots,\mb{D}_P)$.
For simplicity, assume that $\mb{D}$ is nonsingular. 
Minimization of $\psi(\mb{F},\mb{B})$ with respect to $\mb{B}$ for fixed $\mb{F}$ yields $\hat{\mb{B}}=\mb{D}^{-1}\mb{X}^\top \mb{F}$.
Substituting this expression into $\psi(\mb{F},\mb{B})$, we obtain the profiled criterion $\psi^*(\mb{F})=\psi(\mb{F},\hat{\mb{B}})$. 
Then, minimizing $\psi^*(\mb{F})$ over $\mb{F}$ under the constraint $\mb{F}^\top \mb{F} = \mb{I}_K$ leads to the eigenvalue problem
\banum
  \mb{X}\mb{D}^{-1}\mb{X}^\top \mb{F} 
  = \mb{F}\mb{\Lambda},
  \label{eq:homals_eigen}
\eanum
where $\mb{\Lambda}$ is the diagonal matrix of the corresponding eigenvalues.

Next, consider multiple-set canonical correlation analysis (MCCA) for the same data matrices $\mb{X}_1,\ldots,\mb{X}_P$. 
Let $\mb{A}$ be a $J \times K$ weight matrix, partitioned as $\mb{A} = [\mb{A}_1^\top,\ldots,\mb{A}_P^\top]^\top$, where $\mb{A}_p$ is the $J_p \times K$ weight matrix for the $p$th dataset. 
In MCCA, the unnormalized canonical score matrix is given by $\mb{X}\mb{A}$, and we seek $\mb{A}$ that maximizes $\|\mb{X}\mb{A}\|_F^2$ subject to $\mb{A}^\top \mb{D}\mb{A}=\mb{I}_K$.
The corresponding first-order condition yields the generalized eigenvalue problem
\banum
  \mb{X}^\top \mb{X}\mb{A} 
  = \mb{D}\mb{A}\mb{\Delta},
  \label{eq:mcca_g_eigen}
\eanum
where $\mb{\Delta}$ is the diagonal matrix of generalized eigenvalues.

If we write $\mb{\Sigma}=\mb{D}^{\frac{1}{2}}\mb{A}$ and $\mb{V}=\mb{X}^\top\mb{X}$, then \eqref{eq:mcca_g_eigen} is equivalent to the ordinary eigenvalue problem
\banum
  \mb{D}^{-\frac{1}{2}}\mb{V}\mb{D}^{-\frac{1}{2}}\mb{\Sigma} 
  = \mb{\Sigma}\mb{\Delta},
  \label{eq:mcca_eigen}
\eanum
with $\mb{\Sigma}^\top\mb{\Sigma} = \mb{I}_K$. 
The corresponding normalized canonical score matrix is given by 
\banum
  \mb{F} = \mb{X}\mb{A}\mb{\Delta}^{-\frac{1}{2}}.
  \label{eq:mcca_scores}
\eanum

To facilitate the discussion of the infinite-dimensional case in the next section, we rewrite the HOMALS solution in coefficient form.
Since the eigenvectors corresponding to nonzero eigenvalues of $\mb{X}\mb{D}^{-1}\mb{X}^\top$ lie in the column space of $\mb{X}$, the corresponding object score matrix can be written as $\mb{F}=\mb{X}\mb{\Gamma}$ for some coefficient matrix $\mb{\Gamma}=[\bs{\gamma}_1,\dots,\bs{\gamma}_K]$.
In view of \eqref{eq:mcca_scores}, the MCCA solution corresponds to the choice $\mb{\Gamma}=\mb{A}\mb{\Delta}^{-\frac{1}{2}}$.
Substituting $\mb{F}=\mb{X}\mb{\Gamma}$ into \eqref{eq:homals_eigen}, and assuming that $\mb{X}^\top\mb{X}$ is nonsingular, we obtain the following eigenvalue problem:
\banum
  \mb{V}^{\frac{1}{2}}\mb{D}^{-1}\mb{V}^{\frac{1}{2}}\tilde{\mb{\Gamma}}
  =\tilde{\mb{\Gamma}}\mb{\Lambda}.
  \label{eq:homals_eigen_reformulated}
\eanum
Here, $\tilde{\mb{\Gamma}}=[\tilde{\bs{\gamma}}_1,\dots,\tilde{\bs{\gamma}}_K]=\mb{V}^{\frac{1}{2}}\mb{\Gamma}$ and $\tilde{\mb{\Gamma}}^\top\tilde{\mb{\Gamma}}=\mb{I}_K$.  

The equivalence between HOMALS and MCCA follows from the relationship between \eqref{eq:homals_eigen_reformulated} and \eqref{eq:mcca_eigen}.
Indeed, $\mb{V}^{\frac{1}{2}}\mb{D}^{-1}\mb{V}^{\frac{1}{2}}$ and $\mb{D}^{-\frac{1}{2}}\mb{V}\mb{D}^{-\frac{1}{2}}$ share the same nonzero eigenvalues, because the nonzero eigenvalues of both matrices are given by the squared singular values of $\mb{V}^{\frac{1}{2}}\mb{D}^{-\frac{1}{2}}$, equivalently, those of $\mb{X}\mb{D}^{-\frac{1}{2}}$.
In the coefficient formulation above, the corresponding eigenvectors are related through the standard $\mb{C}\mb{C}^\top$ and $\mb{C}^\top\mb{C}$ correspondence, with $\mb{C}=\mb{V}^{\frac{1}{2}}\mb{D}^{-\frac{1}{2}}$.
Therefore, the solutions obtained from HOMALS and MCCA are essentially equivalent in the ordinary finite-dimensional setting.

In the infinite-dimensional functional setting, however, this equivalence cannot be justified by the same argument without additional domain conditions.
This is one of the main differences from the ordinary multivariate case. 
In the next subsection, we formulate functional homogeneity analysis and then examine its relationship with P-FMCCA.

\subsection{Homogeneity analysis for multivariate functional data}
\label{subsec:homogeneity_analysis_functional}

First, we formulate homogeneity analysis for multivariate functional data, which we call functional homogeneity analysis (FHOMALS).
We then show that, unlike in the finite-dimensional case, the usual eigenvalue-based argument does not generally justify an equivalence between FHOMALS and P-FMCCA on the whole space $L_2^P$.
Instead, we establish an equivalence at the level of the truncated canonical representation: the FHOMALS components of $X_{(K)}$ are characterized by the P-FMCCA components derived from the corresponding finite-rank covariance structure.

Recall that the ordinary homogeneity analysis criterion is given by \eqref{eq:homogeneity_criterion}.
In a similar manner, let $\bs{\gamma}\in L_2^P$ define the object score $f=\lan\bs{X},\bs{\gamma}\ran_P$ with $\mbb{E}[f]=0$, and let $\bs{\beta}=(\beta_1,\dots,\beta_P)^\top\in L_2^P$ be a vector of weight functions.
We seek $(\bs{\gamma},\bs{\beta})$ that minimize the following loss function
\ba
 \ell'(\bs{\gamma},\bs{\beta})=\sum_{p=1}^{P}\mbb{E}[(f-\lan X_{p},\beta_{p}\ran)^{2}],
	% \label{eq:cr_GFCCA_homo}
\ea
under the constraint that $\mbb{E}[f^{2}]=1$.

%%%%%%%%%%%%%%%%%%%%%%%%%%%%%%%%%%%%%%%%%%%%%%%
%%   the minimization over beta given gamma
%%%%%%%%%%%%%%%%%%%%%%%%%%%%%%%%%%%%%%%%%%%%%%%
To solve the minimization problem over $\bs{\gamma}$ and $\bs{\beta}$, first, we minimize $\ell'(\bs{\gamma},\bs{\beta})$ over $\bs{\beta}$ given $\bs{\gamma}$. 
The criterion $\ell'(\bs{\gamma},\bs{\beta})$ can be expressed as follows:
\banum
 \ell'(\bs{\gamma},\bs{\beta})
%  &=\sum_{p=1}^{P}\left\{\mbb{E}[f^2]
%  +\mbb{E}[|\left<X_{p},\beta_{p}\right>|^{2}]
%  -2\mbb{E}[f\left<X_{p},\beta_{p}\right>]
%  \right\}\notag\\
%  &=P
%  +\sum_{p=1}^{P}\left(\mbb{E}[|\lan X_p,\beta_p\ran|^2]
%  -2\mbb{E}[\left<\bs{X},\bs{\gamma}\right>_{P}\left<X_{p},\beta_{p}\right>]\right)\notag\\
 &=P+\sum_{p=1}^{P}(\lan\mac{V}_{pp}\beta_p,\beta_p\ran
 -2\lan\tilde{\mac{V}}_{p}\bs{\gamma},\beta_p\ran),
 \label{eq:expansion_ell2}
\eanum
where $\tilde{\mac{V}}_{p}\bs{\gamma}=\mbb{E}[\left<\bs{X},\bs{\gamma}\right>_{P}X_{p}]$.
Thus, for fixed $\bs{\gamma}$, the minimization over $\bs{\beta}$ separates into $P$ independent minimization problems, one for each $\beta_p$.
% From (\ref{eq:expansion_ell2}), it can be seen that we only have to minimize $\ell_{2}$ over each $\beta_{p}$, instead of minimizing over $\bs{\beta}$ simultaneously.
Then, we write the criterion related only to $\beta_{p}$ as $Q(\beta_{p}):=\lan\mac{V}_{pp}\beta_{p},\beta_{p}\ran-2\lan\tilde{\mac{V}}_{p}\bs{\gamma},\beta_{p}\ran$.

Next, we give a sufficient condition to ensure the well-definedness of the above minimization problem.
\begin{Cond}
 \label{cond:cond3}
  For $(p,q)\in[P]^2$ such that $p\neq q$,
  \banum
    \sum_{i=1}^{\infty}\sum_{j=1}^{\infty}\frac{(\mbb{E}[\xi_{pi}\xi_{qj}])^{2}}{\zeta_{pi}^{2}}
    <\infty
    \label{eq:sufficient_condition3}
  \eanum
\end{Cond}
Since Condition~\ref{cond:cond3} is imposed on all ordered pairs $(p,q)$ with $p\ne q$, it also includes the corresponding requirement with $p$ and $q$ interchanged.
Under this condition, for a fixed $\bs{\gamma}$, we can obtain the optimal parameter $\beta_{p}$ minimizing $Q(\beta_p)$ as follows.

\begin{prop}
  \label{prop:beta_minimization}
  Under Condition \ref{cond:cond3}, $\beta_{p}=\mac{V}_{pp}^{-1}\tilde{\mac{V}}_p\bs{\gamma}$ is well-defined, and it is a solution of the minimization problem for $Q(\beta_p)$ over $\beta_{p}$ given $\bs{\gamma}$.
\end{prop}

Since Condition \ref{cond:cond2} implies Condition \ref{cond:cond3}, the optimal parameter $\beta_{p}$ that minimizes $Q(\beta_{p})$ given $\bs{\gamma}$ is well-defined under Condtion \ref{cond:cond2}.

%%%%%%%%%%%%%%%%%%%%%%%%%%%%%%%%%%%%
%%   the minimization over gamma
%%%%%%%%%%%%%%%%%%%%%%%%%%%%%%%%%%%%
Next, we consider the minimization problem of $\ell'$ over $\bs{\gamma}$ given the above representation of $\beta_{p}$.
Following Proposition \ref{prop:beta_minimization}, we use $\beta_{p}=\mac{V}_{pp}^{-1}\tilde{\mac{V}}_p\bs{\gamma}=\mac{V}_{pp}^{-1}\sum_{q=1}^{P}\mac{V}_{pq}\gamma_{q}$.
We denote the loss function by $\ell'(\bs{\gamma})$ when regarded as a function of $\bs{\gamma}$.
Then, under Condition \ref{cond:cond3}, we have
% \ba
%   \mbb{E}[\left<X_{p},\beta_{p}\right>^{2}]
%   % &=\mbb{E}\left[\left<X_{p},\mac{V}_{pp}^{-1}\sum_{q=1}^{P}\mac{V}_{pq}\gamma_{q}\right>^{2}\right]\notag\\
%   % &=\mbb{E}\left[\left<X_{p},\mac{V}_{pp}^{-1}\sum_{q=1}^{P}\mac{V}_{pq}\gamma_{q}\right>
%   % \left<X_{p},\mac{V}_{pp}^{-1}\sum_{r=1}^{P}\mac{V}_{pr}\gamma_{r}\right>\right]\\
%   % &=\left<\mac{V}_{pp}\mac{V}_{pp}^{-1}\sum_{q=1}^{P}\mac{V}_{pq}\gamma_{q},
%   % \mac{V}_{pp}^{-1}\sum_{r=1}^{P}\mac{V}_{pr}\gamma_{r}\right>\\
%   &=\left<\sum_{q=1}^{P}\mac{V}_{pq}\gamma_{q},
%   \mac{V}_{pp}^{-1}\sum_{r=1}^{P}\mac{V}_{pr}\gamma_{r}\right>,
% \ea
% and
% \ba
%   \mbb{E}\left[\left<\mb{X},\bs{\gamma}\right>_{P}\left<X_{p},\beta_{p}\right>\right]
%   % &=\mbb{E}\left[\left<\mb{X},\bs{\gamma}\right>_{P}
%   % \left<X_{p},\mac{V}_{pp}^{-1}\sum_{r=1}^{P}\mac{V}_{pr}\gamma_{r}\right>\right]\notag\\
%   % &=\sum_{q=1}^{P}\mbb{E}\left[\left<X_{q},\gamma_{q}\right>
%   % \left<X_{p},\mac{V}_{pp}^{-1}\sum_{r=1}^{P}\mac{V}_{pr}\gamma_{r}\right>\right]\\
%   % &=\sum_{q=1}^{P}\left<\mac{V}_{pq}\gamma_{q},
%   % \mac{V}_{pp}^{-1}\sum_{r=1}^{P}\mac{V}_{pr}\gamma_{r}\right>\\
%   &=\left<\sum_{q=1}^{P}\mac{V}_{pq}\gamma_{q},
%   \mac{V}_{pp}^{-1}\sum_{r=1}^{P}\mac{V}_{pr}\gamma_{r}\right>.
% \ea

% Therefore, we have
\banum
  \ell'(\bs{\gamma})
  &=P-\sum_{p=1}^{P}\left<\sum_{q=1}^{P}\mac{V}_{pq}\gamma_{q},
  \mac{V}_{pp}^{-1}\sum_{r=1}^{P}\mac{V}_{pr}\gamma_{r}\right>
  % &=P-\sum_{q=1}^{P}\left<\gamma_{q},
  % \sum_{p=1}^{P}\mac{V}_{qp}\mac{V}_{pp}^{-1}\sum_{r=1}^{P}\mac{V}_{pr}\gamma_{r}\right>\\
  =P-\left<\bs{\gamma},\mac{V}\mac{W}^{-1}\mac{V}\bs{\gamma}\right>_{P}.
  \label{eq:ell2_prime}
\eanum
Note that if $\beta_p=\mac{V}_{pp}^{-1}\sum_{q=1}^{P}\mac{V}_{pq}\gamma_{q}$ is well-defined, then $\mac{V}\mac{W}^{-1}\mac{V}\bs{\gamma}$ is also well-defined.
Thus, the minimization of $\ell'(\bs{\gamma})$ under the constraint $\mbb{E}[f^{2}]=1$ is equivalent to the maximization of $\left<\bs{\gamma},\mac{V}\mac{W}^{-1}\mac{V}\bs{\gamma}\right>_{P}$ under the constraint $\lan\bs{\gamma},\mac{V}\bs{\gamma}\ran_P=1$. 
Let $\tilde{\bs{\gamma}}:=\mac{V}^{\frac{1}{2}}\bs{\gamma}$, implying that $\|\tilde{\bs{\gamma}}\|_{P}=1$.
Then, we have
\ba
  \lan\bs{\gamma},\mac{V}\mac{W}^{-1}\mac{V}\bs{\gamma}\ran_P
  % &=\lan\bs{\gamma},\mac{V}^{\frac{1}{2}}\mac{V}^{\frac{1}{2}}\mac{W}^{-1}\mac{V}^{\frac{1}{2}}\mac{V}^{\frac{1}{2}}\bs{\gamma}\ran_P\\
  % &=\left<\mac{V}^{-\frac{1}{2}}\tilde{\bs{\gamma}},
  % \mac{V}\mac{W}^{-1}\mac{V}\mac{V}^{-\frac{1}{2}}\tilde{\bs{\gamma}}\right>_{P}\notag\\
  &=\left<\tilde{\bs{\gamma}},
  \mac{V}^{\frac{1}{2}}\mac{W}^{-1}\mac{V}^{\frac{1}{2}}\tilde{\bs{\gamma}}\right>_{P}.
  % \label{eq:ell2_prime}
\ea
Hence, under Condition~\ref{cond:cond3}, the profiled FHOMALS problem can be rewritten, for admissible $\tilde{\bs{\gamma}}=\mac{V}^{\frac{1}{2}}\bs{\gamma}$, as the constrained maximization of $\lan\tilde{\bs{\gamma}},\mac{V}^{\frac{1}{2}}\mac{W}^{-1}\mac{V}^{\frac{1}{2}}\tilde{\bs{\gamma}}\ran_P$ under the normalization constraint $\|\tilde{\bs{\gamma}}\|_P=1$.

% \ref{cond:cond3}, the minimization of $\ell'(\gamma)$ under the constraint $\mbb{E}[f^2]=1$ is equivalent to the maximization of $\lan\tilde{\bs{\gamma}},\mac{V}^{\frac{1}{2}}\mac{W}^{-1}\mac{V}^{\frac{1}{2}}\tilde{\bs{\gamma}}\ran_P$ under the constraint $\|\tilde{\bs{\gamma}}\|_P=1$.
% We can see that the profiled FHOMALS criterion can formally be written as a Rayleigh quotient of the operator $\mac{V}^{\frac{1}{2}}\mac{W}^{-1}\mac{V}^{\frac{1}{2}}$.
The first FHOMALS component is obtained from this maximization problem.
Subsequent components are obtained by imposing the additional orthogonality constraints $\lan\tilde{\bs{\gamma}},\tilde{\bs{\gamma}}_l\ran_P=0$ for $l=1,\dots,k-1$, when constructing the $k$th component.
If $\tilde{\bs{\gamma}}$ were allowed to range over the whole unit sphere in $L_2^P$, this maximization problem would be characterized by the eigenvalue problem for $\mac{V}^{\frac{1}{2}}\mac{W}^{-1}\mac{V}^{\frac{1}{2}}$.

However, unlike the P-FMCCA case, this representation alone does not justify identifying the above constrained maximization problem with the eigenvalue problem for $\mac{V}^{\frac{1}{2}}\mac{W}^{-1}\mac{V}^{\frac{1}{2}}$.
Indeed, the variable $\tilde{\bs{\gamma}}$ is not an arbitrary element of $L_2^P$ with unit norm.
Rather, it must be representable as $\tilde{\bs{\gamma}}=\mac{V}^{\frac{1}{2}}\bs{\gamma}$, and, in addition, the quantity $\mac{W}^{-1}\mac{V}^{\frac{1}{2}}\tilde{\bs{\gamma}}$ must be well-defined, that is, $\|\mac{W}^{-1}\mac{V}^{\frac{1}{2}}\tilde{\bs{\gamma}}\|_P<\infty$.

Accordingly, the relevant admissible set is given by
\ba
  \mac{G}
  :=
  \left\{
    \tilde{\bs{\gamma}}\in \mathrm{Im}(\mac{V}^{\frac{1}{2}})
    \;\middle|\;
    \|\tilde{\bs{\gamma}}\|_P=1,
    \ \mac{W}^{-1}\mac{V}^{\frac{1}{2}}\tilde{\bs{\gamma}}\ \text{is well-defined}
  \right\}.
  % \label{eq:fhomals_admissible_set}
\ea
Thus, the FHOMALS problem should be understood as the constrained maximization of $\lan \tilde{\bs{\gamma}},\mac{V}^{\frac{1}{2}}\mac{W}^{-1}\mac{V}^{\frac{1}{2}}\tilde{\bs{\gamma}}\ran_P$ over $\tilde{\bs{\gamma}}\in\mac{G}$.
Because $\mac{G}$ need not be the whole unit sphere in $L_2^P$, this constrained problem cannot, in general, be identified directly with the eigenvalue problem for $\mac{V}^{\frac{1}{2}}\mac{W}^{-1}\mac{V}^{\frac{1}{2}}$.

Condition~\ref{cond:cond3} ensures that the profiled FHOMALS criterion is well-defined for admissible $\bs{\gamma}$, namely, for those $\bs{\gamma}\in L_2^P$ such that $\|\mac{V}^{\frac{1}{2}}\bs{\gamma}\|_P=1$ and $\mac{W}^{-1}\mac{V}\bs{\gamma}$ is well-defined.
Equivalently, these are the $\bs{\gamma}$'s for which $\tilde{\bs{\gamma}}=\mac{V}^{\frac{1}{2}}\bs{\gamma}$ belongs to the admissible set $\mac{G}$.
However, this well-definedness alone does not provide a justification for replacing the constrained maximization over $\mac{G}$ by an eigenvalue problem on the whole space $L_2^P$.
Therefore, unlike in the P-FMCCA case, a full-space eigenvalue characterization of FHOMALS is not established here.

% To identify the FHOMALS optimization problem with the eigenvalue problem for $\mac{V}^{\frac{1}{2}}\mac{W}^{-1}\mac{V}^{\frac{1}{2}}$, one would additionally need an admissibility result ensuring that the relevant eigenfunctions of $\mac{V}^{\frac{1}{2}}\mac{W}^{-1}\mac{V}^{\frac{1}{2}}$ belong to $\mac{G}$, or, more generally, an invariance property of the admissible set under the action of this operator.
% At present, Condition \ref{cond:cond3} guarantees the well-definedness of $\beta_p=\mac{V}_{pp}^{-1}\sum_{q=1}^P \mac{V}_{pq}\bs{\gamma}_q$, but it does not by itself imply such an admissibility or invariance property.
% Therefore, in contrast to P-FMCCA, the equivalence between the FHOMALS optimization problem on the whole space $L_2^P$ and the eigenvalue problem for $\mac{V}^{\frac{1}{2}}\mac{W}^{-1}\mac{V}^{\frac{1}{2}}$ remains unproved.

Formally, let $\mac{A}:=\mac{V}^{\frac{1}{2}}\mac{W}^{-\frac{1}{2}}$.
Then, the P-FMCCA and FHOMALS eigenvalue problems can be written, respectively, as
\[
  \mac{A}^*\mac{A}\tilde{\bs{\alpha}}
  =
  \rho \tilde{\bs{\alpha}},
  \qquad
  \mac{A}\mac{A}^*\tilde{\bs{\gamma}}
  =
  \mu \tilde{\bs{\gamma}}.
\]
This is the same formal adjoint relationship as in the ordinary finite-dimensional setting discussed in Section~\ref{subsec:homogeneity_analysis_multivariate}.
Nevertheless, in the present infinite-dimensional setting, this relationship is not sufficient to conclude that FHOMALS on the whole space is equivalent to P-FMCCA.

% The above discussion shows that the eigenanalyses appearing in P-FMCCA and FHOMALS are formally given by
% \banum
%   \mac{W}^{-\frac{1}{2}}\mac{V}\mac{W}^{-\frac{1}{2}}\tilde{\bs{\alpha}}
%   =
%   \rho \tilde{\bs{\alpha}},
%   \label{eq:gfcca_eig_again}
% \eanum
% and
% \banum
%   \mac{V}^{\frac{1}{2}}\mac{W}^{-1}\mac{V}^{\frac{1}{2}}\tilde{\bs{\gamma}}
%   =
%   \mu \tilde{\bs{\gamma}},
%   \label{eq:fhomals_eig_formal}
% \eanum
% respectively.

% If we define $\mac{A}:=\mac{V}^{\frac{1}{2}}\mac{W}^{-\frac{1}{2}}$, then \eqref{eq:gfcca_eig_again} and \eqref{eq:fhomals_eig_formal} can formally be written as
% \ba
%   \mac{A}^*\mac{A}\tilde{\bs{\alpha}}
%   =
%   \rho \tilde{\bs{\alpha}},
%   \qquad
%   \mac{A}\mac{A}^*\tilde{\bs{\gamma}}
%   =
%   \mu \tilde{\bs{\gamma}}.
% \ea
% This observation suggests the same adjoint relationship as in the ordinary finite-dimensional setting as discussed in Section \ref{subsec:homogeneity_analysis_multivariate}.
% Nevertheless, in the present infinite-dimensional setting, this formal relationship is not sufficient to conclude that FHOMALS on the whole space is equivalent to P-FMCCA.

\begin{Rem}
  A formal candidate for the $k$th FHOMALS component induced by the above adjoint relationship is
  \banum
    \tilde{\bs{\gamma}}'
    =\rho_k^{-\frac{1}{2}}\mac{V}^{\frac{1}{2}}\mac{W}^{-\frac{1}{2}}\tilde{\bs{\alpha}}_k.
    \label{eq:gamma_star_candidate}
  \eanum
  This expression is the infinite-dimensional analogue of the finite-dimensional relation $\tilde{\bs{\gamma}}_k=\delta_k^{-\frac{1}{2}} \mb{V}^{\frac{1}{2}}\mb{D}^{-\frac{1}{2}}\bs{\sigma}_k$ where $\bs{\sigma}_k$ is the $k$th column of $\mb{\Sigma}$ in \eqref{eq:mcca_eigen}, and $\delta_k$ is the corresponding eigenvalue.
  % A formal candidate induced by the above adjoint relationship is
  % \banum
  %   \tilde{\bs{\gamma}}^*
  %   =
  %   \rho^{-1/2}\mac{V}^{\frac{1}{2}}\mac{W}^{-\frac{1}{2}}\tilde{\bs{\alpha}}.
  %   \label{eq:gamma_star_candidate}
  % \eanum
  % This expression is analogous to the finite-dimensional relationship between HOMALS and MCCA.
  Moreover, under Condition \ref{cond:cond2} and the normalization adopted above, the candidate in \eqref{eq:gamma_star_candidate} is naturally regarded as belonging to the admissible set.
  However, the essential unresolved issue is not admissibility but optimality.
  More precisely, the present argument does not establish that $\tilde{\bs{\gamma}}'$ solves the FHOMALS optimization problem on the whole space $L_2^P$.
  Nor does it show that the constrained FHOMALS problem is equivalent to the eigenvalue problem for $\mac{V}^{\frac{1}{2}}\mac{W}^{-1}\mac{V}^{\frac{1}{2}}$ on the whole space.
  Hence, unlike in the finite-dimensional case, the whole-space correspondence between FHOMALS and P-FMCCA in the infinite-dimensional setting should be regarded as a formal analogy rather than a proved equivalence.
\end{Rem}

\begin{Rem}
  This distinction is not merely technical.
  For a compact self-adjoint operator $\mac{B}$ on a Hilbert space $\mbb{H}$, maximizing $\lan u,\mac{B}u\ran$ over a proper subset $\mac{G}$ of the unit sphere need not yield the leading eigenvector of $\mac{B}$ on $\mbb{H}$.
  For example, if $\mbb{H}=\ell^2$ with its norm $\|\cdot\|_{\ell^2}$, $\mac{B} e_j=j^{-1}e_j$, and 
  $
  \mac{G}=\{
    u\in \overline{\mathrm{span}\{e_2,e_3,\ldots\}}
    \mid\|u\|_{\ell^2}=1
  \}
  $, then the leading eigenvector $e_1$ is not admissible, and the maximizer over $\mac{G}$ is $e_2$.
  Thus, writing the FHOMALS criterion as a quadratic form does not, by itself, justify replacing the constrained FHOMALS problem by a full-space eigenvalue problem.
\end{Rem}

% \begin{example}
%   The above logical gap can be illustrated at an abstract level.
%   Let $\mac{H}$ be a Hilbert space, let $\mac{S}:=\{u\in \mac{H}\mid \|u\|=1\}$ be the unit sphere, and let $\mac{G}\subset \mac{S}$ be a proper subset.
%   Consider a compact self-adjoint operator $\mac{B}:\mac{H}\to \mac{H}$.
%   Even if one can write an objective function as the Rayleigh quotient $\lan u,\mac{B}u\ran$, the maximizer over $\mac{G}$ need not coincide with the leading eigenvector of $\mac{B}$ on $\mac{H}$.

%   Let $\mac{H}=\ell^2$ with canonical basis $\{e_j\}_{j\ge 1}$, and define
%   \ba
%     \mac{G}
%     =\left\{
%       u\in \overline{\mathrm{span}\{e_2,e_3,\ldots\}}
%       \;\middle|\;
%       \|u\|=1
%     \right\}.
%   \ea
%   Let $\mac{B}$ be the diagonal operator defined by  $\mac{B}e_j=j^{-1}e_j$ for $j\ge 1$.
%   Then $\mac{B}$ is compact and self-adjoint.
%   The leading eigenvector of $\mac{B}$ is $e_1$, but $e_1\notin\mac{G}$.
%   The maximization of $\lan u,\mac{B}u\ran$ over $\mac{G}$ is instead attained at $e_2$.
%   Thus, a quadratic-operator representation alone is not sufficient to replace a constrained maximization over $\mac{G}$ by an eigenvalue problem on the whole space.
% \end{example}

Although a whole-space equivalence between FHOMALS and P-FMCCA is not established by the preceding argument, the truncated canonical representation still yields a meaningful and tractable formulation.
Recall that $\bs{X}_{(K)}$ is the finite-rank truncated canonical representation defined in Section \ref{sec:canonical_decomposition}.
The next theorem considers FHOMALS for this finite-rank process, with the covariance operators $\mac{W}_{(K)}$ and $\mac{V}_{(K)}$ induced by $\bs{X}_{(K)}$.
It shows that the resulting FHOMALS components can be expressed in terms of the P-FMCCA components associated with the same finite-rank covariance structure $(\bs{X}_{(K)},\mac{V}_{(K)},\mac{W}_{(K)})$.
Thus, the theorem should be interpreted as a finite-rank population analogue of the classical HOMALS--MCCA correspondence.

In the following, for any operator $\mac{U}$, $\im({\mac{U}})$ denotes the range of $\mac{U}$.
Let $\mac{V}_{(K),pp}$ be the operator $\mac{V}_{pp}$ corresponding to $\bs{X}_{(K)}$.
Since both $\mac{V}_{(K),pp}$ and $\mac{W}_{(K)}$ are finite-rank operators, their ranges are closed.
Hence, the Moore--Penrose inverses are well-defined as bounded operators on $L_2$ and $L_2^P$, respectively.
In the following, we denote the Moore--Penrose inverse of an operator $\mac{U}$ by $\mac{U}^\dagger$.

For the finite-rank process $\bs{X}_{(K)}$, the block-diagonal covariance operator $\mac{W}_{(K)}$ is generally not invertible.
We therefore formulate the associated P-FMCCA problem using the Moore--Penrose inverse.
If $\bs{\alpha}$ satisfies $\lan\bs{\alpha},\mac{W}_{(K)}\bs{\alpha}\ran_P=1$, then $\tilde{\bs{\alpha}}=\mac{W}_{(K)}^{\frac{1}{2}}\bs{\alpha}$ satisfies $\|\tilde{\bs{\alpha}}\|_P=1$.
Conversely, for $\tilde{\bs{\alpha}}\in\im(\mac{W}_{(K)}^{\frac{1}{2}})$, we use the representative $\bs{\alpha}=(\mac{W}_{(K)}^{\frac{1}{2}})^\dagger\tilde{\bs{\alpha}}$.
This gives the P-FMCCA criterion in the normalized form
\ba
  \lan\tilde{\bs{\alpha}},(\mac{W}_{(K)}^{\frac{1}{2}})^\dagger\mac{V}_{(K)}(\mac{W}_{(K)}^{\frac{1}{2}})^\dagger\tilde{\bs{\alpha}}\ran_P,
  \quad
  \|\tilde{\bs{\alpha}}\|_P=1.
\ea
Hence, the P-FMCCA components are characterized by the eigenvalue problem 
\ba
  (\mac{W}_{(K)}^{\frac{1}{2}})^\dagger\mac{V}_{(K)}(\mac{W}_{(K)}^{\frac{1}{2}})^\dagger\tilde{\bs{\alpha}}=\rho\tilde{\bs{\alpha}}.
\ea

\begin{Theo}
  \label{theo:equivalence_fhomals_fmcca_truncated}
  Under the assumptions and notation of Theorem~\ref{theo:canonical_decomposition}, fix a finite integer $K$ with $1\le K\le r$.
  % Let $\mac{W}_{(K)}$ and $\mac{V}_{(K)}$ be the operators $\mac{W}$ and $\mac{V}$ obtained from $\bs{X}_{(K)}$, respectively.
  % Also, let $\mac{W}_{(K)}^\dagger$ be the Moore--Penrose inverse of $\mac{W}_{(K)}$.
  Let $m_{(K)}$ denote the number of canonical criterion values greater than one for the finite-rank problem based on $(\bs{X}_{(K)},\mac{V}_{(K)},\mac{W}_{(K)})$, counted with multiplicity, and let $\{(\rho_{k}^{(K)},\eta_{k}^{(K)},\bs{\alpha}_{k}^{(K)})\}_{k=1}^{m_{(K)}}$ be the corresponding P-FMCCA components.
  Then, the corresponding homogeneity components of $\bs{X}_{(K)}$ are given, for $k=1,\ldots,m_{(K)}$, by
  \ba
    \bs{\gamma}_k
    =(\rho_{k}^{(K)})^{-\frac{1}{2}}\bs{\alpha}_{k}^{(K)},
    \quad
    \bs{\beta}_k
    =\mac{W}_{(K)}^\dagger\mac{V}_{(K)}\bs{\gamma}_k,
    \quad\text{and}\quad
    f_k
    =\lan \bs{X}_{(K)},\bs{\gamma}_k\ran_P.
  \ea
  These quantities are well-defined.
  If any of these canonical criterion values has multiplicity greater than one, the statement is understood up to the usual choice of an orthonormal basis in the corresponding eigenspaces.
\end{Theo}

Even in the infinite-rank case $r=\infty$, the convergence $\bs{X}_{(K)}\ra\bs{X}_{(\infty)}$ does not by itself imply that the homogeneity components of $\bs{X}_{(K)}$ converge to those of the original process $\bs{X}$.
Indeed, Theorem~\ref{theo:canonical_decomposition} guarantees only that, as $K\ra\infty$, $\bs{X}_{(K)}\ra \bs{X}_{(\infty)}$ in the mean squared norm and that $\bs{X}_{(\infty)}$ and $\bs{X}$ share the same canonical components in the original $\mac{W}$-normalized sense.
Moreover, exact reconstruction of $\bs{X}$ requires the support condition stated in Theorem~\ref{theo:canonical_decomposition}(e), yielding $\bs{X}=\bs{X}_{(r)}$ a.s. if $r<\infty$ and $\bs{X}=\bs{X}_{(\infty)}$ a.s. if $r=\infty$.
Therefore, Theorem~\ref{theo:equivalence_fhomals_fmcca_truncated} does not claim that FHOMALS for $\bs{X}_{(K)}$ converges to FHOMALS for $\bs{X}$ without further assumptions.
Rather, it shows that each truncated process $\bs{X}_{(K)}$ admits a well-defined finite-rank FHOMALS representation through the canonical components.

% This point is important from both theoretical and practical perspectives.
% Theoretically, Theorem \ref{theo:equivalence_fhomals_fmcca_truncated} provides a setting in which the correspondence between P-FMCCA and FHOMALS can be established rigorously.
% Practically, the homogeneity score $f_k$ yields a one-dimensional summary of subject-specific variation, and this summarized score can be easier to interpret than the multivariate collection of canonical variates provided by P-FMCCA.

\begin{Rem}
  Theorem 3 shows more than the existence of FHOMALS components for the finite-rank process $\bs{X}_{(K)}$.
  It also identifies the value of the FHOMALS criterion in terms of the P-FMCCA canonical criterion values.
  For $m\le m_{(K)}$, the components constructed in Theorem~\ref{theo:equivalence_fhomals_fmcca_truncated} minimize the cumulative FHOMALS loss
  \ba
    \sum_{l=1}^m\sum_{p=1}^P \mbb{E}[(f_l-\lan X_{(K),p},\beta_{lp}\ran)^2]
  \ea
  where $X_{(K),p}$ and $\beta_{lp}$ are the $p$th elements of $\bs{X}_{(K)}$ and $\bs{\beta}_l$, respectively.
  The minimum value is $mP-\sum_{j=1}^m\rho_j^{(K)}$.
  % \ba
  %   mP-\sum_{j=1}^m\rho_j^{(K)}.
  % \ea
  Equivalently, the reduction in the cumulative FHOMALS loss is exactly $\sum_{l=1}^m\rho_{l}^{(K)}$.
  Thus, for the truncated canonical representation $\bs{X}_{(K)}$, the amount of homogeneity explained by the FHOMALS scores is measured by the same canonical criterion values that define the P-FMCCA components.
  This provides a specific reason for using the P-FMCCA-based truncated process $\bs{X}_{(K)}$, rather than an arbitrary finite-rank approximation such as a truncated KL expansion.
  For a general finite-rank approximation, FHOMALS components may still be defined, but their criterion values are not necessarily tied to the leading P-FMCCA association structure of the original process.  
\end{Rem}

Hence, although a whole-space equivalence is not established here, Theorem~\ref{theo:equivalence_fhomals_fmcca_truncated} gives a principled way to construct interpretable homogeneity components through the truncated canonical representation.

\section{Concluding Remarks}
\label{sec:concluding-remarks}

In this paper, we formulated functional multiple-set canonical correlation analysis for multivariate functional data at the population level (P-FMCCA) and investigated its theoretical structure.
We provided sufficient conditions for the well-definedness of the proposed framework and clarified how these conditions support the existence of canonical variates and a canonical decomposition.
We also formulated functional homogeneity analysis (FHOMALS) and examined the relationship between P-FMCCA and FHOMALS.
Our analysis clarified that, unlike in the ordinary finite-dimensional case, the usual eigenvalue-based argument does not by itself justify a whole-space equivalence between FHOMALS and P-FMCCA, whereas a well-defined finite-rank correspondence is established for the truncated canonical representation $\bs{X}_{(K)}$.

% One possible direction for extending the present framework is to incorporate regularization.
% \citet{HwangEtAl2012} considered a regularized formulation in the sample-based setting, and it may be possible to investigate analogous regularized versions of P-FMCCA and FHOMALS at the population level.
% In particular, it would be of interest to study how regularization modifies the domain issues caused by inverse covariance operators and how it affects the existence, uniqueness, and structural properties of the resulting solutions.
One possible direction for extending the present framework is to incorporate regularization. 
\cite{HwangEtAl2012} incorporated roughness-penalty regularization into the sample-based FMCCA framework, providing a natural starting point for developing a regularized version of P-FMCCA. 
More broadly, spectral truncation and quadratic regularization are standard approaches to ill-posed inverse problems in functional data analysis \citep{hall2007methodology}, while regularization has also been studied specifically for bivariate FCCA \citep{leurgans1993canonical,cupidon2008some}. 
It would therefore be useful to investigate how these forms of regularization can be incorporated into the multiple-set population setting and how they modify the domain conditions, existence, uniqueness, and structural properties established in the present paper.

Another promising direction is to reformulate the proposed population theory in a reproducing kernel Hilbert space.
As in \citet{eubank2008canonical}, who formulated functional CCA in an RKHS framework, it seems plausible that P-FMCCA could also be formulated in an RKHS framework.
Such an approach may provide a natural way to handle smoother function classes and to connect the present covariance-operator formulation with kernel-based regularization methods.
It would therefore be worthwhile to examine whether the proposed canonical decomposition and the corresponding homogeneity components admit a useful reformulation in such a setting.

There are also several limitations that remain unresolved in the present paper.
First, although the formal relationship between P-FMCCA and FHOMALS was clarified, the equivalence between FHOMALS on the whole space $L_2^P$ and P-FMCCA has not been established in the infinite-dimensional setting.
Second, Theorem~\ref{theo:equivalence_fhomals_fmcca_truncated} provides a well-defined finite-rank characterization of the homogeneity components for the truncated canonical representation $\bs{X}_{(K)}$, but this does not by itself imply that these components converge to those for the original process $\bs{X}$ in the infinite-rank case as $K\ra\infty$.
Third, the present paper does not develop a full inferential or computational theory for regularized estimation in the practically relevant sample-based setting.
These issues should be addressed in future work.

% In particular, it would be valuable to establish sharper conditions under which P-FMCCA and FHOMALS become equivalent, to analyze the convergence of homogeneity components derived from $X_{(K)}$, and to construct regularized estimation procedures with theoretical guarantees.
% We hope that the present population-level formulation provides a useful foundation for such further developments in the analysis of multivariate functional data.

%%%%%%%%%%%%%%%%%%%%%%%%%%%%%%%%%%%%%%%%%%
%%            Appendix
%%%%%%%%%%%%%%%%%%%%%%%%%%%%%%%%%%%%%%%%%%
\appendix

%%%%%%%%%%%%%%%%%%%%%%%%%%%%%%%%%%%%%%%%%%
%%              Proofs
%%%%%%%%%%%%%%%%%%%%%%%%%%%%%%%%%%%%%%%%%%
\section{Proofs of Theorems}

\subsection{Proof of Theorem \ref{theo:optimization-P-FMCCA}}
\label{subsec:proofs-of-theorem_optimization-P-FMCCA}

\noindent
(a) Recall that for a vector $\bs{z}\in L_{2}^{P}$, $(\bs{z})_{p}$ denotes its $p$th component.
Let $\bs{e}_{pj}$ denote the element of $L_2^P$ whose $p$th component is the $j$th eigenfunction $\phi_{pj}$ of $\mac{V}_{pp}$ and whose other components are zero (see the proof of Proposition~\ref{prop:extend_R} in the Supplementary Material \ref{SI-subsec:proof-prop-extend_R}).
Also, let us define $\omega_{pqij}:=\Corr(\xi_{qi},\xi_{pj})=\zeta_{qi}^{-\frac{1}{2}}\zeta_{pj}^{-\frac{1}{2}}\mbb{E}[\xi_{qi}\xi_{pj}]$ for $p,q\in[P]$ and $i,j\in\mbb{N}$.
Since $\tilde{\mac{R}}$ is compact and self-adjoint on $L_2^P$, it admits a spectral decomposition.
Let $(\lambda,\tilde{\bs{\alpha}})$ be an eigenpair of $\tilde{\mac{R}}$ with $\lambda\ne0$.
The eigenfunctions corresponding to positive eigenvalues appearing in the statement of the theorem are particular cases of such eigenfunctions.
Since $\lambda\tilde{\bs{\alpha}}=\tilde{\mac{R}}\tilde{\bs{\alpha}}$, it is enough to show that $\tilde{\mac{R}}\tilde{\bs{\alpha}}\in F_W$.
% Then
% \ba
%   \tilde{\mac{R}}
%   =\sum_{k=1}^\infty\lambda_k\tilde{\bs{\alpha}}_k\otimes_P\tilde{\bs{\alpha}}_k
% \ea
% holds.
Based on the definition of $\tilde{\mac{R}}$ in the proof of Proposition~\ref{prop:extend_R}, we have
% \ba
%   \lambda_k\tilde{\bs{\alpha}}_k
%   =\tilde{\mac{R}}\tilde{\bs{\alpha}}_k
%   =\sum_{p=1}^P\sum_{q\neq p}\sum_{i,j=1}^\infty\omega_{pqij}\lan\tilde{\bs{\alpha}}_k,\bs{e}_{qi}\ran_P\bs{e}_{pj}.
% \ea
\ba
  \tilde{\mac{R}}\tilde{\bs{\alpha}}
  =\sum_{p=1}^P\sum_{q\neq p}\sum_{i,j=1}^\infty\omega_{pqij}\lan\tilde{\bs{\alpha}},\bs{e}_{qi}\ran_P\bs{e}_{pj}.
\ea
Also, for any $\bs{z}\in L_2^P$, we have
\ba
  (\tilde{\mac{R}}\bs{z})_r
  &=\sum_{q\neq r}\sum_{i,j=1}^\infty\omega_{rqij}\lan z_q,\phi_{qi}\ran\phi_{rj}.
\ea
Hence, we have
\ba
  (\tilde{\mac{R}}\tilde{\bs{\alpha}})_r
  =\sum_{q\neq r}\sum_{i,j=1}^\infty\omega_{rqij}\lan \tilde{\alpha}_q,\phi_{qi}\ran\phi_{rj}
  :=h_r.
\ea
% Since $\lambda_k\neq0$, it is enough to show that $\bs{h}_k=(h_{k1},\dots,h_{kP})^\top\in F_W$.
% Therefore, we can observe that $\tilde{\bs{\alpha}}_k\in F_W$ is equivalent to $\bs{h}_k=(h_{k1},\dots,h_{kP})^\top\in F_W$.

To show that $\tilde{\mac{R}}\tilde{\bs{\alpha}}\in F_W$, it remains to verify, for each $r\in[P]$, that
\ba
  Q_r
  :=\sum_{l=1}^{\infty}\zeta_{rl}^{-1}\left|\left<h_{r}, \phi_{rl}\right>\right|^{2}
  <\infty.
\ea

We have
 \begin{align*}
  Q_r
  % &=\sum_{l=1}^{\infty}\zeta_{rl}^{-1}\left|\left<h_{kr}, \phi_{rl}\right>\right|^{2}\\
  % &=\sum_{l=1}^{\infty}\zeta_{rl}^{-1}\left|\left<\sum_{q,r:q\neq r}\sum_{i=1}^\infty\sum_{j=1}^\infty \omega_{rqij}\lan \tilde{\alpha}_{kq},\phi_{qi}\ran\phi_{rj}, \phi_{rl}\right>\right|^{2}\\
  &=\sum_{l=1}^{\infty}\zeta_{rl}^{-1}\left|\sum_{q\neq r}\sum_{i=1}^{\infty}\zeta_{qi}^{-\frac{1}{2}}\zeta_{rl}^{-\frac{1}{2}}\mbb{E}[\xi_{qi}\xi_{rl}]\left<\phi_{qi},\tilde{\alpha}_{q}\right>\right|^{2}\\
  &\le\sum_{l=1}^{\infty}\zeta_{rl}^{-1}\left(\sum_{q\neq r}\sum_{i=1}^{\infty}\zeta_{qi}^{-1}\zeta_{rl}^{-1}(\mbb{E}[\xi_{qi}\xi_{rl}])^{2}\right)
  \left(\sum_{q\neq r}\sum_{i=1}^{\infty}|\left<\phi_{qi},\tilde{\alpha}_{q}\right>|^{2}\right).
 \end{align*}
 Here, $\{\phi_{qj}\}$ is an orthonormal system in $L_2$. 
 Thus, by Bessel's inequality, we obtain
 \begin{equation*}
	\sum_{q\neq r}\sum_{i=1}^{\infty}|\left<\phi_{qi},\tilde{\alpha}_{q}\right>|^{2}
	 \le\sum_{q\neq r}\|\tilde{\alpha}_{q}\|^{2}
   \le\|\tilde{\bs{\alpha}}\|_P^2
   <\infty.
 \end{equation*}
 Therefore,
 \begin{align*}
	Q_{r}
	&\le \|\tilde{\bs{\alpha}}\|_P^2\sum_{l=1}^{\infty}\zeta_{rl}^{-1}\sum_{q\neq r}\sum_{i=1}^{\infty}\zeta_{qi}^{-1}\zeta_{rl}^{-1}(\mbb{E}[\xi_{qi}\xi_{rl}])^{2}
	=\|\tilde{\bs{\alpha}}\|_P^2\sum_{q\neq r}\sum_{l=1}^{\infty}\sum_{i=1}^{\infty}\frac{(\mbb{E}[\xi_{qi}\xi_{rl}])^{2}}{\zeta_{qi}\zeta_{rl}^{2}}.
 \end{align*}
 The last expression is finite by Condition~\ref{cond:cond2}, applied to the ordered pair $(r,q)$ for each $q\neq r$.
 Hence, $Q_r<\infty$ for every $r\in[P]$.
 Moreover, by construction, $h_r$ belongs to the closed linear span of $\{\phi_{rl}\}_{l\ge 1}$, and hence $h_r\perp\ker(\mac{V}_{rr})$.
 Thus, $\tilde{\mac{R}}\tilde{\bs{\alpha}}\in F_W$.
 Since $\lambda\ne0$ and $\lambda\tilde{\bs{\alpha}}=\tilde{\mac{R}}\tilde{\bs{\alpha}}$, it follows that $\tilde{\bs{\alpha}}\in F_W$.
%  Thus, for $(p,q)\in[P]^{2}$ with $p\neq q$, the boundedness of $\sum_{i=1}^{\infty}\sum_{j=1}^{\infty}\frac{(\mbb{E}[\xi_{pi}\xi_{qj}])^{2}}{\zeta_{pi}^{2}\zeta_{qj}}$ ensures Condition \ref{cond:cond2}.

%%%%%%%%%%%%%%%%%%%%%%%%%%%%%%%%%%%%%%%%%%
\vspace{20pt}
\noindent
(b) 
First, we give a lemma required for the proof of the theorem.
The proof of the lemma is given in the Supplementary Material \ref{SI-subsec:proof-lemma_for_theorem_optimization}.\begin{Lemm}
  \label{lemm:lemma_for_theorem_optimization}
  Assume the same conditions as in Theorem \ref{theo:optimization-P-FMCCA}.
  % Let $\Pi=\{\pi\subset L_2^P\mid\dim(\pi^\perp\cap\overline{F_W})=K\}$ be a collection of subspaces in $L_2^P$ with co-dimensions equal to $K$.
  Then, for each $m=0,\dots,K-1$, 
  \ba
    1+\tilde{\rho}_{m+1}
    =\sup_{\tilde{\bs{\alpha}}\in(\myspan\{\tilde{\bs{\alpha}}_1,\dots,\tilde{\bs{\alpha}}_m\}^\perp\cap\overline{F_W})\backslash\{0\}}
    \frac{\lan\tilde{\bs{\alpha}},(I+\tilde{\mac{R}})\tilde{\bs{\alpha}}\ran_P}{\|\tilde{\bs{\alpha}}\|_P^2}.
  \ea
  % \begin{itemize}
  %   \item[(a)] $1+\tilde{\rho}_{K+1}
  %   =\sup_{\tilde{\bs{\alpha}}\in\myspan\{\tilde{\bs{\alpha}}_1,\dots,\tilde{\bs{\alpha}}_K\}^\perp\cap\overline{F_W}}
  %   \frac{\lan\tilde{\bs{\alpha}},(I+\tilde{\mac{R}})\tilde{\bs{\alpha}}\ran_P}{\|\tilde{\bs{\alpha}}\|_P^2}$, and
  %   \item[(b)] $1+\tilde{\rho}_{K+1}
  %   =\inf_{\pi\in\Pi}\sup_{\tilde{\bs{\alpha}}\in\pi}\frac{\lan\tilde{\bs{\alpha}},(I+\tilde{\mac{R}})\tilde{\bs{\alpha}}\ran_P}{\|\tilde{\bs{\alpha}}\|_P^2}$.
  % \end{itemize}
  When $m=0$, the orthogonality constraint is omitted.
\end{Lemm}

\vspace{20pt}
% We now turn to the proof of Theorem \ref{theo:optimization-P-FMCCA} (b).
% Set $\rho_k'=\tilde{\rho}_k$, $\bs{\alpha}_k'=\mac{W}^{-\frac{1}{2}}\tilde{\bs{\alpha}}_k$, and $\eta_k'=\lan\bs{X},\bs{\alpha}_k'\ran_P$ for $k\in[K]$.
Set $\bs{\alpha}_k'=\mac{W}^{-\frac{1}{2}}\tilde{\bs{\alpha}}_k$ and $\eta_k'=\lan\bs{\alpha}_k',\bs{X}\ran_P$ for $k\in[K]$.
From $\mbb{E}[\bs{X}]=0$, we have $\mbb{E}[\eta_k']=0$ and
\ba
  \Var(\eta_k')
  % &=\mbb{E}[\eta_k'^2]
  % =\mbb{E}[\lan\bs{X},\bs{\alpha}_k'\ran_P^2]
  =\lan\bs{\alpha}_k',\mac{V}\bs{\alpha}_k'\ran_P
  % &=\lan\mac{W}^{-\frac{1}{2}}\tilde{\bs{\alpha}}_k,\mac{V}\mac{W}^{-\frac{1}{2}}\tilde{\bs{\alpha}}_k\ran_P
  % =\lan\tilde{\bs{\alpha}}_k,\mac{W}^{-\frac{1}{2}}\mac{V}\mac{W}^{-\frac{1}{2}}\tilde{\bs{\alpha}}_k\ran_P\\
  =1+\lan\tilde{\bs{\alpha}}_k,\mac{W}^{-\frac{1}{2}}\mac{V}_{\star}\mac{W}^{-\frac{1}{2}}\tilde{\bs{\alpha}}_k\ran_P
  % =1+\lan\tilde{\bs{\alpha}}_k,\tilde{\rho}_k\tilde{\bs{\alpha}}_k\ran_P
  =1+\tilde{\rho}_k.
\ea

Thus, it remains to show that these candidate weight functions solve the P-FMCCA optimization problem.
First, consider the case $k=1$.
For any $\bs{\alpha}\in F_W^{-1}$ with $\lan\bs{\alpha},\mac{W}\bs{\alpha}\ran_P=1$, define $\tilde{\bs{\alpha}}=\mac{W}^{\frac{1}{2}}\bs{\alpha}$.
Note that $\|\tilde{\bs{\alpha}}\|_P^2=1$.
% \ba
%   \|\tilde{\bs{\alpha}}\|_P^2
%   =\lan\mac{W}^{\frac{1}{2}}\bs{\alpha},\mac{W}^{\frac{1}{2}}\bs{\alpha}\ran_P
%   =\lan\bs{\alpha},\mac{W}\bs{\alpha}\ran_P
%   =1.
% \ea
Then, from Lemma \ref{lemm:lemma_for_theorem_optimization} with $m=0$, we have
\ba
  \lan\bs{\alpha},\mac{V}\bs{\alpha}\ran_P
  % &=\lan
  %   \mac{W}^{-\frac{1}{2}}\tilde{\bs{\alpha}},
  %   \mac{V}\mac{W}^{-\frac{1}{2}}\tilde{\bs{\alpha}}
  % \ran_P
  &=\lan
    \tilde{\bs{\alpha}},
    \mac{W}^{-\frac{1}{2}}\mac{V}\mac{W}^{-\frac{1}{2}}\tilde{\bs{\alpha}}
  \ran_P
  =\lan\tilde{\bs{\alpha}},(\mac{I}+\tilde{\mac{R}})\tilde{\bs{\alpha}}\ran_P\\
  % \le 1+\tilde{\rho}_1\|\tilde{\bs{\alpha}}\|_P^2
  &\le 1+\tilde{\rho}_1
  =1+\lan\tilde{\bs{\alpha}}_1,\tilde{\mac{R}}\tilde{\bs{\alpha}}_1\ran_P
  % =\lan\tilde{\bs{\alpha}}_1,(\mac{I}+\tilde{\mac{R}})\tilde{\bs{\alpha}}_1\ran_P\\
  % =\lan\tilde{\bs{\alpha}}_1,\mac{W}^{-\frac{1}{2}}\mac{V}\mac{W}^{-\frac{1}{2}}\tilde{\bs{\alpha}}_1\ran_P
  =\lan\bs{\alpha}_1',\mac{V}\bs{\alpha}_1'\ran_P.
\ea
Taking the supremum over all such $\bs{\alpha}$, that is, over all $\bs{\alpha}\in F_W^{-1}$ such that $\lan\bs{\alpha},\mac{W}\bs{\alpha}\ran_P=1$, we obtain
\ba
  \sup_{\bs{\alpha}\in F_W^{-1}:\lan\bs{\alpha},\mac{W}\bs{\alpha}\ran_P=1}\lan\bs{\alpha},\mac{V}\bs{\alpha}\ran_P
  \le \lan\bs{\alpha}_1',\mac{V}\bs{\alpha}_1'\ran_P.
\ea
Using $F_W^{-1}=\overline{F_W}$, this is equivalent to 
\banum
  \sup_{\bs{\alpha}\in \overline{F_W}:\lan\bs{\alpha},\mac{W}\bs{\alpha}\ran_P=1}\lan\bs{\alpha},\mac{V}\bs{\alpha}\ran_P
  \le \lan\bs{\alpha}_1',\mac{V}\bs{\alpha}_1'\ran_P.
  \label{eq:proof-theorem3-1}
\eanum

Here, we have $\im(\mac{V})\subset\overline{F_W}$.
For any $\bs{\alpha}\in L_2^P$, using the decomposition $\bs{\alpha}=\bs{u}+\bs{u}^{\perp}$ with $\bs{u}\in\overline{F_W}$ and $\bs{u}^{\perp}\in\overline{F_W}^\perp$, we have
\ba
  \lan\bs{\alpha},\mac{V}\bs{\alpha}\ran_P
  &=\lan\bs{u},\mac{V}\bs{\alpha}\ran_P
  =\lan\mac{V}\bs{u},\bs{\alpha}\ran_P
  =\lan\mac{V}\bs{u},\bs{u}\ran_P.
\ea
Also, using $\overline{F_W}=(\ker(\mac{W}))^\perp$, we have $\lan\bs{\alpha},\mac{W}\bs{\alpha}\ran_P=\lan\bs{u},\mac{W}\bs{u}\ran_P$.
Therefore, 
\ba
  \sup_{\bs{\alpha}\in L_2^P:\lan\bs{\alpha},\mac{W}\bs{\alpha}\ran_P=1}\lan\bs{\alpha},\mac{V}\bs{\alpha}\ran_P
  % =\sup_{\bs{\alpha}\in L_2^P}\lan\bs{u},\mac{V}\bs{u}\ran_P
  =\sup_{\bs{u}\in\overline{F_W}:\lan\bs{u},\mac{W}\bs{u}\ran_P=1}\lan\bs{u},\mac{V}\bs{u}\ran_P.
  % \label{eq:proof-theorem3-2}
\ea
From this fact and \eqref{eq:proof-theorem3-1}, we have
\ba
  \sup_{\bs{\alpha}\in L_2^P:\lan\bs{\alpha},\mac{W}\bs{\alpha}\ran_P=1}\lan\bs{\alpha},\mac{V}\bs{\alpha}\ran_P
  \le\lan\bs{\alpha}_1',\mac{V}\bs{\alpha}_1'\ran_P.
\ea

Since $\bs{\alpha}_1'\in F_W^{-1}\subset L_2^P$ and
$\lan\bs{\alpha}_1',\mac{W}\bs{\alpha}_1'\ran_P=1$,
the reverse inequality is trivial.
Therefore, we have
\ba
  \sup_{\bs{\alpha}\in L_2^P:\lan\bs{\alpha},\mac{W}\bs{\alpha}\ran_P=1}\lan\bs{\alpha},\mac{V}\bs{\alpha}\ran_P
  =\lan\bs{\alpha}_1',\mac{V}\bs{\alpha}_1'\ran_P.
\ea
Then, from the definition of the P-FMCCA components $\rho_1$ and $\bs{\alpha}_1$, we have $\rho_1=1+\tilde{\rho}_1$ and $\bs{\alpha}_1$ can be chosen as $\bs{\alpha}_1'=\mac{W}^{-\frac{1}{2}}\tilde{\bs{\alpha}}_1$.

Next, we prove the assertion for $k=2,\ldots,K$ by mathematical induction.
Fix $k\in\{2,\dots,K\}$ and assume that the first $k-1$ canonical components can be chosen as $\{(1+\tilde{\rho}_l,\eta_l',\bs{\alpha}_l')\}_{l=1,\dots,k-1}$.
% Again, set $\tilde{\rho}_k'=\tilde{\rho}_k$ and
Again, set $\bs{\alpha}_k'=\mac{W}^{-\frac{1}{2}}\tilde{\bs{\alpha}}_k$ and $\eta_k'=\lan\bs{\alpha}_k',\bs{X}\ran_P$.
For $l\in[k-1]$, the candidate component satisfies
$\mac{V}\bs{\alpha}_l'=(1+\tilde{\rho}_l)\mac{W}\bs{\alpha}_l'$.
Hence, for any $\bs{\alpha}\in L_2^P$, we have
$\mbb{E}[\eta(\bs{\alpha})\eta_l']=(1+\tilde{\rho}_l)\lan\bs{\alpha},\mac{W}\bs{\alpha}_l'\ran_P$.
Since $1+\tilde{\rho}_l>0$, the constraint
$\mbb{E}[\eta(\bs{\alpha})\eta_l']=0$ is equivalent to
$\lan\bs{\alpha},\mac{W}\bs{\alpha}_l'\ran_P=0$.

For any $\bs{\alpha}\in F_W^{-1}$ such that $\lan\bs{\alpha},\mac{W}\bs{\alpha}\ran_P=1$ and $\lan\bs{\alpha},\mac{W}\bs{\alpha}_l'\ran_P=0$ for all $l\in[k-1]$, define $\tilde{\bs{\alpha}}=\mac{W}^{\frac{1}{2}}\bs{\alpha}$.
Similar to the case of $k=1$, Lemma~\ref{lemm:lemma_for_theorem_optimization} with $m=k-1$ gives $\lan\bs{\alpha},\mac{V}\bs{\alpha}\ran_P \le \lan\bs{\alpha}_k',\mac{V}\bs{\alpha}_k'\ran_P$.
% \ba
%   |\lan\bs{\alpha},\mac{V}\bs{\alpha}\ran_P|
%   &=|1+\lan\tilde{\bs{\alpha}},\mac{R}\tilde{\bs{\alpha}}\ran_P|
%   \le 1+\|\tilde{\bs{\alpha}}\|_P\|\mac{R}\tilde{\bs{\alpha}}\|_P\\
%   &\le 1+\rho_k\|\tilde{\bs{\alpha}}\|_P
%   =1+\lan\tilde{\bs{\alpha}}_K,\mac{R}\tilde{\bs{\alpha}}_K\ran_P\\
%   &=\lan\bs{\alpha}_K,\mac{V}\bs{\alpha}_K\ran_P.
% \ea
% We apply Lemma \ref{lemm:lemma_for_theorem_optimization} (a) with $K$ replaced by $K-1$ to obtain the second line from the first.
Define the set $\mcr{A}_k(B)=\{\bs{\alpha}\in B\mid\lan\bs{\alpha},\mac{W}\bs{\alpha}\ran_P=1,\ \lan\bs{\alpha},\mac{W}\bs{\alpha}_l'\ran_P=0\ \text{for all } l\in[k-1]\}$ for $B\subset L_2^P$.
Thus, we obtain
\ba
  \sup_{\bs{\alpha}\in \mcr{A}_k(F_W^{-1})}\lan\bs{\alpha},\mac{V}\bs{\alpha}\ran_P
  \le\lan\bs{\alpha}_k',\mac{V}\bs{\alpha}_k'\ran_P.
\ea
By the same decomposition argument as in the case $k=1$, the constraints defining $\mac{A}_k(L_2^P)$ are preserved after projection onto $\overline{F_W}$.
Hence,
\ba
  \sup_{\bs{\alpha}\in \mcr{A}_k(L_2^P)}\lan\bs{\alpha},\mac{V}\bs{\alpha}\ran_P
  =\sup_{\bs{\alpha}\in \mcr{A}_k(\overline{F_W})}\lan\bs{\alpha},\mac{V}\bs{\alpha}\ran_P
  =\sup_{\bs{\alpha}\in \mcr{A}_k(F_W^{-1})}\lan\bs{\alpha},\mac{V}\bs{\alpha}\ran_P
  \le\lan\bs{\alpha}_k',\mac{V}\bs{\alpha}_k'\ran_P.
\ea

Since $\bs{\alpha}_k'\in\mac{A}_k(F_W^{-1})\subset\mcr{A}_k(L_2^P)$, the reverse inequality is trivial.
Hence,
\ba
  \sup_{\bs{\alpha}\in \mcr{A}_k(L_2^P)}\lan\bs{\alpha},\mac{V}\bs{\alpha}\ran_P
  =\lan\bs{\alpha}_k',\mac{V}\bs{\alpha}_k'\ran_P.
\ea
Therefore, for each $k=2,\ldots,K$, the $k$th canonical component can be chosen so that $\rho_k=1+\tilde{\rho}_k$ and $\bs{\alpha}_k=\bs{\alpha}_k'=\mac{W}^{-\frac{1}{2}}\tilde{\bs{\alpha}}_k$.
Together with the case $k=1$, this proves the assertion for all $k\in[K]$.
% $(\rho_k,\bs{\alpha}_k)=(1+\rho_k',\bs{\alpha}_k')$ for all $k\in[K]$. 
% That is, for all $k\in[K]$, $\Var(\eta_k)=\rho_k=1+\tilde{\rho}_k$ and $\bs{\alpha}_k=\mac{W}^{-\frac{1}{2}}\tilde{\bs{\alpha}}_k$ hold.

%%%%%%%%%%%%%%%%%%%%%%%%%%%%%%%%%%%%%%%%%%
\vspace{10pt}
\noindent
(c) Observe that, for $k,l\in[K]$,
\ba
  \Corr(\eta_k',\eta_l')
  % &=\frac{\Cov(\eta_k',\eta_l')}{\sqrt{\Var(\eta_k')\Var(\eta_l')}}
  =\frac{\mbb{E}[\eta_k'\eta_l']}{\sqrt{(1+\tilde{\rho}_k)(1+\tilde{\rho}_l)}}
  % &=\frac{1}{\sqrt{(1+\tilde{\rho}_k)(1+\tilde{\rho}_l)}}\lan\bs{\alpha}_k',\mac{V}\bs{\alpha}_l'\ran_P
  % =\frac{1}{\sqrt{(1+\tilde{\rho}_k)(1+\tilde{\rho}_l)}}\lan\mac{W}^{-\frac{1}{2}}\tilde{\bs{\alpha}}_k,\mac{V}\mac{W}^{-\frac{1}{2}}\tilde{\bs{\alpha}}_l\ran_P\\
  % &=\frac{1}{\sqrt{(1+\tilde{\rho}_k)(1+\tilde{\rho}_l)}}\lan\tilde{\bs{\alpha}}_k,\mac{W}^{-\frac{1}{2}}\mac{V}\mac{W}^{-\frac{1}{2}}\tilde{\bs{\alpha}}_l\ran_P\\
  =\frac{1}{\sqrt{(1+\tilde{\rho}_k)(1+\tilde{\rho}_l)}}\lan\tilde{\bs{\alpha}}_k,(1+\tilde{\rho}_{l})\tilde{\bs{\alpha}}_l\ran_P
  % &=\begin{cases}
  %   0 & \text{if } k\neq l,\\
  %   \frac{1}{\sqrt{(1+\tilde{\rho}_k)^2}}(1+\tilde{\rho}_k)\lan\tilde{\bs{\alpha}}_k,\tilde{\bs{\alpha}}_k\ran_P & \text{if } k=l.
  % \end{cases}\\
  % &=\begin{cases}
  %   0 & \text{if } k\neq l,\\
  %   1 & \text{if } k=l.
  % \end{cases}
  =\delta_{kl}.
\ea
From (b), we have $\eta_k=\eta_k'$, and hence statement (c) follows.

\subsection{Proof of Theorem \ref{theo:canonical_decomposition}}
\label{subsec:proofs_of_theorem_canonical-decomposition}

\vspace{10pt}
\noindent

We first state a lemma necessary for the proof of the theorem, which is proved in \ref{SI-subsec:proof-lemma_for_theorem_canonical-decomposition}.
\begin{Lemm}
	\label{lemma:lemma_for_theorem_canonical-decomposition}
	\noindent
  For any $\bs{u}\in F_W^{-1}$, the following statements hold.
	% \begin{enumerate}
	%  \item[(1)]
	% 					 $\left<\bs{u},\bs{X}_{(K)}\right>_{P}=\left<\bs{u},\bs{X}\right>_{P}$
	% 					 and $\left<\bs{u},\bs{X}_{(K)}^{\perp}\right>_{P}=0$ if
	% 					 $\mac{W}^{\frac{1}{2}}\bs{u}\in\Span\{\tilde{\bs{\alpha}}_{k},k\in[K]\}$, and
	%  \item[(2)] $\left<\bs{u},\bs{X}_{(K)}\right>_{P}=0$ and
	% 						$\left<\bs{u},\bs{X}_{(K)}^{\perp}\right>_{P}=\left<\bs{u},\bs{X}\right>_{P}$
	% 						if
	% 					 $\mac{W}^{\frac{1}{2}}\bs{u}\in\Span\{\tilde{\bs{\alpha}}_{k},k\in[K]\}^{\perp}$.
	% \end{enumerate}
	\begin{enumerate}
	 \item[(1)] If $\mac{W}^{\frac{1}{2}}\bs{u}\in\Span\{\tilde{\bs {\alpha}}_{k},k\in[K]\}$, then $\left<\bs{u},\bs{X}_{(K)}\right>_{P}=\left<\bs{u},\bs{X}\right>_{P}$ a.s. and $\left<\bs{u},\bs{X}_{(K)}^{\perp}\right>_{P}=0$ a.s.,
	 \item[(2)] If $\mac{W}^{\frac{1}{2}}\bs{u}\in\Span\{\tilde{\bs{\alpha}}_{k},k\in[K]\}^{\perp}$, then $\left<\bs{u},\bs{X}_{(K)}\right>_{P}=0$ a.s. and	$\left<\bs{u},\bs{X}_{(K)}^{\perp}\right>_{P}=\left<\bs{u},\bs{X}\right>_{P}$ a.s.
	\end{enumerate}
\end{Lemm}

\noindent
(a) For $\bs{u}\in F_W^{-1}$, we consider a decomposition $\bs{u}=\mac{W}^{-\frac{1}{2}}\mac{P}_{K}\mac{W}^{\frac{1}{2}}\bs{u}+(\mac{I}-\mac{W}^{-\frac{1}{2}}\mac{P}_{K}\mac{W}^{\frac{1}{2}})\bs{u}=\bs{u}_{1}+\bs{u}_{2}$. 
 Clearly, $\mac{W}^{\frac{1}{2}}\bs{u}_{1}\in\Span\{\tilde{\bs{\alpha}}_{k}, k\in[K]\}$ and $\mac{W}^{\frac{1}{2}}\bs{u}_{2}\in\Span\{\tilde{\bs{\alpha}}_{k}, k\in[K]\}^{\perp}$. 
%  By Lemma \ref{lemma:lemma_for_theorem_canonical-decomposition},
%  \begin{equation*}
% 	\left<\bs{u},\bs{X}_{(K)}\right>_P
%   =\left<\bs{u}_{1},\bs{X}_{(K)}\right>_P
%   =\left<\bs{u}_{1},\bs{X}\right>_P.
%  \end{equation*}

First, let us confirm that the first canonical components of $\bs{X}_{(K)}$ and $\bs{X}$ coincide with each other.
We write $\mac{T}=\mac{W}^{-\frac{1}{2}}\mac{P}_K\mac{W}^{\frac{1}{2}}$ and define $\mcr{A}_1=\{\bs{u}\in F_W^{-1}\mid\lan\bs{u},\mac{W}\bs{u}\ran_P=1\}$.
By Lemma \ref{lemma:lemma_for_theorem_canonical-decomposition}, 
$
\left<\bs{u},\bs{X}_{(K)}\right>_P
  =\left<\bs{u}_{1},\bs{X}_{(K)}\right>_P
  =\left<\bs{u}_{1},\bs{X}\right>_P
$ a.s.
% \ba
%   \left<\bs{u},\bs{X}_{(K)}\right>_P
%   =\left<\bs{u}_{1},\bs{X}_{(K)}\right>_P
%   =\left<\bs{u}_{1},\bs{X}\right>_P.
% \ea
Then, we have
\ba
  \sup_{\bs{u}\in\mcr{A}_1}\mbb{E}[\left<\bs{u},\bs{X}_{(K)}\right>_{P}^{2}]
  =\sup_{\bs{u}\in\mcr{A}_1}\mbb{E}[\left<\bs{u}_{1},\bs{X}\right>_{P}^{2}]
  =\sup_{\bs{u}_1\in\mcr{A}_2}\lan\bs{u}_1,\mac{V}\bs{u}_1\ran_P
\ea
where $\mcr{A}_2=\mac{T}\mcr{A}_1=\{\mac{T}\bs{u}\mid\bs{u}\in\mcr{A}_1\}$.
Noting that $\lan\bs{u},\mac{W}\bs{u}\ran_P=1$ implies $\|\mac{W}^{\frac{1}{2}}\bs{u}\|_P^2=\lan\mac{W}^{\frac{1}{2}}\bs{u},\mac{W}^{\frac{1}{2}}\bs{u}\ran_P=1$, we have, for any $\bs{u}\in\mcr{A}_1$, 
\ba
 \lan\mac{T}\bs{u},\mac{W}\mac{T}\bs{u}\ran_P
%  &=\lan\mac{W}^{-\frac{1}{2}}\mac{P}_K\mac{W}^{\frac{1}{2}}\bs{u},\mac{W}\mac{W}^{-\frac{1}{2}}\mac{P}_K\mac{W}^{\frac{1}{2}}\bs{u}\ran_P\\
%  &=\lan\mac{W}^{-\frac{1}{2}}\mac{P}_K\mac{W}^{\frac{1}{2}}\bs{u},\mac{W}^{\frac{1}{2}}\mac{P}_K\mac{W}^{\frac{1}{2}}\bs{u}\ran_P\\
 &=\lan\mac{P}_K\mac{W}^{\frac{1}{2}}\bs{u},\mac{P}_K\mac{W}^{\frac{1}{2}}\bs{u}\ran_P
%  &=\|\mac{P}_K\mac{W}^{\frac{1}{2}}\bs{u}\|_P^2\\
 \le\|\mac{W}^{\frac{1}{2}}\bs{u}\|_P^2
 =1.
\ea
Since $\mac{T}$ is an operator from $L_2^P$ to $F_W^{-1}$, any $\bs{w}\in\mcr{A}_2$ belongs to $F_W^{-1}$.
Therefore, $\mcr{A}_2\subset\{\bs{v}\in F_W^{-1}\mid\lan\bs{v},\mac{W}\bs{v}\ran_P\le1\}=:\mcr{A}_3$.
Hence, 
$\sup_{\bs{u}_1\in\mcr{A}_2}\lan\bs{u}_1,\mac{V}\bs{u}_1\ran_P
 \le\sup_{\bs{v}\in\mcr{A}_3}\lan\bs{v},\mac{V}\bs{v}\ran_P
$.
% \ba
%  \sup_{\bs{u}_1\in\mcr{A}_2}\lan\bs{u}_1,\mac{V}\bs{u}_1\ran_P
%  &\le\sup_{\bs{v}\in\mcr{A}_3}\lan\bs{v},\mac{V}\bs{v}\ran_P.
% \ea
In addition, since $\mcr{A}_1\subset\mcr{A}_3$, we obtain $\sup_{\bs{v}\in\mcr{A}_1}\varphi(\bs{v})\le\sup_{\bs{v}\in\mcr{A}_3}\varphi(\bs{v})$ with $\varphi(\bs{v})=\lan\bs{v},\mac{V}\bs{v}\ran_P$.
Furthermore, we can show that $\sup_{\bs{v}\in\mcr{A}_1}\varphi(\bs{v})\ge\sup_{\bs{v}\in\mcr{A}_3}\varphi(\bs{v})$.
Thus, we have $\sup_{\bs{v}\in\mcr{A}_1}\varphi(\bs{v})=\sup_{\bs{v}\in\mcr{A}_3}\varphi(\bs{v})$.
Collecting the results, we obtain the following:
\ba
 \sup_{\bs{u}\in\mcr{A}_1}\mbb{E}[\left<\bs{u},\bs{X}_{(K)}\right>_{P}^{2}]
 &=\sup_{\bs{u}_1\in\mcr{A}_2}\lan\bs{u}_1,\mac{V}\bs{u}_1\ran_P
 \le\sup_{\bs{v}\in\mcr{A}_3}\lan\bs{v},\mac{V}\bs{v}\ran_P
 =\sup_{\bs{v}\in\mcr{A}_1}\lan\bs{v},\mac{V}\bs{v}\ran_P\\
 &=\sup_{\bs{u}\in\mcr{A}_1}\mbb{E}[\left<\bs{u},\bs{X}\right>_{P}^{2}].
\ea
The right-hand side of the last line of the above equations is the first canonical criterion value of $\bs{X}$. 
In addition, the weight function $\bs{\alpha}_{1}\in\mcr{A}_{1}$, and
\begin{equation*}
  \lan\bs{\alpha}_{1},\bs{X}_{(K)}\ran_P
  =\left\lan\mac{W}^{\frac{1}{2}}\bs{\alpha}_{1},\sum_{l=1}^{K}\eta_{l}\mac{W}^{\frac{1}{2}}\bs{\alpha}_{l}\right\ran_{P}
  =\sum_{l=1}^{K}\eta_{l}\lan\tilde{\bs{\alpha}}_{1},\tilde{\bs{\alpha}}_{l}\ran_{P}
  =\eta_{1}.
\end{equation*}
Therefore,
\begin{equation*}
  \mbb{E}[\lan\bs{\alpha}_{1},\bs{X}_{(K)}\ran_{P}^{2}]
  =\mbb{E}[\eta_{1}^{2}]
  =\rho_{1}
  =\sup_{\bs{u}\in\mcr{A}_1}\mbb{E}[\left<\bs{u},\bs{X}\right>_{P}^{2}].
\end{equation*}
This proves that the first canonical component $(\rho_{1},\eta_{1},\bs{\alpha}_{1})$ of $\bs{X}$ is also a solution for $\bs{X}_{(K)}$. 
For $k=2,\ldots,K$, we repeat the same argument after adding the orthogonality constraints with respect to the first $k-1$ canonical components to the admissible sets $\mcr{A}_1$, $\mcr{A}_2$, and $\mcr{A}_3$.
Hence, the first $K$ canonical components of $\bs{X}_{(K)}$ can be chosen to coincide with the first $K$ canonical components of $\bs{X}$.

\vspace{10pt}
\noindent (b) Similar to the proof of (a), for $\bs{u}\in F_W^{-1}$, consider the decomposition $\bs{u}=\bs{u}_{1}+\bs{u}_{2}$, where
$
  \bs{u}_{1}=\mac{W}^{-\frac{1}{2}}\mac{P}_{K}\mac{W}^{\frac{1}{2}}\bs{u}
$, and 
$
  \bs{u}_{2}=(\mac{I}-\mac{W}^{-\frac{1}{2}}\mac{P}_{K}\mac{W}^{\frac{1}{2}})\bs{u}
$.

First, we show that the first canonical component of $\bs{X}_{(K)}^{\perp}$ coincides with the $(K+1)$th canonical component of $\bs{X}$.
We define $\mac{T}':F_W^{-1}\to F_W^{-1}$ as $\mac{T}'=\mac{I}-\mac{W}^{-\frac{1}{2}}\mac{P}_{K}\mac{W}^{\frac{1}{2}}$ and define $\mcr{B}_1=\{\bs{u}\in F_W^{-1}\mid\lan\bs{u},\mac{W}\bs{u}\ran_P=1\}$.
%% \mac{T}のdomainはL_2^PではなくF_W^{-1}であることで良さそう。
By Lemma \ref{lemma:lemma_for_theorem_canonical-decomposition}, we have
\ba
  \left<\bs{u}_{1},\bs{X}_{(K)}^{\perp}\right>_{P}
  =\lan\bs{u}_1,\bs{X}\ran_P-\lan\bs{u}_1,\bs{X}_{(K)}\ran_P
  =\lan\bs{u}_1,\bs{X}\ran_P-\lan\bs{u}_1,\bs{X}\ran_P
  =0\ a.s.,
\ea
and
\ba
  \left<\bs{u},\bs{X}_{(K)}^{\perp}\right>_{P}
    =\left<\bs{u}_{2},\bs{X}_{(K)}^{\perp}\right>_{P}
    =\left<\bs{u}_{2},\bs{X}\right>_{P}\ a.s.
\ea
Then, we have
\ba
  \sup_{\bs{u}\in\mcr{B}_1}\mbb{E}\left[\left<\bs{u},\bs{X}_{(K)}^{\perp}\right>_{P}^{2}\right]
  =\sup_{\bs{u}\in\mcr{B}_1}\mbb{E}[\left<\bs{u}_{2},\bs{X}\right>_{P}^{2}]
  =\sup_{\bs{u}_2\in\mcr{B}_2}\lan\bs{u}_2,\mac{V}\bs{u}_2\ran_P
\ea
where $\mcr{B}_2=\mac{T}'\mcr{B}_1=\{\mac{T}'\bs{u}\mid\bs{u}\in\mcr{B}_1\}$.
Noting that $\lan\bs{u},\mac{W}\bs{u}\ran_P=1$ implies $\|\mac{W}^{\frac{1}{2}}\bs{u}\|_P^2=\lan\mac{W}^{\frac{1}{2}}\bs{u},\mac{W}^{\frac{1}{2}}\bs{u}\ran_P=1$, we have, for any $\bs{u}\in\mcr{B}_1$, 
\ba
 \lan\mac{T}'\bs{u},\mac{W}\mac{T}'\bs{u}\ran_P
%  &=\lan(\mac{I}-\mac{W}^{-\frac{1}{2}}\mac{P}_K\mac{W}^{\frac{1}{2}})\bs{u},\mac{W}(\mac{I}-\mac{W}^{-\frac{1}{2}}\mac{P}_K\mac{W}^{\frac{1}{2}})\bs{u}\ran_P\\
%  &=\lan\mac{W}^{\frac{1}{2}}(\mac{I}-\mac{W}^{-\frac{1}{2}}\mac{P}_K\mac{W}^{\frac{1}{2}})\bs{u},\mac{W}^{\frac{1}{2}}(\mac{I}-\mac{W}^{-\frac{1}{2}}\mac{P}_K\mac{W}^{\frac{1}{2}})\bs{u}\ran_P\\
%  &=\lan(\mac{W}^{\frac{1}{2}}-\mac{P}_K\mac{W}^{\frac{1}{2}})\bs{u},(\mac{W}^{\frac{1}{2}}-\mac{P}_K\mac{W}^{\frac{1}{2}})\bs{u}\ran_P\\
%  &=\lan(\mac{I}-\mac{P}_K)\mac{W}^{\frac{1}{2}}\bs{u},(\mac{I}-\mac{P}_K)\mac{W}^{\frac{1}{2}}\bs{u}\ran_P\\
 &=\|(\mac{I}-\mac{P}_K)\mac{W}^{\frac{1}{2}}\bs{u}\|_P^2
%  &\le\|\mac{W}^{\frac{1}{2}}\bs{u}\|_P^2\\
 \le1.
\ea
In addition, we observe that, for $l\in[K]$, 
\ba
  \lan\mac{T}'\bs{u},\mac{W}\bs{\alpha}_{l}\ran_P
  % &=\lan\mac{W}^{\frac{1}{2}}\mac{T}'\bs{u},\mac{W}^{\frac{1}{2}}\bs{\alpha}_{l}\ran_P\\
  % &=\lan\mac{W}^{\frac{1}{2}}(\mac{I}-\mac{W}^{-\frac{1}{2}}\mac{P}_K\mac{W}^{\frac{1}{2}})\bs{u},\mac{W}^{\frac{1}{2}}\bs{\alpha}_{l}\ran_P\\
  % &=\lan(\mac{I}-\mac{P}_K)\mac{W}^{\frac{1}{2}}\bs{u},\tilde{\bs{\alpha}}_{l}\ran_P\\
  &=\lan\mac{W}^{\frac{1}{2}}\bs{u},(\mac{I}-\mac{P}_K)\tilde{\bs{\alpha}}_{l}\ran_P
  =0
\ea
Therefore, $\mcr{B}_2\subset\{\bs{v}\in F_W^{-1}\mid\lan\bs{v},\mac{W}\bs{v}\ran_P\le1\wedge\lan\bs{v},\mac{W}\bs{\alpha}_{l}\ran_P=0,l\in[K]\}=:\mcr{B}_3$.
Hence, 
\ba
 \sup_{\bs{u}_2\in\mcr{B}_2}\lan\bs{u}_2,\mac{V}\bs{u}_2\ran_P
 &\le\sup_{\bs{u}_2\in\mcr{B}_3}\lan\bs{u}_2,\mac{V}\bs{u}_2\ran_P.
\ea
Define $\mcr{B}_4=\{\bs{u}\in F_W^{-1}\mid\lan\bs{u},\mac{W}\bs{u}\ran_P=1\wedge\lan\bs{u},\mac{W}\bs{\alpha}_{l}\ran_P=0,l\in[K]\}$.
Just as in the proof of (a), we can show that $\sup_{\bs{u}_2\in\mcr{B}_3}\lan\bs{u}_2,\mac{V}\bs{u}_2\ran_P=\sup_{\bs{u}\in\mcr{B}_4}\lan\bs{u},\mac{V}\bs{u}\ran_P$.
Collecting the results, we obtain the following:
\ba
 \sup_{\bs{u}\in\mcr{B}_1}\mbb{E}\left[\left<\bs{u},\bs{X}_{(K)}^{\perp}\right>_{P}^{2}\right]
 &=\sup_{\bs{u}_2\in\mcr{B}_2}\lan\bs{u}_2,\mac{V}\bs{u}_2\ran_P
 \le\sup_{\bs{v}\in\mcr{B}_3}\lan\bs{v},\mac{V}\bs{v}\ran_P
%  =\sup_{\bs{v}\in\mcr{B}_4}\lan\bs{v},\mac{V}\bs{v}\ran_P\\
 =\sup_{\bs{v}\in\mcr{B}_4}\mbb{E}[\left<\bs{v},\bs{X}\right>_{P}^{2}].
\ea

The right-hand side of the above equations is the $(K+1)$th canonical criterion value of $\bs{X}$. 
In addition, $\bs{\alpha}_{K+1}\in\mcr{B}_{1}$ and
\begin{align*}
  \lan\bs{\alpha}_{K+1},\bs{X}_{(K)}^{\perp}\ran_{P}
  &=\lan\bs{\alpha}_{K+1},\bs{X}\ran_{P}-\lan\bs{\alpha}_{K+1},\bs{X}_{(K)}\ran_{P}
  % &=\eta_{K+1}
    % -\lan\mac{W}^{\frac{1}{2}}\bs{\alpha}_{K+1},\sum_{l=1}^{K}\eta_{l}\mac{W}^{\frac{1}{2}}\bs{\alpha}_{l}\ran_{P}\\
  =\eta_{K+1}
    -\sum_{l=1}^{K}\eta_{l}\lan\tilde{\bs{\alpha}}_{K+1},\tilde{\bs{\alpha}}_{l}\ran_{P}
  =\eta_{K+1}.
\end{align*}
Then,
\begin{equation*}
  \mbb{E}[\lan\bs{\alpha}_{K+1},\bs{X}_{(K)}^{\perp}\ran_{P}^{2}]
  =\mbb{E}[\eta_{K+1}^{2}]
  =\rho_{K+1}
  =\sup_{\bs{u}\in\mcr{B}_{4}}\mbb{E}[\lan\bs{u},\bs{X}\ran_{P}^{2}].
\end{equation*}
Therefore, the first canonical component of $\bs{X}_{(K)}^{\perp}$ coincides with the $(K+1)$th canonical component of $\bs{X}$.
For $k\ge2$, the statement follows by mathematical induction.
In the induction step, it suffices to repeat the preceding argument with the admissible sets $\mcr{B}_1$, $\mcr{B}_2$, $\mcr{B}_3$, and $\mcr{B}_4$ further restricted by the orthogonality constraints with respect to the first $K+k-1$ canonical components of $\bs{X}$.

\vspace{10pt}
\noindent (c) For any $\bs{u},\bs{v}\in L_2^P$, observe that
\ba
  \lan\mbb{E}[\bs{X}_{(K)}\otimes_P\bs{X}_{(K)}^{\perp}]\bs{u}, \bs{v}\ran_{P}
  % =\mbb{E}[\lan\bs{u},\bs{X}_{(K)}\ran_{P}\lan\bs{v},\bs{X}_{(K)}^{\perp}\ran_{P}]
  =\Cov(\lan\bs{u},\bs{X}_{(K)}\ran_{P},\lan\bs{v},\bs{X}_{(K)}^{\perp}\ran_{P}).
\ea
Thus, it is sufficient to prove that $\Cov(\lan\bs{u},\bs{X}_{(K)}\ran_{P},\lan\bs{v},\bs{X}_{(K)}^{\perp}\ran_{P})=0$ for any $\bs{u},\bs{v}\in L_2^P$.

Note that
\begin{align*}
  &\Cov(\lan\bs{u},\bs{X}_{(K)}\ran_{P},\lan\bs{v},\bs{X}_{(K)}^{\perp}\ran_{P})\\
  % &=\Cov(\lan\bs{u},\bs{X}_{(K)}\ran_{P},\lan\bs{v},\bs{X}-\bs{X}_{(K)}\ran_{P})\\
  &=\Cov\left(\left\lan\bs{u},\sum_{k=1}^{K}\lan\bs{\alpha}_{k},\bs{X}\ran_{P}\mac{W}\bs{\alpha}_{k}\right\ran_{P},\lan\bs{v},\bs{X}\ran_{P}\right)
  -\Cov\left(\left\lan\bs{u},\sum_{k=1}^{K}\eta_{k}\mac{W}\bs{\alpha}_{k}\right\ran_{P},
  \left\lan\bs{v},\sum_{l=1}^{K}\eta_{l}\mac{W}\bs{\alpha}_{l}\right\ran_{P}\right).
\end{align*}
Under Condition~\ref{cond:cond2}, the first term of the last line is expanded as follows:
\begin{align*}
  \Cov\left(\left<\bs{u},\sum_{k=1}^{K}\lan\bs{\alpha}_{k},\bs{X}\ran_{P}\mac{W}\bs{\alpha}_{k}\right>_{P},\lan\bs{v},\bs{X}\ran_{P}\right)
  % &=\sum_{k=1}^{K}\mbb{E}\left[\lan\bs{\alpha}_{k},\bs{X}\ran_{P}\lan\bs{v},\bs{X}\ran_{P}\right]\lan\bs{u},\mac{W}\bs{\alpha}_{k}\ran_{P}\\
  &=\sum_{k=1}^{K}\lan\bs{v},\mac{V}\bs{\alpha}_{k}\ran_{P}\lan\bs{u},\mac{W}\bs{\alpha}_{k}\ran_{P}\\
  % &=\sum_{k=1}^{K}\lan\mac{W}^{\frac{1}{2}}\bs{v},\mac{W}^{-\frac{1}{2}}\mac{V}\mac{W}^{-\frac{1}{2}}\tilde{\bs{\alpha}}_{k}\ran_{P}
  % \lan\mac{W}^{\frac{1}{2}}\bs{u},\tilde{\bs{\alpha}}_{k}\ran_{P}\\
  % &=\sum_{k=1}^{K}\lan\mac{W}^{\frac{1}{2}}\bs{v},\rho_{k}\tilde{\bs{\alpha}}_{k}\ran_{P}
  % \lan\mac{W}^{\frac{1}{2}}\bs{u},\tilde{\bs{\alpha}}_{k}\ran_{P}\\
  &=\sum_{k=1}^K\rho_{k}\lan\mac{W}^{\frac{1}{2}}\bs{u},\tilde{\bs{\alpha}}_{k}\ran_{P}\lan\mac{W}^{\frac{1}{2}}\bs{v},\tilde{\bs{\alpha}}_{k}\ran_{P}
  % &=\rho_{k}\lan\mac{W}^{\frac{1}{2}}\bs{u},\mac{P}_{K}\mac{W}^{\frac{1}{2}}\bs{v}\ran_{P}.
\end{align*}
On the other hand, the second term can be expanded as
\begin{align*}
 \Cov\left(\left<\bs{u},\sum_{k=1}^{K}\eta_{k}\mac{W}\bs{\alpha}_{k}\right>_{P},
	\left<\bs{v},\sum_{l=1}^{K}\eta_{l}\mac{W}\bs{\alpha}_{l}\right>_{P}\right)
 &=\sum_{k=1}^{K}\sum_{l=1}^{K}\mbb{E}[\eta_{k}\eta_{l}]
 \lan\mac{W}^{\frac{1}{2}}\bs{u},\tilde{\bs{\alpha}}_{k}\ran_{P}
 \lan\mac{W}^{\frac{1}{2}}\bs{v},\tilde{\bs{\alpha}}_{l}\ran_{P}\\
 &=\sum_{k=1}^K\rho_{k}\lan\mac{W}^{\frac{1}{2}}\bs{u},\tilde{\bs{\alpha}}_{k}\ran_{P}\lan\mac{W}^{\frac{1}{2}}\bs{v},\tilde{\bs{\alpha}}_{k}\ran_{P}
%  \lan\mac{W}^{\frac{1}{2}}\bs{u},\mac{P}_{k}\mac{W}^{\frac{1}{2}}\bs{v}\ran_{P}.
\end{align*}
Thus, we have $\Cov(\lan\bs{u},\bs{X}_{(K)}\ran_{P},\lan\bs{v},\bs{X}_{(K)}^{\perp}\ran_{P})=0$ for all $\bs{u},\bs{v}\in L_2^P$, which implies that $\mbb{E}[\bs{X}_{(K)}\otimes_P\bs{X}_{(K)}^{\perp}]=0$.

%%%%%%%%%%%%%%%%%%%%%%%%%%%%%%%%%%%%%%%%%%%%%%%%%%%%
%%                   (d)の証明
%%%%%%%%%%%%%%%%%%%%%%%%%%%%%%%%%%%%%%%%%%%%%%%%%%%%
\vspace{10pt}
\noindent (d) 
% For $n,m\in\mbb{N}:n>m$,
% \ba
%   \mbb{E}\left[\left\|\sum_{k=m+1}^{n}\eta_{k}\mac{W}\bs{\alpha}_{k}\right\|^{2}_{P}\right]
%   %  &=\mbb{E}\left[\left<\sum_{k=m+1}^{n}\eta_{k}\mac{W}\bs{\alpha}_{k},\sum_{l=m+1}^{n}\eta_{l}\mac{W}\bs{\alpha}_{l}\right>_{P}\right]\\
%   &=\sum_{k=m+1}^{n}\sum_{l=m+1}^{n}\mbb{E}[\eta_{k}\eta_{l}]\lan\mac{W}\bs{\alpha}_{k},\mac{W}\bs{\alpha}_{l}\ran_{P}
%   %  &=\sum_{k=m+1}^{n}\mbb{E}\left[\lan\eta_{k}\mac{W}\bs{\alpha}_k,\eta_{k}\mac{W}\bs{\alpha}_k\ran_{P}\right]\\
%   =\sum_{k=m+1}^{n}\rho_k\|\mac{W}\bs{\alpha}_k\|_{P}^{2}.
% \ea
For any $K$, $\rho_k=1+\tilde{\rho}_k$ for all $k\in[K]$ by Theorem \ref{theo:optimization-P-FMCCA}(b), and hence, we have
\ba
  \sum_{k=1}^K\mbb{E}[\|\eta_{k}\mac{W}\bs{\alpha}_{k}\|_{P}^{2}]
  % &=\sum_{k=1}^K\mbb{E}[\lan\eta_k\mac{W}\bs{\alpha}_k,\eta_k\mac{W}\bs{\alpha}_k\ran_P]\\
  % &=\sum_{k=1}^K\mbb{E}[\eta_k^2]\lan\mac{W}\bs{\alpha}_k,\mac{W}\bs{\alpha}_k\ran_P]\\
  =\sum_{k=1}^{K}\rho_{k}\lan\mac{W}\tilde{\bs{\alpha}}_k,\tilde{\bs{\alpha}}_k\ran_P
  %  &=\sum_{k=1}^K(1+\tilde{\rho_{k}})\lan\mac{W}\tilde{\bs{\alpha}}_k,\tilde{\bs{\alpha}}_k\ran_{P}
  \le\sum_{k=1}^K(1+\|\tilde{\mac{R}}\|)\lan\mac{W}\tilde{\bs{\alpha}}_k,\tilde{\bs{\alpha}}_k\ran_{P}
  %  &=(1+\|\tilde{\mac{R}}\|)\sum_{k=1}^K\lan\mac{W}\tilde{\bs{\alpha}}_k,\tilde{\bs{\alpha}}_k\ran_{P}\\
  \le(1+\|\tilde{\mac{R}}\|)\tr(\mac{W}).
\ea

Since $\mac{W}$ is a trace-class operator by Proposition \ref{prop:W-HS}, $s_K:=\sum_{k=1}^{K}\mbb{E}[\|\eta_{k}\mac{W}\bs{\alpha}_{k}\|_{P}^{2}]$ is bounded above, which implies that
 $\{s_{K}\}$ is a convergent sequence since it is nondecreasing.
Then, $\{s_{K}\}$ is a Cauchy sequence.
Thus, for any $\epsilon>0$, there exists $n_{0}\in\mbb{N}$, such that for any $n,m\in\mbb{N}$, if $n>m\ge{n}_{0}$, then $\sum_{k=m+1}^{n}\mbb{E}[\|\eta_{k}\mac{W}\bs{\alpha}_{k}\|_{P}^{2}]<\epsilon$. 
Note that
\ba
  \mbb{E}\left[\left\|\sum_{k=m+1}^{n}\eta_{k}\mac{W}\bs{\alpha}_{k}\right\|^{2}_{P}\right]
  %  &=\mbb{E}\left[\left<\sum_{k=m+1}^{n}\eta_{k}\mac{W}\bs{\alpha}_{k},\sum_{l=m+1}^{n}\eta_{l}\mac{W}\bs{\alpha}_{l}\right>_{P}\right]\\
  &=\sum_{k=m+1}^{n}\sum_{l=m+1}^{n}\mbb{E}[\eta_{k}\eta_{l}]\lan\mac{W}\bs{\alpha}_{k},\mac{W}\bs{\alpha}_{l}\ran_{P}
  %  &=\sum_{k=m+1}^{n}\mbb{E}\left[\lan\eta_{k}\mac{W}\bs{\alpha}_k,\eta_{k}\mac{W}\bs{\alpha}_k\ran_{P}\right]\\
  =\sum_{k=m+1}^{n}\mbb{E}[\|\eta_k\mac{W}\bs{\alpha}_k\|_{P}^{2}].
\ea
Therefore, if $n>m\ge{n}_{0}$, then $\mbb{E}[\|\sum_{k=m+1}^{n}\eta_{k}\mac{W}\bs{\alpha}_{k}\|_{P}^{2}]<\epsilon$. 
Let $\mbb{L}^2(\Omega,L_2^P)$ be the Bochner space with norm $\|x\|_{\mbb{L}^2}=(\mbb{E}[\|x\|_P^2])^{\frac{1}{2}}$.
From the above discussion, $\tilde{s}_K:=\sum_{k=1}^K\eta_k\mac{W}\alpha_k$ is a Cauchy sequence in $\mbb{L}^2(\Omega,L_2^P)$.
Then, by the completeness, $\tilde{s}_K$ converges in $\mbb{L}^2(\Omega,L_2^P)$.
Thus, when $K\ra\infty$, we have
\ba
  \bs{X}_{(K)}
  % =\mac{W}^{\fr}\tilde{\mac{P}}_{K}\bs{X}
  =\sum_{k=1}^{K}\eta_{k}\mac{W}\bs{\alpha}_{k}
  \ra
  \sum_{k=1}^{\infty}\eta_{k}\mac{W}\bs{\alpha}_{k}
  % =\mac{W}^{\fr}\tilde{\mac{P}}_{\infty}\bs{X}
  =\bs{X}_{(\infty)}.
\ea

% Next, we define the extension of $\mac{P}_{\infty}\mac{W}^{-\frac{1}{2}}$ to $L_2^P$.
% From the proof of Proposition \ref{prop:canonical_decomposition}, we have, for any $k\ge1$ and $\bs{z}\in F_W$,
% \ba
%   (\tilde{\bs{\alpha}}_k\otimes_P\tilde{\bs{\alpha}}_k)\mac{W}^{-\frac{1}{2}}\bs{z}
%   =(\bs{\alpha}_k\otimes_P\tilde{\bs{\alpha}}_k)\bs{z}.
% \ea
% Thus, on the set $F_W$, we have
% \ba
%   \mac{P}_{\infty}\mac{W}^{-\frac{1}{2}}
%   =\sum_{k=1}^{\infty}(\tilde{\bs{\alpha}}_k\otimes_P\tilde{\bs{\alpha}}_k)\mac{W}^{-\frac{1}{2}}
%   =\sum_{k=1}^{\infty}(\bs{\alpha}_k\otimes_P\tilde{\bs{\alpha}}_k).
% \ea

% We define an operator $\tilde{\mac{P}}_{\infty}$ on $L_2^P$ by
% \ba
%   \tilde{\mac{P}}_{\infty}
%   :=\sum_{k=1}^{\infty}(\bs{\alpha}_k\otimes_P\tilde{\bs{\alpha}}_k).
% \ea
% Then, we have
% \ba
%   \mac{W}^{\frac{1}{2}}\tilde{\mac{P}}_{\infty}\bs{X}
%   =\mac{W}^{\frac{1}{2}}\sum_{k=1}^{\infty}\lan\bs{X},\bs{\alpha}_k\ran_P\tilde{\bs{\alpha}}_k
%   =\mac{W}^{\frac{1}{2}}\sum_{k=1}^{\infty}\eta_k\mac{W}^{\frac{1}{2}}\bs{\alpha}_k
%   =\sum_{k=1}^{\infty}\eta_k\mac{W}\bs{\alpha}_k
%   =\bs{X}_{(\infty)}.
% \ea

The same argument as the proofs of (a) and (b) shows that $\bs{X}_{(\infty)}$ and $\bs{X}$ share the same canonical components.
Also, similar to the proof of (c), we can prove that $\bs{X}_{(\infty)}$ and $\bs{X}_{(\infty)}^{\perp}$ are uncorrelated.

%%%%%%%%%%%%%%%%%%%%%%%%%%%%%%%%%%%%%%%%%%%%%%%%%%%%
%%               (e)の証明
%%%%%%%%%%%%%%%%%%%%%%%%%%%%%%%%%%%%%%%%%%%%%%%%%%%%
\vspace{10pt}
\noindent (e) 
Let $I_r=[r]$ if $r<\infty$ and let $I_r=\mbb{N}$ if $r=\infty$.
Let $\mac{P}_{I_r}$ denote the orthogonal projection operator from $L_2^P$ onto $\overline{\myspan\{\tilde{\bs{\alpha}}_{k}, k\in I_r\}}$.
By assumption, $\bs{X}\in F_W$ a.s. and $\mac{W}^{-\frac{1}{2}}\bs{X}\in\overline{\myspan\{\tilde{\bs{\alpha}}_{k}, k\in I_r\}}$ a.s.
Hence, $\mac{P}_{I_r}\mac{W}^{-\frac{1}{2}}\bs{X}=\mac{W}^{-\frac{1}{2}}\bs{X}$ a.s.

First, suppose that $r<\infty$.
Then, $I_r=[r]$ , and hence, $\mac{P}_{I_r}$ coincides with the projection operator $\mac{P}_r$ defined in Section~\ref{sec:canonical_decomposition}.
Therefore, we have the following almost surely:
\ba
	\bs{X}_{(r)}
  &=\sum_{k=1}^{r}\eta_{k}\mac{W}\bs{\alpha}_{k}
  =\mac{W}^{\frac{1}{2}}\sum_{k=1}^{r}\lan\bs{\alpha}_{k},\bs{X}\ran_{P}\mac{W}^{\frac{1}{2}}\bs{\alpha}_{k}
  % &=\mac{W}^{\frac{1}{2}}\sum_{k=1}^{r}(\tilde{\bs{\alpha}}_k\otimes_P\tilde{\bs{\alpha}}_k)\mac{W}^{-\frac{1}{2}}\bs{X}
  =\mac{W}^{\frac{1}{2}}\mac{P}_{I_r}\mac{W}^{-\frac{1}{2}}\bs{X}
  =\bs{X}.
\ea

Next, suppose that $r=\infty$.
Then, $\mac{P}_{I_r}=\mac{P}_{\infty}$, the orthogonal projection operator onto $\overline{\myspan\{\tilde{\bs{\alpha}}_{k}, k\in\mbb{N}\}}$.
Using the definition of $\bs{X}_{(\infty)}$, we similarly obtain $\bs{X}_{(\infty)}=\bs{X}$ a.s.
% \ba
% 	\bs{X}_{(\infty)}
%   &=\sum_{k=1}^{\infty}\eta_{k}\mac{W}\bs{\alpha}_{k}
%   =\mac{W}^{\frac{1}{2}}\sum_{k=1}^{\infty}\lan\bs{X},\bs{\alpha}_{k}\ran_{P}\mac{W}^{\frac{1}{2}}\bs{\alpha}_{k}\\
%   % &=\mac{W}^{\frac{1}{2}}\sum_{k=1}^{\infty}(\tilde{\bs{\alpha}}_k\otimes_P\tilde{\bs{\alpha}}_k)\mac{W}^{-\frac{1}{2}}\bs{X}
%   &=\mac{W}^{\frac{1}{2}}\mac{P}_{\infty}\mac{W}^{-\frac{1}{2}}\bs{X}
%   =\bs{X}
% \ea
% almost surely.

Therefore, the canonical representation reconstructs $\bs{X}$: if $r<\infty$, then $\bs{X}_{(r)}=\bs{X}$ a.s., and if $r=\infty$, then $\bs{X}_{(\infty)}=\bs{X}$ a.s.

\subsection{Proof of Theorem~\ref{theo:equivalence_fhomals_fmcca_truncated}}
\label{subsec:proofs-of-theorem4}

% In the following, for any operator $\mac{U}$, $\im({\mac{U}})$ denotes the range of $\mac{U}$.
% Let $\mac{V}_{(K),pp}$ be the operator $\mac{V}_{pp}$ corresponding to $\bs{X}_{(K)}$.
% Since both $\mac{V}_{(K),pp}$ and $\mac{W}_{(K)}$ are finite-rank operators, their ranges are closed.
% Hence, the Moore--Penrose inverses are well-defined as bounded operators on $L_2$ and $L_2^P$, respectively.

For any $\bs{\gamma}\in L_2^P$, a solution of the minimization problem of $Q(\beta_p)$ over $\beta_p$ is given by $\beta_p=\mac{V}_{(K),pp}^\dagger\tilde{\mac{V}}_{(K),p}\bs{\gamma}$ where $\tilde{\mac{V}}_{(K),p}$ is the operator corresponding to $\tilde{\mac{V}}_p$.
To see this, we need to verify that $\tilde{\mac{V}}_{(K),p}\bs{\gamma}\in\im(\mac{V}_{(K),pp})$.
Indeed, if $h\in\ker(\mac{V}_{(K),pp})$, then
\ba
  \mbb{E}[\lan h,X_{(K),p}\ran^2]
  =\lan h,\mac{V}_{(K),pp}h\ran
  =0,
\ea
and so, $\lan h,X_{(K),p}\ran=0$ a.s.
Hence, 
\ba
  \lan h,\tilde{\mac{V}}_{(K),p}\bs{\gamma}\ran
  =\mbb{E}[\lan\bs{\gamma},\bs{X}_{(K)}\ran_P\lan h,X_{(K),p}\ran]
  =0.
\ea
Thus, $\tilde{\mac{V}}_{(K),p}\bs{\gamma}\in(\ker(V_{(K),pp}))^\perp=\im(\mac{V}_{(K),pp})$, where the last equality follows from the fact that $\im(\mac{V}_{(K),pp})$ is closed.
Consequently, by the same argument as in the proof of Proposition~\ref{prop:beta_minimization}, we can verify that $\beta_p=\mac{V}_{(K),pp}^\dagger\tilde{\mac{V}}_{(K),p}\bs{\gamma}$ is a solution of the minimization problem of $Q(\beta_p)$ over $\beta_p$.

Note that we can write the solution as $\bs{\beta}=\mac{W}_{(K)}^\dagger\mac{V}_{(K)}\bs{\gamma}$.
Then, from \eqref{eq:ell2_prime}, the profiled FHOMALS criterion can be written as
\ba
  \ell'(\bs{\gamma})
  =P-\lan\bs{\gamma},\mac{V}_{(K)}\mac{W}_{(K)}^\dagger\mac{V}_{(K)}\bs{\gamma}\ran_P
  % \mbb{E}[\lan\bs{\gamma}_{k},\bs{X}_{(K)}\ran_{P}\lan\bs{\beta}_{k},\bs{X}_{(K)}\ran_{P}]
  % &=\lan\mac{W}_{(K)}^\dagger\mac{V}_{(K)}\bs{\gamma}_{k},\mac{V}_{(K)}\bs{\gamma}_{k}\ran_{P}\\
  =P-\lan\tilde{\bs{\gamma}},\mac{V}_{(K)}^{\fr}\mac{W}_{(K)}^\dagger\mac{V}_{(K)}^{\fr}\tilde{\bs{\gamma}}\ran_{P}
\ea
where $\tilde{\bs{\gamma}}=\mac{V}_{(K)}^{\frac{1}{2}}\bs{\gamma}$.
We write $\mac{S}_{(K)}:=\mac{V}_{(K)}^{\fr}\mac{W}_{(K)}^\dagger\mac{V}_{(K)}^{\fr}$ and let $\mac{M}_K:=\im(\mac{V}_{(K)}^{\fr})$.
Since $\tilde{\bs{\gamma}}$ belongs to $\mac{M}_K$, the minimization of $\ell'(\bs{\gamma})$ under the constraint $\lan\tilde{\bs{\gamma}},\tilde{\bs{\gamma}}\ran_P=1$ is equivalent to maximizing $\lan\tilde{\bs{\gamma}},\mac{S}_{(K)}\tilde{\bs{\gamma}}\ran_P$ over $\tilde{\bs{\gamma}}\in\mac{M}_K$ subject to the same constraint.

Because $\mac{V}_{(K)}$ and $\mac{W}_{(K)}$ are finite-rank self-adjoint nonnegative operators, $\mac{S}_{(K)}$ is also finite-rank, compact, self-adjoint, and nonnegative.
Since $\mac{V}_{(K)}$ is finite-rank, $\mac{M}_K$ is a finite-dimensional closed subspace of $L_2^P$.
Moreover, $\im(\mac{S}_{(K)})\subset\mac{M}_K$.
Hence, $\mac{S}_{(K)}$ leaves $\mac{M}_K$ invariant.
Therefore, its restriction $\mac{S}_{(K)}|_{\mac{M}_K}$ is a compact self-adjoint nonnegative operator on $\mac{M}_K$, and the above maximization problem is characterized by the eigenvalue problem for $\mac{S}_{(K)}|_{\mac{M}_K}$.
%% \mac{M}_K上で\gammaを求める必要があるが、もしinvarianceがなければ、\mac{S}_{(K)}\gammaは\mac{M}_Kに属さない可能性があるので、\mac{M}_K上での固有値問題を解くことができない。

Let the positive eigenvalues of $\mac{S}_{(K)}|_{\mac{M}_K}$ be arranged in nonincreasing order as $\mu_1^{(K)}\ge\mu_2^{(K)}\ge\cdots\ge\mu_{s_{(K)}}^{(K)}>0$ where $s_{(K)}$ is the number of positive eigenvalues.
% Since $\mac{V}_{(K)}$ is finite-rank, $\im(\mac{V}_{(K)}^{\fr})$ is a finite-dimensional closed subspace of $L_2^P$.
% Then, using the fact that the restriction of $\mac{S}_{(K)}$ to $\im(\mac{V}_{(K)}^{\fr})$ is a nonnegative compact operator, 
By the Courant--Fischer minimax principle, for $k=1,\dots,s_{(K)}$,
\ba
  \mu_k^{(K)}
  =\max_{\mac{L}\subset\mac{M}_K:\dim(\mac{L})=k}
  \min_{\tilde{\bs{\gamma}}\in\mac{L}:\|\tilde{\bs{\gamma}}\|_P=1}\lan\tilde{\bs{\gamma}},\mac{S}_{(K)}\tilde{\bs{\gamma}}\ran_P.
\ea
Equivalently, the successive maximizers under the orthogonality constraints are the normalized eigenfunctions of $\mac{S}_{(K)}|_{\mac{M}_K}$ corresponding to these positive eigenvalues.

Finally, the eigenspaces of $\mac{S}_{(K)}$ corresponding to nonzero eigenvalues coincide with those of the restricted operator $\mac{S}_{(K)}|_{\mac{M}_K}$.
%% ここでの eigenspace というのは、各固有値に対応する固有関数からなる線形空間を指す。具体的には、ker(\mac{S}_{(K)}-\mu\mac{I}) である。
Indeed, if $\mu\neq0$ and $\mac{S}_{(K)}\tilde{\bs{\gamma}}=\mu\tilde{\bs{\gamma}}$, then
\ba
  \tilde{\bs{\gamma}}
  =\mu^{-1}\mac{S}_{(K)}\tilde{\bs{\gamma}}
  \in\im(\mac{S}_{(K)})
  \subset\mac{M}_K.
\ea
Thus, the eigenvalue equation $\mac{S}_{(K)}\tilde{\bs{\gamma}}=\mu\tilde{\bs{\gamma}}$ with $\mu\neq0$ may be considered either on $L_2^P$ or on $\mac{M}_K$.

We write the Moore--Penrose inverse of $\mac{W}_{(K)}^{\frac{1}{2}}$ as $(\mac{W}_{(K)}^{\frac{1}{2}})^\dagger$, which is a self-adjoint operator from $L_2^P$ into $L_2^P$.
Let $\mac{A}_{(K)}=\mac{V}_{(K)}^{\fr}(\mac{W}_{(K)}^{\fr})^\dagger$, which is also finite-rank.
Thus, $\mac{A}_{(K)}$ is a compact operator.
Then, using $\mac{W}_{(K)}^\dagger=((\mac{W}_{(K)}^{\fr})^\dagger)^2$, we have $\mac{A}_{(K)}\mac{A}_{(K)}^*=\mac{V}_{(K)}^{\fr}\mac{W}_{(K)}^\dagger\mac{V}_{(K)}^{\fr}=\mac{S}_{(K)}$, whereas $\mac{A}_{(K)}^*\mac{A}_{(K)}=(\mac{W}_{(K)}^{\fr})^\dagger\mac{V}_{(K)}(\mac{W}_{(K)}^{\fr})^\dagger$.

The eigenspaces of $\mac{A}_{(K)}^*\mac{A}_{(K)}$ and $\mac{A}_{(K)}\mac{A}_{(K)}^*$ corresponding to positive eigenvalues are related through $\mac{A}_{(K)}$ and $\mac{A}_{(K)}^*$.
Indeed, suppose that $\mac{A}_{(K)}^*\mac{A}_{(K)}\bs{a}=\rho\bs{a}$ with $\rho>0$ under the constraint $\lan\bs{a},\bs{a}\ran_P=1$.
Define $\bs{b}:=\rho^{-\fr}\mac{A}_{(K)}\bs{a}$.
Then, 
\ba
  \|\bs{b}\|_P^2
  =\rho^{-1}\lan\bs{a},\mac{A}_{(K)}^*\mac{A}_{(K)}\bs{a}\ran_P
  =1,
\ea
and $\mac{A}_{(K)}\mac{A}_{(K)}^*\bs{b}=\rho\bs{b}$.
Conversely, suppose that $\mac{A}_{(K)}\mac{A}_{(K)}^*\bs{b}=\rho\bs{b}$ with $\rho>0$ under the constraint $\lan\bs{b},\bs{b}\ran_P=1$.
Define $\bs{a}:=\rho^{-\fr}\mac{A}_{(K)}^*\bs{b}$.
Then, 
\ba
  \|\bs{a}\|_P^2
  =\rho^{-1}\lan\bs{b},\mac{A}_{(K)}\mac{A}_{(K)}^*\bs{b}\ran_P
  =1,
\ea
and $\mac{A}_{(K)}^*\mac{A}_{(K)}\bs{a}=\rho\bs{a}$.
Moreover, the maps $\bs{a}\mapsto\rho^{-\fr}\mac{A}_{(K)}\bs{a}$ and $\bs{b}\mapsto\rho^{-\fr}\mac{A}_{(K)}^*\bs{b}$ are inverse to each other between the eigenspaces of the same positive eigenvalue $\rho$.
Hence, these eigenspaces have the same dimension.
Consequently, $\mac{A}_{(K)}^*\mac{A}_{(K)}$ and $\mac{A}_{(K)}\mac{A}_{(K)}^*$ have the same positive eigenvalues with the same multiplicities.
In particular, each canonical criterion value $\rho_k^{(K)}>1$, $k\in[m_{(K)}]$, is a positive eigenvalue of both operators with the same multiplicity.

Now, the P-FMCCA optimization problem for $\bs{X}_{(K)}$ is characterized by the following eigenvalue problem for $\mac{A}_{(K)}^*\mac{A}_{(K)}$:
\ba
  (\mac{W}_{(K)}^{\fr})^\dagger\mac{V}_{(K)}(\mac{W}_{(K)}^{\fr})^\dagger\tilde{\bs{\alpha}}
  =\rho\tilde{\bs{\alpha}}
\ea
with the constraint $\lan\tilde{\bs{\alpha}},\tilde{\bs{\alpha}}\ran_P=1$, and the corresponding weight function is $\bs{\alpha}=(\mac{W}_{(K)}^{\fr})^\dagger\tilde{\bs{\alpha}}$.
For an eigenfunction $\tilde{\bs{\alpha}}$ of $\mac{A}_{(K)}^*\mac{A}_{(K)}$ corresponding to a positive eigenvalue $\rho$, we have $\tilde{\bs{\alpha}}=\rho^{-1}\mac{A}_{(K)}^*\mac{A}_{(K)}\tilde{\bs{\alpha}}\in\im(\mac{A}_{(K)}^*)\subset\im((\mac{W}_{(K)}^{\fr})^\dagger)$.
Since $\mac{W}_{(K)}^{\fr}$ is finite-rank, self-adjoint, and nonnegative, its Moore--Penrose inverse satisfies $\im((\mac{W}_{(K)}^{\fr})^\dagger)=\im(\mac{W}_{(K)}^{\fr})$ and $\mac{W}_{(K)}^{\fr}(\mac{W}_{(K)}^{\fr})^\dagger$ is the orthogonal projection onto $\im(\mac{W}_{(K)}^{\fr})$.
%% MP逆のスペクトル分解に基づく表現を考えれば、\mac{W}_{(K)}^1/2 とそのMP逆が同じ固有関数の方向に沿って作用することがわかる。
Therefore, $\bs{\alpha}=(\mac{W}_{(K)}^{\fr})^\dagger\tilde{\bs{\alpha}}$ is well-defined and satisfies $\mac{W}_{(K)}^{\fr}\bs{\alpha}=\tilde{\bs{\alpha}}$.

% Now, similar to the case of $\mac{W}^{-\frac{1}{2}}$ in \eqref{eq:F_W_univariate} the domain of $\mac{W}_{(K)}^{-\frac{1}{2}}$, say $F_{W_{(K)}}$, is defined.
% In the finite-rank case, the boundedness condition in $F_{W_{(K)}}$ is automatically satisfied, and hence, $F_{W_{(K)}}=\ker(\mac{W}_{(K)}^{\frac{1}{2}})^\perp=\im(\mac{W}_{(K)}^{\frac{1}{2}})$.
% Therefore, the restriction of $(\mac{W}_{(K)}^{\frac{1}{2}})^\dagger$ to $\im(\mac{W}_{(K)}^{\frac{1}{2}})$ is equivalent to $\mac{W}_{(K)}^{-\frac{1}{2}}$.

Finally, we show that the FHOMALS components of $\bs{X}_{(K)}$ are given by the P-FMCCA components of $\bs{X}_{(K)}$.
Fix $k\in[m_{(K)}]$, and let $(\rho_k^{(K)},\eta_k^{(K)},\bs{\alpha}_k^{(K)})$ be the $k$th P-FMCCA component for the finite-rank problem based on $(\bs{X}_{(K)},\mac{V}_{(K)},\mac{W}_{(K)})$ with $\rho_k^{(K)}$ exceeding the baseline value one.
Write $\tilde{\bs{\alpha}}_k^{(K)}=\mac{W}_{(K)}^{\fr}\bs{\alpha}_k^{(K)}$.
Then, $\mac{A}_{(K)}^*\mac{A}_{(K)}\tilde{\bs{\alpha}}_k^{(K)}=\rho_k^{(K)}\tilde{\bs{\alpha}}_k^{(K)}$.
Since $\rho_k^{(K)}>0$, define $\tilde{\bs{\gamma}}_k=(\rho_k^{(K)})^{-\fr}\mac{A}_{(K)}\tilde{\bs{\alpha}}_k^{(K)}$.
Then, 
\ba
  \mac{S}_{(K)}\tilde{\bs{\gamma}}_k
  =\mac{A}_{(K)}\mac{A}_{(K)}^*(\rho_k^{(K)})^{-\fr}\mac{A}_{(K)}\tilde{\bs{\alpha}}_k^{(K)}
  =\rho_k^{(K)}\tilde{\bs{\gamma}}_k.
\ea
Moreover,
\ba
  \|\tilde{\bs{\gamma}}_k\|_P^2
  =(\rho_k^{(K)})^{-1}\left\lan\tilde{\bs{\alpha}}_k^{(K)},\mac{A}_{(K)}^*\mac{A}_{(K)}\tilde{\bs{\alpha}}_k^{(K)}\right\ran_P
  =1.
\ea
Therefore, $\tilde{\bs{\gamma}}_k$ is a normalized eigenfunction of the FHOMALS operator $\mac{S}_{(K)}$ corresponding to the positive eigenvalue $\rho_k^{(K)}$.

Since $\mac{A}_{(K)}\tilde{\bs{\alpha}}_k^{(K)}=\mac{V}_{(K)}^{\fr}(\mac{W}_{(K)}^{\fr})^\dagger\tilde{\bs{\alpha}}_k^{(K)}=\mac{V}_{(K)}^{\fr}\bs{\alpha}_k^{(K)}$, we have 
\ba
  \tilde{\bs{\gamma}}_k
  =\mac{V}_{(K)}^{\fr}(\rho_k^{(K)})^{-\fr}\bs{\alpha}_k^{(K)}.
\ea
Hence, choosing $\bs{\gamma}_k=(\rho_k^{(K)})^{-\fr}\bs{\alpha}_k^{(K)}$ is a solution of the FHOMALS optimization problem.

The corresponding minimizing $\bs{\beta}$ is then given by $\bs{\beta}_k=\mac{W}_{(K)}^\dagger\mac{V}_{(K)}\bs{\gamma}_k$.
Finally, the corresponding homogeneity score is $f_k=\lan\bs{X}_{(K)},\bs{\gamma}_k\ran_P$.
It remains to verify the normalization and orthogonality of the homogeneity scores.
For $k,l=1,\dots,m_{(K)}$, we have
\ba
  \mbb{E}[f_kf_l]
  =\lan\bs{\gamma}_k,\mac{V}_{(K)}\bs{\gamma}_l\ran_P
  =\lan\mac{V}_{(K)}^{\fr}\bs{\gamma}_k,\mac{V}_{(K)}^{\fr}\bs{\gamma}_l\ran_P
  =\delta_{kl}.
\ea
In particular, $\mbb{E}[f_k^2]=1$ for each $k$.
Therefore, the constructed scores satisfy the normalization and orthogonality constraints required for the FHOMALS components.

If a canonical criterion value exceeding the baseline value one has multiplicity greater than one, the above construction gives one possible orthonormal choice of the corresponding eigenfunctions.

\section*{Acknowledgements}

The authors gratefully acknowledge \textit{JSPS KAKENHI Grants (JP26730016, JP23K28042, JP26K02613, JP24K14855, JP25K15032, JP25K21806), 
the MEXT Project for Seismology Toward Research Innovation with Data of Earthquakes (STAR-E, JPJ010217), and
the MEXT STAR-E NEXT Project (JPJ013735)}.
The authors used ChatGPT (OpenAI) to assist with English language editing of the manuscript.

\section*{Supplementary Materials}

Supplementary materials for this article are available online.
The supplementary document contains the proofs of all propositions and lemmas.

\section*{Conflicts of Interest}

The authors declare that they have no conflicts of interest.

\section*{Data availability}

No empirical data analysis was conducted in this article; therefore, no data were used.

\bibliographystyle{apalike}
\bibliography{myrefs}

@article{hall2007methodology,
  title={Methodology and convergence rates for functional linear regression},
  author={Hall, Peter and Horowitz, Joel L},
  journal={The Annals of Statistics},
  volume={35},
  number={1},
  pages={70--91},
  year={2007}
}

@article{yang2011functional,
  title={Functional singular component analysis},
  author={Yang, Wenjing and M{\"u}ller, Hans-Georg and Stadtm{\"u}ller, Ulrich},
  journal={Journal of the Royal Statistical Society Series B: Statistical Methodology},
  volume={73},
  number={3},
  pages={303--324},
  year={2011},
  publisher={Oxford University Press}
}

@article{cupidon2008some,
  title={Some properties of canonical correlations and variates in infinite dimensions},
  author={Cupidon, J and Eubank, R and Gilliam, D and Ruymgaart, F},
  journal={Journal of Multivariate Analysis},
  volume={99},
  number={6},
  pages={1083--1104},
  year={2008},
  publisher={Elsevier}
}

@article{leurgans1993canonical,
  title={Canonical correlation analysis when the data are curves},
  author={Leurgans, Sue E and Moyeed, Rana A and Silverman, Bernard W},
  journal={Journal of the Royal Statistical Society Series B: Statistical Methodology},
  volume={55},
  number={3},
  pages={725--740},
  year={1993},
  publisher={Oxford University Press}
}

@article{ChiouMuller2014,
  author  = {Chiou, Jeng-Min and M{\"u}ller, Hans-Georg},
  title   = {Linear manifold modelling of multivariate functional data},
  journal = {Journal of the Royal Statistical Society: Series B (Statistical Methodology)},
  year    = {2014},
  volume  = {76},
  number  = {3},
  pages   = {605--626},
  doi     = {10.1111/rssb.12038}
}

@article{HappGreven2018,
  author  = {Happ, Clara and Greven, Sonja},
  title   = {Multivariate Functional Principal Component Analysis for Data Observed on Different (Dimensional) Domains},
  journal = {Journal of the American Statistical Association},
  year    = {2018},
  volume  = {113},
  number  = {522},
  pages   = {649--659},
  doi     = {10.1080/01621459.2016.1273115}
}

@article{eubank2008canonical,
  title={Canonical correlation for stochastic processes},
  author={Eubank, RL and Hsing, Tailen},
  journal={Stochastic Processes and their Applications},
  volume={118},
  number={9},
  pages={1634--1661},
  year={2008},
  publisher={Elsevier}
}

@book{gohberg2012basic,
  title={Basic classes of linear operators},
  author={Gohberg, Israel and Goldberg, Seymour and Kaashoek, Marinus},
  year={2012},
  publisher={Birkh{\"a}user}
}

@article{van1988homogeneity,
  title={Homogeneity analysis with k sets of variables: An alternating least squares method with optimal scaling features},
  author={Van der Burg, Eeke and De Leeuw, Jan and Verdegaal, Renee},
  journal={Psychometrika},
  volume={53},
  number={2},
  pages={177--197},
  year={1988},
  publisher={Springer-Verlag}
}

@article{de2009gifi,
  title={Gifi methods for optimal scaling in R: The package homals},
  author={De Leeuw, Jan and Mair, Patrick},
  journal={Journal of Statistical Software},
  volume={31},
  pages={1--21},
  year={2009}
}

@article{he2003functional,
  title={Functional canonical analysis for square integrable stochastic processes},
  author={He, Guozhong and M{\"u}ller, Hans-Georg and Wang, Jane-Ling},
  journal={Journal of Multivariate Analysis},
  volume={85},
  number={1},
  pages={54--77},
  year={2003},
  publisher={Elsevier}
}

@book{RamsaySilverman2005,
  author    = {Ramsay, James O. and Silverman, Bernard W.},
  title     = {Functional Data Analysis},
  edition   = {2nd},
  series    = {Springer Series in Statistics},
  publisher = {Springer},
  address   = {New York},
  year      = {2005},
  doi       = {10.1007/b98888}
}

@book{HorvathKokoszka2012,
  author    = {Horv{\'a}th, Lajos and Kokoszka, Piotr},
  title     = {Inference for Functional Data with Applications},
  series    = {Springer Series in Statistics},
  publisher = {Springer},
  address   = {New York},
  year      = {2012},
  doi       = {10.1007/978-1-4614-3655-3}
}

@article{HwangEtAl2012,
  author  = {Hwang, Heungsun and Jung, Kwanghee and Takane, Yoshio and Woodward, Todd S.},
  title   = {Functional Multiple-Set Canonical Correlation Analysis},
  journal = {Psychometrika},
  volume  = {77},
  number  = {1},
  pages   = {48--64},
  year    = {2012},
  doi     = {10.1007/s11336-011-9234-4}
}

@article{GoreckiEtAl2020,
  author  = {G{\'o}recki, Tomasz and Krzy{\'s}ko, Miros{\l}aw and Wo{\l}y{\'n}ski, Waldemar},
  title   = {Generalized Canonical Correlation Analysis for Functional Data},
  journal = {Biometrical Letters},
  volume  = {57},
  number  = {1},
  pages   = {1--12},
  year    = {2020},
  doi     = {10.2478/bile-2020-0001}
}

@article{SortEtAl2024,
  author  = {Sort, Lucas and Le Brusquet, Laurent and Tenenhaus, Arthur},
  title   = {Functional Generalized Canonical Correlation Analysis for Studying Multiple Longitudinal Variables},
  journal = {Biometrics},
  volume  = {80},
  number  = {4},
  pages   = {ujae113},
  year    = {2024},
  doi     = {10.1093/biomtc/ujae113}
}

@article{Hotelling1936,
  author  = {Hotelling, Harold},
  title   = {Relations Between Two Sets of Variates},
  journal = {Biometrika},
  volume  = {28},
  number  = {3--4},
  pages   = {321--377},
  year    = {1936},
  doi     = {10.1093/biomet/28.3-4.321}
}

@article{Horst1961,
  author  = {Horst, Paul},
  title   = {Relations Among $m$ Sets of Measures},
  journal = {Psychometrika},
  volume  = {26},
  number  = {2},
  pages   = {129--149},
  year    = {1961},
  doi     = {10.1007/BF02289710}
}

@inproceedings{Carroll1968,
  author    = {Carroll, J. Douglas},
  title     = {Generalization of Canonical Correlation Analysis to Three or More Sets of Variables},
  booktitle = {Proceedings of the 76th Annual Convention of the American Psychological Association},
  volume    = {3},
  pages     = {227--228},
  publisher = {American Psychological Association},
  address   = {Washington, DC},
  year      = {1968}
}

@article{Kettenring1971,
  author  = {Kettenring, Jon R.},
  title   = {Canonical Analysis of Several Sets of Variables},
  journal = {Biometrika},
  volume  = {58},
  number  = {3},
  pages   = {433--451},
  year    = {1971},
  doi     = {10.1093/biomet/58.3.433}
}

@book{Gifi1990,
  author    = {Gifi, Albert},
  title     = {Nonlinear Multivariate Analysis},
  series    = {Wiley Series in Probability and Mathematical Statistics},
  publisher = {Wiley},
  address   = {Chichester},
  year      = {1990}
}

@article{TakaneEtAl2008,
  author  = {Takane, Yoshio and Hwang, Heungsun and Abdi, Herv{\'e}},
  title   = {Regularized Multiple-Set Canonical Correlation Analysis},
  journal = {Psychometrika},
  volume  = {73},
  number  = {4},
  pages   = {753--775},
  year    = {2008},
  doi     = {10.1007/s11336-008-9065-0}
}

%%%%%%%%%%%%%%%%%%%%%%%%%%%%%%%%%%%%%%%%%%%%%%%%%%%%
%%
%%         Supplementary Inforamtion
%%
%%%%%%%%%%%%%%%%%%%%%%%%%%%%%%%%%%%%%%%%%%%%%%%%%%%%
\clearpage

% セクション・数式・定理環境・図表の番号を(S.1)のようにする
\renewcommand{\theequation}{S.\arabic{equation}}
\setcounter{equation}{0} % 数式番号をリセット

\renewcommand{\thesection}{S.\arabic{section}}
\setcounter{section}{0} % セクション番号をリセット

\renewcommand{\theTheo}{S.\arabic{Theo}}
\setcounter{Theo}{0} % 定理環境の番号をリセット

\renewcommand{\thefigure}{S.\arabic{figure}}
\setcounter{figure}{0} % 図の番号をリセット

\renewcommand{\thetable}{S.\arabic{table}}
\setcounter{table}{0} % 表の番号をリセット

\begin{center}
    {\Large Functional Multiple-Set Canonical Correlation Analysis Revisited: From Finite-Dimensional Samples to Infinite-Dimensional Populations}
\end{center}
\begin{center}
    {\Large Supplementary Materials}
\end{center}
\begin{center}
    {\large Michio Yamamoto, Yoshikazu Terada}
\end{center}

All notation is used with the same meanings as in the main text.

\section{Proofs of Lemmas}
\label{SI-sec:Lemmas}

\subsection{Proof of Lemma~\ref*{lemm:lemma_for_theorem_optimization}}
\label{SI-subsec:proof-lemma_for_theorem_optimization}

% Let $\Pi=\{\pi\subset L_2^P\mid\dim(\pi^\perp\cap\overline{F_W})=m\}$ be a collection of subspaces in $L_2^P$ with codimensions equal to $m$.
% Let $\Pi=\{\pi\subset \overline{F_W}\mid\text{$\pi$ is a closed subspace and }\dim(\pi^\perp\cap\overline{F_W})=m\}$ be a collection of closed subspaces in $\overline{F_W}$ with codimensions equal to $m$.
Let $\tilde{\bs{\alpha}}\in\myspan\{\tilde{\bs{\alpha}}_1,\dots,\tilde{\bs{\alpha}}_m\}^\perp\cap\overline{F_W}\backslash\{0\}$.
Since $\tilde{\mac{R}}$ is a compact self-adjoint operator on $L_2^P$, we may extend the orthonormal family of positive eigenfunctions $\{\tilde{\bs{\alpha}}_k\}_{k=1}^r$ by adding eigenfunctions corresponding to the negative eigenvalues, and if necessary, a CONS of $\ker(\tilde{\mac{R}})$, so that the resulting system is a CONS of $L_2^P$.
Let $\{\bs{\psi}_a\}_{a\in \mcr{A}}$ denote these additional eigenfunctions, and let $\lambda_a\le 0$ be the corresponding eigenvalues.

Since $\tilde{\bs{\alpha}}$ is orthogonal to $\tilde{\bs{\alpha}}_1,\dots,\tilde{\bs{\alpha}}_m$, it can be written as
\ba
  \tilde{\bs{\alpha}}
  =\sum_{i>m}^r\lan\tilde{\bs{\alpha}},\tilde{\bs{\alpha}}_i\ran_P\tilde{\bs{\alpha}}_i
  +\sum_{a\in \mcr{A}}\lan\tilde{\bs{\alpha}},\bs{\psi}_a\ran_P\bs{\psi}_a.
\ea
Hence,
\ba
  \tilde{\mac{R}}\tilde{\bs{\alpha}}
  % &=\sum_{i=1}^r\lan\tilde{\mac{R}}\tilde{\bs{\alpha}},\tilde{\bs{\alpha}}_i\ran_P\tilde{\bs{\alpha}}_i
  % +\sum_{a\in \mcr{A}}\lan\tilde{\mac{R}}\tilde{\bs{\alpha}},\bs{\psi}_a\ran_P\bs{\psi}_a
  % =\sum_{i=1}^\infty\lan\sum_{j=1}^\infty\tilde{\rho}_j\lan\tilde{\bs{\alpha}},\tilde{\bs{\alpha}}_j\ran_P\tilde{\bs{\alpha}}_j,\tilde{\bs{\alpha}}_i\ran_P\tilde{\bs{\alpha}}_i\\
  % =\sum_{i=1}^r\tilde{\rho}_i\lan\tilde{\bs{\alpha}},\tilde{\bs{\alpha}}_i\ran_P\tilde{\bs{\alpha}}_i + \sum_{a\in \mcr{A}}\lambda_a\lan\tilde{\bs{\alpha}},\bs{\psi}_a\ran_P\bs{\psi}_a\\
  =\sum_{i>m}^r\tilde{\rho}_i\lan\tilde{\bs{\alpha}},\tilde{\bs{\alpha}}_i\ran_P\tilde{\bs{\alpha}}_i
  +\sum_{a\in \mcr{A}}\lambda_a\lan\tilde{\bs{\alpha}},\bs{\psi}_a\ran_P\bs{\psi}_a.
\ea
% Because $\tilde{\mac{R}}$ is a self-adjoint Hilbert--Schmidt operator on $L_2^P$ (Proposition~\ref{prop:extend_R}) and the criterion~\eqref{eq:criterion_population} is non-negative for any weight functions, $\tilde{\mac{R}}$ possesses eigenvalues $\tilde{\rho}_1\ge\tilde{\rho}_2\ge\cdots$ with $1+\tilde{\rho}_i\ge0$ for all $i\in\mbb{N}$.
% The correspponding eigenfunctions, augmented with a complete orthonormal system (CONS) for $\ker(\mac{R})$, yield $\{\tilde{\bs{\alpha}}_k\}_{k\ge1}$ which forms a CONS for $L_2^P$.
Therefore,
\ba
  \lan\tilde{\bs{\alpha}},(\mac{I}+\tilde{\mac{R}})\tilde{\bs{\alpha}}\ran_P
  &=\lan\tilde{\bs{\alpha}},\sum_{i>m}^r(1+\tilde{\rho}_i)\lan\tilde{\bs{\alpha}},\tilde{\bs{\alpha}}_i\ran_P\tilde{\bs{\alpha}}_i\ran_P
  +\lan\tilde{\bs{\alpha}},\sum_{a\in \mcr{A}}(1+\lambda_a)\lan\tilde{\bs{\alpha}},\bs{\psi}_a\ran_P\bs{\psi}_a\ran_P\\
  &=\sum_{i>m}^r(1+\tilde{\rho}_i)\left|\lan\tilde{\bs{\alpha}},\tilde{\bs{\alpha}}_i\ran_P\right|^2
  +\sum_{a\in \mcr{A}}(1+\lambda_a)\left|\lan\tilde{\bs{\alpha}},\bs{\psi}_a\ran_P\right|^2\\
  &\le(1+\tilde{\rho}_{m+1})\left(\sum_{i>m}^r\left|\lan\tilde{\bs{\alpha}},\tilde{\bs{\alpha}}_i\ran_P\right|^2
  +\sum_{a\in \mcr{A}}\left|\lan\tilde{\bs{\alpha}},\bs{\psi}_a\ran_P\right|^2\right)\\
  &=(1+\tilde{\rho}_{m+1})\|\tilde{\bs{\alpha}}\|_P^2.
\ea
Then,
\banum
  1+\tilde{\rho}_{m+1}
  % \ge\sup_{\tilde{\bs{\alpha}}\in\myspan\{\tilde{\bs{\alpha}}_1,\dots,\tilde{\bs{\alpha}}_m\}^\perp\backslash\{0\}}
  % \frac{\lan\tilde{\bs{\alpha}},(\mac{I}+\tilde{\mac{R}})\tilde{\bs{\alpha}}\ran_P}{\|\tilde{\bs{\alpha}}\|_P^2}
  \ge\sup_{\tilde{\bs{\alpha}}\in\myspan\{\tilde{\bs{\alpha}}_1,\dots,\tilde{\bs{\alpha}}_m\}^\perp\cap\overline{F_W}\backslash\{0\}}
  \frac{\lan\tilde{\bs{\alpha}},(\mac{I}+\tilde{\mac{R}})\tilde{\bs{\alpha}}\ran_P}{\|\tilde{\bs{\alpha}}\|_P^2}
  \label{SI-eq:lemma_for_theorem_optimization-1}
\eanum
Next, we show the other direction.
Since $\tilde{\bs{\alpha}}_{m+1}\in\myspan\{\tilde{\bs{\alpha}}_1,\dots,\tilde{\bs{\alpha}}_m\}^\perp\cap\overline{F_W}\backslash\{0\}$, we can choose $\tilde{\bs{\alpha}}=\tilde{\bs{\alpha}}_{m+1}$ in the supremum.
Then, 
\ba
  \frac{\lan\tilde{\bs{\alpha}}_{m+1},(\mac{I}+\tilde{\mac{R}})\tilde{\bs{\alpha}}_{m+1}\ran_P}{\|\tilde{\bs{\alpha}}_{m+1}\|_P^2}
  =1+\tilde{\rho}_{m+1}.
\ea
Hence,
\ba
  1+\tilde{\rho}_{m+1}
  \le\sup_{\tilde{\bs{\alpha}}\in\myspan\{\tilde{\bs{\alpha}}_1,\dots,\tilde{\bs{\alpha}}_m\}^\perp\cap\overline{F_W}\backslash\{0\}}
  \frac{\lan\tilde{\bs{\alpha}},(\mac{I}+\tilde{\mac{R}})\tilde{\bs{\alpha}}\ran_P}{\|\tilde{\bs{\alpha}}\|_P^2}.
\ea
Together with \eqref{SI-eq:lemma_for_theorem_optimization-1}, we have established this lemma.

%%%%%%%%%%%%%%%%%%%%%%%%%%%%%%%%%%%%%%%%%%%%%%%%%%%
%%           XXXXXXXXXXXXXXXXXXXXXX
%%%%%%%%%%%%%%%%%%%%%%%%%%%%%%%%%%%%%%%%%%%%%%%%%%%
\subsection{Proof of Lemma~\ref*{lemma:lemma_for_theorem_canonical-decomposition}}
\label{SI-subsec:proof-lemma_for_theorem_canonical-decomposition}

Throughout this proof, all equalities involving inner products with $\bs{X}$, $\bs{X}_{(K)}$, or $\bs{X}_{(K)}^\perp$ are understood as equalities of real-valued random variables and hold almost surely.

(1) If $\mac{W}^{\frac{1}{2}}\bs{u}\in\Span\{\tilde{\bs{\alpha}}_{k},k\in[K]\}$, then there exist $a_{k}\in\mbb{R},k\in[K]$, such that $\mac{W}^{\frac{1}{2}}\bs{u}=\sum_{k=1}^{K}a_{k}\tilde{\bs{\alpha}}_{k}$.
Since $\bs{\alpha}_{k}=\mac{W}^{-\frac{1}{2}}\tilde{\bs{\alpha}}_{k}$, we have
\begin{equation*}
  \bs{u}
  =\sum_{k=1}^{K}a_{k}\mac{W}^{-\frac{1}{2}}\tilde{\bs{\alpha}}_{k}
  =\sum_{k=1}^{K}a_{k}\bs{\alpha}_{k}\in\Span\{\bs{\alpha}_{k},k\in[K]\}.
\end{equation*}
Observe that for $1\le{k}\le{K}$,
\begin{equation*}
\left<\bs{\alpha}_{k},\bs{X}_{(K)}\right>_{P}
=\left<\mac{W}^{-\frac{1}{2}}\tilde{\bs{\alpha}}_{k},\sum_{l=1}^{K}\eta_{l}\mac{W}^{\frac{1}{2}}\tilde{\bs{\alpha}}_{l}\right>_{P}
=\sum_{l=1}^{K}\eta_{l}\left<\tilde{\bs{\alpha}}_{k},\tilde{\bs{\alpha}}_{l}\right>_{P}
=\eta_{k}
=\left<\bs{\alpha}_{k},\bs{X}\right>_{P}.
\end{equation*}
Therefore, we have
$\left<\bs{u},\bs{X}_{(K)}\right>_{P}=\left<\bs{u},\bs{X}\right>_{P}$.
In addition,
\begin{equation*}
 \left<\bs{\alpha}_{k},\bs{X}_{(K)}^{\perp}\right>_{P}
=\left<\bs{\alpha}_{k},\bs{X}-\bs{X}_{(K)}\right>_{P}
=\left<\bs{\alpha}_{k},\bs{X}\right>_{P}-\left<\bs{\alpha}_{k},\bs{X}_{(K)}\right>_{P}
=0.
\end{equation*}
Thus, $\left<\bs{u},\bs{X}_{(K)}^{\perp}\right>_{P}=0$.

\noindent (2) If $\mac{W}^{\frac{1}{2}}\bs{u}\in\Span\{\tilde{\bs{\alpha}}_{k},k\in[K]\}^{\perp}$,
\begin{equation*}
  \left<\bs{u},\bs{X}_{(K)}\right>_{P}
  =\left<\bs{u},\sum_{k=1}^{K}\eta_{k}\mac{W}^{\frac{1}{2}}\tilde{\bs{\alpha}}_{k}\right>_{P}
  =\sum_{k=1}^{K}\eta_{k}\left<\mac{W}^{\frac{1}{2}}\bs{u},\tilde{\bs{\alpha}}_{k}\right>_{P}
  =0,
\end{equation*}
and
\begin{equation*}
  \left<\bs{u},\bs{X}_{(K)}^{\perp}\right>_{P}
  =\left<\bs{u},\bs{X}-\bs{X}_{(K)}\right>_{P}
  =\left<\bs{u},\bs{X}\right>_{P}
  -\left<\bs{u},\bs{X}_{(K)}\right>_{P}
  =\left<\bs{u},\bs{X}\right>_{P}.
\end{equation*}

\section{Proofs of propositions}
\label{SI-sec:proofs}

%%%%%%%%%%%%%%%%%%%%%%%%%%%%%%%%%%%%%%%%%%%%%%%%%%%
%%           V is Hilbert-Schmidt
%%%%%%%%%%%%%%%%%%%%%%%%%%%%%%%%%%%%%%%%%%%%%%%%%%%
\subsection{Proof of Proposition~\ref*{prop:V-HS}}
\label{SI-subsec:proof-prop-V-HS}

Let $\{e_j\}_{j\ge1}$ be an orthonormal basis of $L_2(\mac{T})$ and let $\{\bs{e}_{pj}\}_{p\in[P],j\ge1}$ be an orthonormal basis of $L_2^P$ such that the $p$th component of $\bs{e}_{pj}$ is $e_j$ and the others are zero:
\ba
\bs{e}_{pj}
=(0,\dots,0,e_j,0,\dots,0)^\top
\ea
where $e_j$ appears in the $p$th position.

Each diagonal block $\mac{V}_{pp}$, $p\in[P]$, is a trace-class operator provided that $\mbb{E}[\|X_p\|^2]<\infty$.
Indeed, by Parseval's identity and Fubini's theorem, we have
\ba
  \tr(\mac{V}_{pp})
  &=\sum_{j=1}^\infty\lan\mac{V}_{pp}e_{j},e_{j}\ran
  =\sum_{j=1}^\infty\mbb{E}[\lan X_p,e_{j}\ran^2]
  =\mbb{E}[\|X_p\|^2]
  <\infty.
\ea
Therefore, $\mac{V}_{pp}$ is trace-class.

Next, we show that $\mac{V}$ is trace-class.
Note that, for any $\bs{z}\in L_2^P$,
\ba
  \mac{V}\bs{z}
  &=\left(\sum_{q=1}^P\mac{V}_{1q}z_q,\dots,\sum_{q=1}^P\mac{V}_{Pq}z_q\right)^\top.
\ea
Then, for any $\bs{e}_{pj}$ and for $p'\in[P]$, we have
\ba
  (\mac{V}\bs{e}_{pj})_{p'}
  &=\sum_{q=1}^P\mac{V}_{p'q}(e_{pj})_q
  =\mac{V}_{p'p}e_j
  =\mbb{E}[\lan X_{p},e_j\ran X_{p'}].
\ea
Note that, for any $\bs{z}\in L_2^P$, we have
\ba
  \lan\bs{z},\bs{e}_{pj}\ran
  =\sum_{p'=1}^P\lan z_{p'},(e_{pj})_{p'}\ran
  % =\sum_{p=1}^P\lan z_p,e_j\ran\delta_{pp}
  =\lan z_p,e_j\ran.
\ea
Then, we have
\ba
  \lan\mac{V}\bs{e}_{pj},\bs{e}_{pj}\ran
  &=\lan(\mac{V}\bs{e}_{pj})_p,e_j\ran
  =\lan\mbb{E}[\lan X_p,e_j\ran X_p],e_j\ran
  =\mbb{E}[\lan X_p,e_j\ran^2].
\ea
Hence,
\ba
  \tr(\mac{V})
  &=\sum_{p=1}^P\sum_{j=1}^\infty\lan\mac{V}\bs{e}_{pj},\bs{e}_{pj}\ran
  =\sum_{p=1}^P\sum_{j=1}^\infty\mbb{E}[\lan X_p,e_j\ran^2]
  =\sum_{p=1}^P\mbb{E}[\|X_p\|^2]
  =\mbb{E}[\|\bs{X}\|_P^2]
  <\infty.
\ea
This shows that $\mac{V}$ is trace-class.

%% 以下のtrace-classの議論は、Kostenko (2019) や Muger (2022) のlecture note を参考にした．
Since $\mathcal V$ is trace-class on $L_2^P$, each block operator $\mac{V}_{pq}$ is also trace-class. 
Indeed, if $\iota_q:L_2\to L_2^P$ denotes the canonical embedding into the $q$th component and $\pi_p:L_2^P\to L_2$ denotes the projection onto the $p$th component, then $\mac{V}_{pq}=\pi_p\mac{V}\iota_q$. 
Since trace-class operators form a two-sided ideal under composition with bounded operators and since $\pi_p$ and $\iota_q$ are bounded, $\mac{V}_{pq}$ is trace-class.
Moreover, writing $\|\cdot\|_{op}$ for the operator norm and $\|\cdot\|_{TR}$ for the trace norm, we have
\ba
  \|\mac{V}_{pq}\|_{TR}
  =\|\pi_p\mac{V}\iota_q\|_{TR}
  \le\|\pi_p\|_{op}\|\mac{V}\|_{TR}\|\iota_q\|_{op}
  =\|\mac{V}\|_{TR}
  <\infty.
\ea

\subsection{Proof of Proposition~\ref*{prop:W-HS}}
\label{SI-subsec:proof-prop-W-HS}

Let $\{\bs{e}_{pj}\}_{p\in[P],j\ge1}$ be the same as in the proof of Proposition \ref{prop:V-HS}.
Note that, for any $\bs{e}_{pj}$, we have
\ba
  \mac{W}\bs{e}_{pj}
  &=\left(\mac{V}_{11}(\bs{e}_{pj})_1,\dots,\mac{V}_{PP}(\bs{e}_{pj})_P\right)^\top
  =(0,\dots,0,\mac{V}_{pp}e_j,0,\dots,0)^\top.
\ea
Then, we have
\ba
  \tr(\mac{W})
  &=\sum_{p=1}^P\sum_{j=1}^\infty\lan\mac{W}\bs{e}_{pj},\bs{e}_{pj}\ran
  =\sum_{p=1}^P\sum_{j=1}^\infty\lan\mac{V}_{pp}e_j,e_j\ran
  =\sum_{p=1}^P\tr(\mac{V}_{pp})
  <\infty.
\ea
Therefore, $\mac{W}$ is a trace-class operator.

% For $p\in[P], define an operator $\mac{W}^{(p)}:L_2^P\ra L_2$ as $\mac{W}^{(p)}\bs{h}(\bs{t})=\mac{V}_{pp} h_p(t_p)$.
% Similar to the proof of Proposition \ref{prop:V-HS}, we can see that when all $\mac{W}^{(p)}$'s are Hilbert-Schmidt, $\mac{W}$ is Hilbert-Schmidt. 
% Also, we observe that
% \ba
%   \|\mac{W}^{(p)}\|_{HS}^2
%   &=\sum_{k=1}^\infty\sum_{j=1}^P\lan\mac{W}^{(p)}e_{jk},\mac{W}^{(p)}e_{jk}\ran\\
%   &=\sum_{k=1}^\infty\sum_{j=1}^P\lan\mac{V}_{pp}(e_{jk})_{p},(\mac{V}_{pp}(e_{jk})_{p}\ran\\
%   &=\sum_{k=1}^\infty\lan\mac{V}_{pp}(e_{pk})_{p},\mac{V}_{pp}(e_{pk})_{p}\ran\\
%   &=\sum_{k=1}^\infty\lan\mac{V}_{pp}e_k,\mac{V}_{pp}e_k\ran\\
%   &=\|\mac{V}_{pp}\|_{HS}^2.
% \ea
% From Proposition \ref{prop:V-HS}, $\mac{V}_{pp}$ is Hilbert-Schmidt, i.e., $\mac{W}^{(p)}$ is Hilbert-Schmidt.
% Therefore, we conclude that $\mac{W}$ is Hilbert-Schmidt.

\subsection{Proof of Proposition~\ref*{prop:existence_condition1}}
\label{SI-subsec:proof-prop-existence_condition1}

Note that, for $\bs{z}=(z_{1},\dots,z_{P})^{\top}\in F_{W}$, we have
\begin{equation*}
  \mac{V}_{pp}^{-\frac{1}{2}}z_{p}
  =\sum_{i=1}^{\infty}\zeta_{pi}^{-\frac{1}{2}}\left<\phi_{pi},z_{p}\right>\phi_{pi}.
\end{equation*}
% \begin{equation*} \mac{V}_{pq}y
%   =\sum_{i=1}^{\infty}\sum_{j=1}^{\infty}\mbb{E}[\xi_{pi}\xi_{qj}]\left<\phi_{qj},y\right>\phi_{pi},
% \end{equation*}
% specifically if $p=q$,
% \begin{equation*}
%   \mac{V}_{pp}y=\sum_{i=1}^{\infty}\zeta_{pi}\left<\phi_{pi},y\right>\phi_{pi}.
% \end{equation*}
Thus, 
% for $\bs{z}=(z_{1},\dots,z_{P})^{\top}\in F_{W}$, we have
% \begin{equation*}
%   \mac{V}_{pp}^{-\frac{1}{2}}z_{p}
%   =\sum_{i=1}^{\infty}\zeta_{pi}^{-\frac{1}{2}}\left<\phi_{pi},z_{p}\right>\phi_{pi}.
% \end{equation*}
 using $\mac{W}^{-\frac{1}{2}}\bs{z}=(\mac{V}_{11}^{-\frac{1}{2}}z_{1},\dots,\mac{V}_{PP}^{-\frac{1}{2}}z_{P})^{\top}$,
\begin{equation*}
  \mac{V}_\star\mac{W}^{-\frac{1}{2}}\bs{z}
  =\mac{V}_\star
  \begin{bmatrix}
    \mac{V}_{11}^{-\frac{1}{2}}z_{1}\\
    \vdots\\
    \mac{V}_{PP}^{-\frac{1}{2}}z_{P}
  \end{bmatrix}
  =\begin{bmatrix}
    \sum_{q\neq 1}\mac{V}_{1q}\mac{V}_{qq}^{-\frac{1}{2}}z_{q}\\
    \vdots\\
    \sum_{q\neq P}\mac{V}_{Pq}\mac{V}_{qq}^{-\frac{1}{2}}z_{q}
  \end{bmatrix}.
\end{equation*}

We derive a sufficient condition that guarantees $\mac{V}_\star\mac{W}^{-\frac{1}{2}}(F_W)\subset F_W$.
Fix an arbitrary $\bs{z}\in F_W$. 
First, we show that for each $p$, the following holds:
\begin{align}
  \sum_{l=1}^{\infty}\zeta_{pl}^{-1}\left|\lan(\mac{V}_\star\mac{W}^{-\frac{1}{2}}\bs{z})_p,\phi_{pl}\ran\right|^2
  <\infty.
  \label{eq:condition1-alt}
\end{align}
Now, since
\banum
  (\mac{V}_\star\mac{W}^{-\frac{1}{2}}\bs{z})_p
  =\sum_{q\neq p}\mac{V}_{pq}\mac{V}_{qq}^{-\frac{1}{2}}z_q
  \label{eq:Proposition3-proof-1}
\eanum
and
\begin{align*}
  \mac{V}_{pp}^{-\frac{1}{2}}y
  =\sum_{l=1}^{\infty}\zeta_{pl}^{-\frac{1}{2}}\lan y,\phi_{pl}\ran\phi_{pl},
\end{align*}
we obtain
\begin{align*}
  &\left\|\sum_{q\neq p}\mac{V}_{pp}^{-\frac{1}{2}}\mac{V}_{pq}\mac{V}_{qq}^{-\frac{1}{2}}z_q\right\|^2\\
  &=\lan\mac{V}_{pp}^{-\frac{1}{2}}(\mac{V}_\star\mac{W}^{-\frac{1}{2}}\bs{z})_p,
  \mac{V}_{pp}^{-\frac{1}{2}}(\mac{V}_\star\mac{W}^{-\frac{1}{2}}\bs{z})_p\ran\\
  &=\lan\sum_{l=1}^{\infty}\zeta_{pl}^{-\frac{1}{2}}\lan(\mac{V}_\star\mac{W}^{-\frac{1}{2}}\bs{z})_p,\phi_{pl}\ran\phi_{pl},
  \sum_{l=1}^{\infty}\zeta_{pl}^{-\frac{1}{2}}\lan(\mac{V}_\star\mac{W}^{-\frac{1}{2}}\bs{z})_p,\phi_{pl}\ran\phi_{pl}\ran\\
  &=\sum_{l=1}^{\infty}\zeta_{pl}^{-1}\left|\lan(\mac{V}_\star\mac{W}^{-\frac{1}{2}}\bs{z})_p,\phi_{pl}\ran\right|^2.
\end{align*}
Thus, \eqref{eq:condition1-alt} is equivalent to the finiteness of $\|\sum_{q\neq p}\mac{V}_{pp}^{-\frac{1}{2}}\mac{V}_{pq}\mac{V}_{qq}^{-\frac{1}{2}}z_q\|$ for each $p$.
% This shows that the boundedness of $\sum_{q\neq p}\mac{V}_{pp}^{-\frac{1}{2}}\mac{V}_{pq}\mac{V}_{qq}^{-\frac{1}{2}}z_q$ is equivalent to \eqref{eq:condition1-alt}.
% equivalent to the requirement that, for any $\bs{z}\in F_{W}$ and for any $p\in[P]$, $\|\mac{V}_{pp}^{-\frac{1}{2}}\sum_{q\neq{p}}\mac{V}_{pq}\mac{V}_{qq}^{-\frac{1}{2}}z_{q}\|^{2}<\infty$.
% \begin{equation*}
%   \|\mac{V}_{pp}^{-\frac{1}{2}}\sum_{q\neq{p}}\mac{V}_{pq}\mac{V}_{qq}^{-\frac{1}{2}}z_{q}\|^{2}<\infty.
% \end{equation*}
Note that
\ba
  \left\|\sum_{q\neq p}\mac{V}_{pp}^{-\frac{1}{2}}\mac{V}_{pq}\mac{V}_{qq}^{-\frac{1}{2}}z_{q}\right\|
  \le\sum_{q\neq p}\|\mac{V}_{pp}^{-\frac{1}{2}}\mac{V}_{pq}\mac{V}_{qq}^{-\frac{1}{2}}z_{q}\|
  \le\sum_{q\neq p}\|\mac{V}_{pp}^{-\frac{1}{2}}\mac{V}_{pq}\mac{V}_{qq}^{-\frac{1}{2}}\|\,\|z_{q}\|.
\ea
% \begin{align*}
%   &\left\|\sum_{q\neq p}\mac{V}_{pp}^{-\frac{1}{2}}\mac{V}_{pq}\mac{V}_{qq}^{-\frac{1}{2}}z_{q}\right\|^{2}\\
%   &=\sum_{q\neq p}\|\mac{V}_{pp}^{-\frac{1}{2}}\mac{V}_{pq}\mac{V}_{qq}^{-\frac{1}{2}}z_{q}\|^{2}
%   +\sum_{q\neq p}\sum_{r\neq q}\left<\mac{V}_{pp}^{-\frac{1}{2}}\mac{V}_{pq}\mac{V}_{qq}^{-\frac{1}{2}}z_{q},\mac{V}_{pp}^{-\frac{1}{2}}\mac{V}_{pr}\mac{V}_{rr}^{-\frac{1}{2}}z_{r}\right>\\
%   % &\le\sum_{q\neq p}\|\mac{V}_{pp}^{-\frac{1}{2}}\mac{V}_{pq}\mac{V}_{qq}^{-\frac{1}{2}}z_{q}\|^{2}
%   % +\sum_{q\neq p}\sum_{r\neq q}\left|\left<\mac{V}_{pp}^{-\frac{1}{2}}\mac{V}_{pq}\mac{V}_{qq}^{-\frac{1}{2}}z_{q},\mac{V}_{pp}^{-\frac{1}{2}}\mac{V}_{pr}\mac{V}_{rr}^{-\frac{1}{2}}z_{r}\right>\right|\\
%   &\le\sum_{q\neq p}\|\mac{V}_{pp}^{-\frac{1}{2}}\mac{V}_{pq}\mac{V}_{qq}^{-\frac{1}{2}}z_{q}\|^{2}
%   +\sum_{q\neq p}\sum_{r\neq q}\|\mac{V}_{pp}^{-\frac{1}{2}}\mac{V}_{pq}\mac{V}_{qq}^{-\frac{1}{2}}z_{q}\|\cdot\|\mac{V}_{pp}^{-\frac{1}{2}}\mac{V}_{pr}\mac{V}_{rr}^{-\frac{1}{2}}z_{r}\|.
% \end{align*}
Therefore, if  $\|\mac{V}_{pp}^{-\frac{1}{2}}\mac{V}_{pq}\mac{V}_{qq}^{-\frac{1}{2}}z_{q}\|<\infty$ for $(p,q)\in[P]^2$ with $p\neq q$, then \eqref{eq:condition1-alt} holds. 
%% He et al. (2003) を引用せずに、ここで示してしまっておく。(2026/06/28)
% Actually, the operator $\mac{V}_{pp}^{-\frac{1}{2}}\mac{V}_{pq}\mac{V}_{qq}^{-\frac{1}{2}}$ is same as the one for the ordinary functional canonical correlation analysis (FCCA) with the pair of square-integrable stochastic processes $(z_{p},z_{q})$. 
% Thus, Proposition 4.2 of \citet{he2003functional} guarantees condition \eqref{eq:condition1-alt} under the assumption of the proposition.
For $j\ge1$, we have
\ba
  \mac{V}_{pp}^{-\frac{1}{2}}\mac{V}_{pq}\mac{V}_{qq}^{-\frac{1}{2}}\phi_{qj}
  =\sum_{i=1}^{\infty}\zeta_{pi}^{-\frac{1}{2}}\zeta_{qj}^{-\frac{1}{2}}\mbb{E}[\xi_{pi}\xi_{qj}]\phi_{pi}.
\ea
Hence, under Condition~\ref{cond:cond1}, we have
\ba
  \|\mac{V}_{pp}^{-\frac{1}{2}}\mac{V}_{pq}\mac{V}_{qq}^{-\frac{1}{2}}\|_{HS}^2
  =\sum_{j=1}^\infty\|\mac{V}_{pp}^{-\frac{1}{2}}\mac{V}_{pq}\mac{V}_{qq}^{-\frac{1}{2}}\phi_{qj}\|^2
  =\sum_{j=1}^\infty\sum_{i=1}^{\infty}\zeta_{pi}^{-1}\zeta_{qj}^{-1}(\mbb{E}[\xi_{pi}\xi_{qj}])^2
  <\infty.
\ea
Since $\{\phi_{qj}\}_{j\ge1}$ is a CONS of $(\ker(\mac{V}_{qq}))^{\perp}$, the preceding calculation shows that $\mac{V}_{pp}^{-\frac{1}{2}}\mac{V}_{pq}\mac{V}_{qq}^{-\frac{1}{2}}$ is a Hilbert--Schmidt operator on $(\ker(\mac{V}_{qq}))^{\perp}$.
We then extend $\mac{V}_{pp}^{-\frac{1}{2}}\mac{V}_{pq}\mac{V}_{qq}^{-\frac{1}{2}}$ to the whole space $L_2$ by defining it to be zero on $\ker(\mac{V}_{qq})$.
With this extension, $\mac{V}_{pp}^{-\frac{1}{2}}\mac{V}_{pq}\mac{V}_{qq}^{-\frac{1}{2}}$ is a Hilbert--Schmidt operator, and hence a bounded operator, from $L_2$ to $L_2$.
 
Next, we confirm that for each $p$, $(\mac{V}_\star\mac{W}^{-\frac{1}{2}}\bs{z})_p\perp\ker(\mac{V}_{pp})$.
For any $x\in\ker(\mac{V}_{pp})$, we have $\lan x,\phi_{pi}\ran=0$ for all $i$.
Thus, we have
\ba
  \mac{V}_{qp}x
  =\sum_{i=1}^\infty\sum_{j=1}^\infty\mbb{E}[\xi_{qi}\xi_{pj}]\lan x,\phi_{pj}\ran\phi_{qi}
  =0.
\ea
Therefore, 
\ba
  \lan(\mac{V}_\star\mac{W}^{-\frac{1}{2}}\bs{z})_p,x\ran
  &=\sum_{q\neq p}\lan\mac{V}_{pq}\mac{V}_{qq}^{-\frac{1}{2}}z_q,x\ran
  =\sum_{q\neq p}\lan\mac{V}_{qq}^{-\frac{1}{2}}z_q,\mac{V}_{qp}x\ran
  =0.
\ea
Consequently, $(\mac{V}_\star\mac{W}^{-\frac{1}{2}}\bs{z})_p\perp\ker(\mac{V}_{pp})$ for all $p$.
This completes the proof.

\subsection{Proof of Proposition~\ref*{prop:extend_R}}
\label{SI-subsec:proof-prop-extend_R}

For any $\bs{z}\in F_W$, we have
\ba
  (\mac{R}\bs{z})_{p}
  % &=(\mac{W}^{-\frac{1}{2}}\mac{V}_\star\mac{W}^{-\frac{1}{2}}\bs{z})_{p}\\
  &=\sum_{q\neq p}\sum_{i=1}^\infty\sum_{j=1}^\infty\zeta_{qi}^{-\frac{1}{2}}\zeta_{pj}^{-\frac{1}{2}}\mbb{E}[\xi_{qi}\xi_{pj}]\lan z_q,\phi_{qi}\ran\phi_{pj}\\
  &=\sum_{q\neq p}\sum_{i=1}^\infty\sum_{j=1}^\infty \omega_{pqij}\lan z_q,\phi_{qi}\ran\phi_{pj}
\ea
where $\omega_{pqij}:=\zeta_{qi}^{-\frac{1}{2}}\zeta_{pj}^{-\frac{1}{2}}\mbb{E}[\xi_{qi}\xi_{pj}]$.
% \ba
%   \omega_{pqij}
%   &:=\zeta_{qi}^{-\frac{1}{2}}\zeta_{pj}^{-\frac{1}{2}}\mbb{E}[\xi_{qi}\xi_{pj}].
% \ea

Let $\bs{e}_{jk}\in L_2^P$ denote the element whose $j$th component is $\phi_{jk}$ and whose other components are zero.
Then, the collection $\{\bs{e}_{jk}\}_{j\in[P],k\ge1}$ forms an orthonormal system.
We have
\ba
  \mac{R}\bs{z}
  % &=\mac{W}^{-\frac{1}{2}}\mac{V}_\star\mac{W}^{-\frac{1}{2}}\bs{z}\\
  &=\begin{bmatrix}
    \sum_{q\neq 1}\sum_{i=1}^\infty\sum_{j=1}^\infty \omega_{1qij}\lan z_q,\phi_{qi}\ran\phi_{1j}\\
    \vdots\\
    \sum_{q\neq P}\sum_{i=1}^\infty\sum_{j=1}^\infty \omega_{Pqij}\lan z_q,\phi_{qi}\ran\phi_{Pj}
  \end{bmatrix}\\
  &=\sum_{i=1}^\infty\sum_{j=1}^\infty
  \begin{bmatrix}
    \sum_{q\neq 1}\omega_{1qij}\lan z_q,\phi_{qi}\ran\phi_{1j}\\
    \vdots\\
    \sum_{q\neq P}\omega_{Pqij}\lan z_q,\phi_{qi}\ran\phi_{Pj}
  \end{bmatrix}\\
  &=\sum_{i=1}^\infty\sum_{j=1}^\infty\sum_{q\neq 1}\omega_{1qij}\lan z_q,\phi_{qi}\ran
  \begin{bmatrix}
    \phi_{1j}\\
    0\\
    \vdots\\
    0
  \end{bmatrix}
  +\cdots
  +\sum_{i=1}^\infty\sum_{j=1}^\infty\sum_{q\neq P}\omega_{Pqij}\lan z_q,\phi_{qi}\ran
  \begin{bmatrix}
    0\\
    \vdots\\
    0\\
    \phi_{Pj}
  \end{bmatrix}\\
  &=\sum_{p=1}^P\sum_{q\neq p}\sum_{i=1}^\infty\sum_{j=1}^\infty \omega_{pqij}\lan z_q,\phi_{qi}\ran\bs{e}_{pj}\\
  &=\sum_{p=1}^P\sum_{q\neq p}\sum_{j=1}^\infty\sum_{i=1}^\infty \omega_{pqij}\lan \bs{z},\bs{e}_{qi}\ran_P\bs{e}_{pj}.
\ea
We define the operator $\tilde{\mac{R}}$ on the closed span $\overline{\Span\{\bs{e}_{pj},p\in[P],j\ge1\}}$ or for short $\overline{\Span\{\bs{e}_{pj}\}}$ by
\ba
  \tilde{\mac{R}}
  &:=\sum_{p=1}^P\sum_{q\neq p}\sum_{j=1}^\infty\sum_{i=1}^\infty \omega_{pqij}\bs{e}_{qi}\otimes_P\bs{e}_{pj}.
\ea

Note that $\{\bs{e}_{pj}\}_{p\in[P],j\ge1}$ is a CONS of $\overline{\Span\{\bs{e}_{pj}\}}$.
Then, we have
\ba
  \|\tilde{\mac{R}}\|_{HS}^2
  &=\sum_{k=1}^P\sum_{l=1}^\infty\|\tilde{\mac{R}}\bs{e}_{kl}\|_P^2\\
  &=\sum_{k=1}^P\sum_{l=1}^\infty\left\|\sum_{p=1}^P\sum_{q\neq p}\sum_{i=1}^\infty\sum_{j=1}^\infty \omega_{pqij}\lan\bs{e}_{kl},\bs{e}_{qi}\ran_P\bs{e}_{pj}\right\|_P^2\\
  &=\sum_{k=1}^P\sum_{l=1}^\infty\lan\sum_{p,q,i,j}\omega_{pqij}\lan\bs{e}_{kl},\bs{e}_{qi}\ran_P\bs{e}_{pj},
  \sum_{p',q',i',j'}\omega_{p'q'i'j'}\lan\bs{e}_{kl},\bs{e}_{q'i'}\ran_P\bs{e}_{p'j'}\ran_P\\
  &=\sum_{k=1}^P\sum_{l=1}^\infty\lan\sum_{(p,k):p\neq k}\sum_{j=1}^\infty \omega_{pklj}\bs{e}_{pj},
  \sum_{(p',k):p'\neq k}\sum_{j'=1}^\infty \omega_{p'klj'}\bs{e}_{p'j'}\ran_P\\
  &=\sum_{k=1}^P\sum_{l=1}^\infty\sum_{(p,k):p\neq k}\sum_{j=1}^\infty\lan \omega_{pklj}\bs{e}_{pj},
  \omega_{pklj}\bs{e}_{pj}\ran_P\\
  &=\sum_{(p,k):p\neq k}\sum_{l=1}^\infty\sum_{j=1}^\infty \omega_{pklj}^2.
\ea
Therefore, we conclude that $\tilde{\mac{R}}$ is a Hilbert--Schmidt operator on $\overline{\Span\{\bs{e}_{pj}\}}$ if the series $\sum_{(p,k):p\neq k}\sum_{l=1}^\infty\sum_{j=1}^\infty \omega_{pklj}^2$ converges, which is equivalent to Condition~\ref{cond:cond1}.
We extend $\dom(\tilde{\mac{R}})$ to $L_2^P$ such that $\tilde{\mac{R}}|_{\overline{\Span\{\bs{e}_{pj}\}}^\perp}=0$, thus ensuring that $\tilde{\mac{R}}$ is a Hilbert--Schmidt operator on $L_2^P$ with $\tilde{\mac{R}}|_{F_W}=\mac{R}$.

Finally, the symmetry of $\omega_{pqij}$ implies that $\tilde{\mac{R}}$ is self-adjoint.
This can be seen as follows.
First, by the definition of $\omega_{pqij}$,
\ba
  \omega_{pqij}
  =\zeta_{qi}^{-\frac{1}{2}}\zeta_{pj}^{-\frac{1}{2}}\mbb{E}[\xi_{qi}\xi_{pj}]
  =\zeta_{pj}^{-\frac{1}{2}}\zeta_{qi}^{-\frac{1}{2}}\mbb{E}[\xi_{pj}\xi_{qi}]
  =\omega_{qpji}.
\ea
Moreover, $(\bs{e}_{qi}\otimes_P\bs{e}_{pj})^*=\bs{e}_{pj}\otimes_P\bs{e}_{qi}$.
Therefore, by interchanging the indices $(p,q,i,j)$ and $(q,p,j,i)$ in the series representation of $\tilde{\mac{R}}$, we obtain $\tilde{\mac{R}}^*=\tilde{\mac{R}}$.
Hence, $\tilde{\mac{R}}$ is self-adjoint.

\subsection{Proof of Proposition \ref{prop:beta_minimization}}

First, we show that under Condition \ref{cond:cond3}, $\beta_p=\mac{V}_{pp}^{-1}\tilde{\mac{V}}_p\bs{\gamma}$ is well-defined.
Observe that $\mac{V}_{pp}^{-1}\tilde{\mac{V}}_p\bs{\gamma}=\mac{V}_{pp}^{-1}\sum_{q=1}^P\mac{V}_{pq}\gamma_q$.
Therefore, it suffices to show that $\mac{V}_{pp}^{-1}\mac{V}_{pq}\gamma_q\in L_2$ for each $q$.
For $q=p$, we have 
\ba
  \|\mac{V}_{pp}^{-1}\mac{V}_{pp}\gamma_p\|
  =\left\|\sum_{i=1}^\infty\lan\gamma_p,\phi_{pi}\ran\phi_{pi}\right\|
  \le\|\gamma_p\|
  <\infty.
\ea
On the other hand, for $q\neq p$, we have
\ba
 \|\mac{V}_{pp}^{-1}\mac{V}_{pq}\gamma_q\|^2
 &=\lan\sum_{i=1}^\infty\zeta_{pi}^{-1}\sum_{j=1}^\infty\mbb{E}[\xi_{pi}\xi_{qj}]\lan \gamma_q,\phi_{qj}\ran\phi_{pi},
 \sum_{i'=1}^\infty\zeta_{pi'}^{-1}\sum_{j'=1}^\infty\mbb{E}[\xi_{pi'}\xi_{qj'}]\lan \gamma_q,\phi_{qj'}\ran\phi_{pi'}\ran\\
 &=\sum_{i=1}^\infty\zeta_{pi}^{-2}\sum_{j=1}^\infty\sum_{j'=1}^\infty\mbb{E}[\xi_{pi}\xi_{qj}]\mbb{E}[\xi_{pi}\xi_{qj'}]\lan \gamma_q,\phi_{qj}\ran\lan \gamma_q,\phi_{qj'}\ran\\
 &=\sum_{i=1}^\infty\zeta_{pi}^{-2}\left(
  \sum_{j=1}^\infty\mbb{E}[\xi_{pi}\xi_{qj}]\lan \gamma_q,\phi_{qj}\ran
 \right)^2\\
 &\le\sum_{i=1}^\infty\zeta_{pi}^{-2}\left\{
  \left(\sum_{j=1}^\infty(\mbb{E}[\xi_{pi}\xi_{qj}])^2\right)^{\frac{1}{2}}
  \left(\sum_{j'=1}^\infty\lan \gamma_q,\phi_{qj'}\ran^2\right)^{\frac{1}{2}}
 \right\}^2\\
 &\le\sum_{i=1}^\infty\zeta_{pi}^{-2}\left(\sum_{j=1}^\infty(\mbb{E}[\xi_{pi}\xi_{qj}])^2\right)\|\gamma_q\|^2\\
 &=\|\gamma_q\|^2\sum_{i=1}^\infty\sum_{j=1}^\infty\frac{(\mbb{E}[\xi_{pi}\xi_{qj}])^2}{\zeta_{pi}^{2}}.
\ea
Thus, given $\bs{\gamma}$, under Condition \ref{cond:cond3}, we have $\mac{V}_{pp}^{-1}\mac{V}_{pq}\gamma_q\in L_2$ for each $q$.

Define $\tilde{\beta}_p=\mac{V}_{pp}^{-1}\tilde{\mac{V}}_{p}\bs{\gamma}$.
Next, we show that $\tilde{\beta}_p$ minimizes the criterion $Q(\beta_p)$ given $\bs{\gamma}$.
Now, $\mac{V}_{pp}^{-1}\tilde{\mac{V}}_{p}\bs{\gamma}$ is well-defined, implying that $\|\mac{V}_{pp}^{-\frac{1}{2}}(\mac{V}_{pp}\beta_{p}-\tilde{\mac{V}}_{p}\bs{\gamma})\|^{2}<\infty$.
Thus, we have
\ba
 0
 &\le\left<\mac{V}_{pp}^{-\frac{1}{2}}(\mac{V}_{pp}\beta_{p}-\tilde{\mac{V}}_{p}\bs{\gamma}),
  \mac{V}_{pp}^{-\frac{1}{2}}(\mac{V}_{pp}\beta_{p}-\tilde{\mac{V}}_{p}\bs{\gamma})\right>
  =\lan\mac{V}_{pp}\beta_p-\tilde{\mac{V}}_p\bs{\gamma},
    \beta_p-\mac{V}_{pp}^{-1}\tilde{\mac{V}}_p\bs{\gamma}\ran\\
  &=\lan\mac{V}_{pp}\beta_p,\beta_p\ran
  -\lan\mac{V}_{pp}\beta_p,\mac{V}_{pp}^{-1}\tilde{\mac{V}}_p\bs{\gamma}\ran
  -\lan\tilde{\mac{V}}_p\bs{\gamma},\beta_p\ran
  +\lan\tilde{\mac{V}}_p\bs{\gamma},\mac{V}_{pp}^{-1}\tilde{\mac{V}}_p\bs{\gamma}\ran\\
  &=Q(\beta_p)
  +\lan\tilde{\mac{V}}_p\bs{\gamma},\mac{V}_{pp}^{-1}\tilde{\mac{V}}_p\bs{\gamma}\ran.
\ea
Therefore, $Q(\beta_{p})$ has the following lower bound
\begin{equation*}
  Q(\beta_{p})\ge
  -\left<\mac{V}_{pp}^{-1}\tilde{\mac{V}}_{p}\bs{\gamma},\tilde{\mac{V}}_{p}\bs{\gamma}\right>.
\end{equation*}
This lower bound is attained by choosing $\beta_{p}=\mac{V}_{pp}^{-1}\tilde{\mac{V}}_{p}\bs{\gamma}$.
Thus, $\tilde{\beta}_{p}$ defined above is a solution of the minimization problem over $\beta_{p}$.

\end{document}